\documentclass[a4paper,11pt]{article}
\pdfoutput=1 

\usepackage{jheppub} 

\usepackage[T1]{fontenc} 
\usepackage{diagbox}
\usepackage{multirow}
\usepackage{slashed}

\usepackage{graphicx}

\usepackage{longtable,pdflscape}
\usepackage{stackengine}
\usepackage[export]{adjustbox}

\usepackage{xcolor,colortbl}

\definecolor{Gray}{gray}{0.95}
\definecolor{RGray}{gray}{0.85}
\definecolor{CGray}{gray}{0.92}

\allowdisplaybreaks

\def\be{\begin{equation}}
\def\ee{\end{equation}}
\def\beq{\begin{eqnarray}}
\def\eeq{\end{eqnarray}}

\def\ba{\begin{array}{c}}
\def\bat{\begin{array}{cc}}
\def\ea{\end{array}}

\def\nn{\nonumber}

\def\gev{{\rm GeV}}

\usepackage{tikz}
\usetikzlibrary{calc}
\usetikzlibrary{trees}
\usetikzlibrary{decorations.pathmorphing}
\usetikzlibrary{decorations.markings}
\usetikzlibrary{arrows.meta}

\tikzset{
    vector/.style={decorate, decoration={snake}, draw},
	provector/.style={decorate, decoration={snake,amplitude=2.5pt}, draw},
	antivector/.style={decorate, decoration={snake,amplitude=-2.5pt}, draw},
    fermion/.style={draw=black, postaction={decorate},
        decoration={markings,mark=at position .5 with {\arrow[draw=black]{>}}}},
    fermionbar/.style={draw=black, postaction={decorate},
        decoration={markings,mark=at position .5 with {\arrow[draw=black]{<}}}},
    fermionnoarrow/.style={draw=black},
    gluon/.style={decorate, draw=black,
        decoration={coil,amplitude=4pt, segment length=5pt}},
    scalar/.style={dashed,draw=black, postaction={decorate},
        decoration={markings,mark=at position .5 with {\arrow[draw=black]{>}}}},
    scalarbar/.style={dashed,draw=black, postaction={decorate},
        decoration={markings,mark=at position .5 with {\arrow[draw=black]{<}}}},
    scalarnoarrow/.style={dashed,draw=black},
    electron/.style={draw=black, postaction={decorate},
        decoration={markings,mark=at position .5 with {\arrow[draw=black]{>}}}},
	bigvector/.style={decorate, decoration={snake,amplitude=4pt}, draw},
}

\title{\bf \boldmath Rare top-quark decays $t \to cg(g)$ in the aligned two-Higgs-doublet model}

\author[a]{Fang-Min Cai,}
\author[a]{Shuichiro Funatsu,}
\author[a,1]{Xin-Qiang Li\note{Corresponding author.},}
\author[a,b]{and Ya-Dong Yang}

\affiliation[a]{Institute of Particle Physics and Key Laboratory of Quark and Lepton Physics~(MOE), Central China Normal University, Wuhan, Hubei 430079, China}
\affiliation[b]{School of Physics and Microelectronics, Zhengzhou University, Zhengzhou, Henan 450001, China}

\emailAdd{caifangmin@mails.ccnu.edu.cn}
\emailAdd{funatsu@mail.ccnu.edu.cn}
\emailAdd{xqli@mail.ccnu.edu.cn}
\emailAdd{yangyd@mail.ccnu.edu.cn}

\abstract{We update the Standard Model (SM) predictions for the branching ratios of the rare top-quark decays $t \to cg(g)$, and evaluate the maximum rates that can be reached in the aligned as well as in the four conventional two-Higgs-doublet models (2HDMs) with $\mathcal{Z}_2$ symmetries. Taking into account the relevant constraints on the model parameters resulting from a global fit obtained at the $95.5\%$ confidence level, we find that the branching ratios of $t \to cg$ and $t \to cgg$ decays can reach up to $3.36\times 10^{-9}$ and $2.95\times 10^{-9}$ respectively, being therefore of the same order, in the aligned 2HDM (A2HDM). This is obviously different from the SM case, where the predicted branching ratio of the three-body decay $t \to cgg$ is about two orders of magnitude larger than that of the two-body decay $t \to cg$. On the other hand, compared with the SM predictions, no significant enhancements are observed in the four conventional 2HDMs with $\mathcal{Z}_2$ symmetries for the branching ratios of these two decays. Nevertheless, the predicted branching ratios of $t \to cg$ and $t \to cgg$ decays in the A2HDM are still out of the expected sensitivities of the future high-luminosity Large Hadron Collider and the Future Circular Collider in hadron-hadron mode.} 

\begin{document}
\maketitle
\flushbottom

\section{Introduction}

The top quark, as the heaviest elementary particle in the Standard Model (SM), can decay into other light quarks associated with gauge or scalar bosons. Among all the possible decay modes, the flavour-changing neutral-current (FCNC) decays of the top quark, which are absent at tree level and suppressed at loop level by the Glashow-Iliopoulos-Maiani (GIM) mechanism~\cite{Glashow:1970gm} in the SM, are very sensitive to contributions from new physics (NP)~\cite{Eilam:1990zc,Deshpande:1991pn,Hou:1991un,Luke:1993cy,Atwood:1996vj,Bejar:2000ub,Diaz:2001vj,Iltan:2001yt,Arhrib:2005nx,Bar-Shalom:2005ldb,Abbas:2015cua,Hou:2020ciy,Balaji:2020qjg,Li:1993mg,deDivitiis:1997sh,Guasch:1999jp,Eilam:2001dh,Liu:2004qw,Frank:2005vd,Eilam:2006rb,Cao:2007dk,Heng:2009wr,Cao:2014udj,Dedes:2014asa,Eilam:1989zm,AguilarSaavedra:2002ns,Han:2011xd,Gao:2013fxa,Dey:2016cve,Diaz-Furlong:2016ril,Chiang:2018oyd,Altmannshofer:2019ogm,Liu:2021crr,AguilarSaavedra:2004wm,TopQuarkWorkingGroup:2013hxj}. For the two-body FCNC top-quark decays $t\to qV$ and $t\to qh$, with $q=c,u$, $V=\gamma,Z,g$, and $h$ the SM Higgs, their branching ratios are predicted to be of the orders of $10^{-17}$ to $10^{-12}$ within the SM, but could be enhanced by several orders of magnitude in many NP scenarios (see, \textit{e.g.}, Refs.~\cite{AguilarSaavedra:2004wm,TopQuarkWorkingGroup:2013hxj} and references therein). Among the three-body loop-induced top-quark decays $t \to cVV$ and $t \to u_1 \bar{u}_2 u_2$ with $u_i=u,c$, the decay $t\to cgg$ has the largest branching ratio, reaching up to about $1.02\times 10^{-9}$ within the SM~\cite{Diaz-Cruz:1999wcs,Cordero-Cid:2004pco,Eilam:2006uh}. Interestingly, it is found that the branching ratio of the three-body decay $t \to cgg$ is about two orders of magnitude larger than that of the two-body decay $t \to cg$ within the SM~\cite{Eilam:2006uh}, a phenomenon that will be dubbed ``higher-order dominance'' and has also been revealed in the bottom- and charm-quark decays~\cite{Hou:1988wt,Liu:1989pc,Simma:1990nr,Hou:1987vd, Hou:1990js,Greub:2000sy,Greub:1996wn}. Although their SM rates are predicted to be far below the detectable level of present and future colliders, these rare FCNC top-quark decays have been widely studied in various NP models, and significant enhancements are observed relative to the SM expectations; the most prominent examples are the two-Higgs-doublet model (2HDM)~\cite{Eilam:1990zc,Deshpande:1991pn,Hou:1991un,Luke:1993cy,Atwood:1996vj,Bejar:2000ub,Diaz:2001vj,Iltan:2001yt,Arhrib:2005nx,Bar-Shalom:2005ldb,Abbas:2015cua,Hou:2020ciy,Balaji:2020qjg} and the supersymmetry~\cite{Li:1993mg,deDivitiis:1997sh,Guasch:1999jp,Eilam:2001dh,Liu:2004qw,Frank:2005vd,Eilam:2006rb,Cao:2007dk,Heng:2009wr,Cao:2014udj,Dedes:2014asa}, within which these rare top-quark decays might be even detectable at the future colliders. Especially, it is interesting to check if the ``higher-order dominance'' phenomenon observed in $t\to cg(g)$ decays within the SM~\cite{Eilam:2006uh} is still valid for a viable NP model. 

Among the various NP scenarios, the 2HDM~\cite{Lee:1973iz} provides a minimal extension of the SM scalar sector by adding a second scalar doublet with hypercharge $Y=1/2$, and can naturally satisfy the electroweak precision tests, giving rise at the same time to a very rich phenomenology~\cite{Gunion:1989we,Branco:2011iw}. In a generic 2HDM, the non-diagonal couplings of the neutral scalars to the SM fermions would unavoidably lead to non-vanishing tree-level FCNC interactions. To avoid these unwanted tree-level FCNC couplings, an \textit{ad-hoc} discrete $\mathcal{Z}_2$ symmetry is often imposed on the Yukawa sector of the model~\cite{Glashow:1976nt}. Depending on the $\mathcal{Z}_2$ charge assignments to the scalars and fermions, this results in four types of 2HDMs (types I, II, X, Y)~\cite{Branco:2011iw,Gunion:1989we} under the hypothesis of natural flavour conservation~\cite{Glashow:1976nt}. In the aligned 2HDM (A2HDM)~\cite{Pich:2009sp}, on the other hand, the absence of tree-level FCNC interactions is automatically guaranteed by assuming the alignment in flavour space of the two Yukawa matrices for each type of the right-handed fermions. Interestingly, the A2HDM can recover, as particular cases, all the known specific implementations of the conventional 2HDMs based on $\mathcal{Z}_2$ symmetries. The model is also featured by possible new sources of CP violation beyond that of the Cabibbo-Kobayashi-Maskawa (CKM) quark-mixing matrix~\cite{Cabibbo:1963yz,Kobayashi:1973fv}. Thus, the A2HDM has been widely studied from the various theoretical points of view~\cite{Ferreira:2010xe,Bijnens:2011gd,Ferreira:2015rha,Botella:2015yfa,Penuelas:2017ikk,Gori:2017qwg}, as well as for its phenomenologies in the low-energy flavour physics~\cite{Jung:2010ik,Jung:2010ab,Jung:2012vu,Chang:2015rva,Cho:2017jym,Cree:2011uy,Celis:2012dk,Li:2014fea,Enomoto:2015wbn,Wang:2016ggf,Hu:2016gpe,Hu:2017qxj,Arnan:2017lxi,DelleRose:2019ukt,Wang:2020kov,Eberhardt:2020dat}, the electric dipole moments of the leptons, the neutron and various atoms~\cite{Dekens:2014jka,Jung:2013hka,Gisbert:2019ftm,Enomoto:2021dkl}, the anomalous magnetic moment of the muon~\cite{Han:2015yys,Ilisie:2015tra,Han:2016bvl,Cherchiglia:2016eui,Cherchiglia:2017uwv,DelleRose:2020oaa}, and the high-energy collider physics~\cite{Carena:2012rw,Altmannshofer:2012ar,Bai:2012ex,Barger:2013ofa,Lopez-Val:2013yba,Wang:2013sha,Celis:2013rcs,Celis:2013ixa,Duarte:2013zfa,Ilisie:2014hea,Abbas:2015cua,Berge:2015nua,Wang:2016rvz,Ayala:2016djv,Abbas:2018pfp,Kanemura:2021atq,Ilisie:2021jek}. Recently, assuming that there are no new sources of CP violation beyond the CKM quark-mixing matrix, and taking into account the most constraining processes and observables mentioned above, together with the theoretical requirements of perturbativity and positivity of the scalar potential, a detailed global fit has been performed to the A2HDM~\cite{Eberhardt:2020dat}.  

In this paper, we shall perform a detailed study of the rare top-quark decay $t \to cgg$ and its comparison with the two-body decay $t \to cg$ in the A2HDM as well as in the four conventional 2HDMs with $\mathcal{Z}_2$ symmetries. Taking into account the relevant constraints on the model parameters from a global fit obtained at the $95.5\%$ confidence level (CL)~\cite{Eberhardt:2020dat}, we find that the branching ratios of $t \to cg$ and $t \to cgg$ decays can reach up to $3.36\times 10^{-9}$ and $2.95\times 10^{-9}$ respectively, being therefore of the same order, in the A2HDM. This is obviously different from the SM case, where the predicted branching ratio of $t \to cgg$ is about two orders of magnitude larger than that of $t \to cg$~\cite{Eilam:2006uh}. On the other hand, compared with the SM predictions, no significant enhancements are observed in the four conventional 2HDMs with $\mathcal{Z}_2$ symmetries for the branching ratios of these two decays. Searches for the decay $t\to cg$ have been performed at the Large Hadron Collider (LHC), with the upper limit on its branching ratio being of $\mathcal{O}(10^{-4})$~\cite{ATLAS:2015iqc,CMS:2016uzc}, which is still much higher than the model predictions. These rare FCNC decays will also be studied at the future experiments, such as the high-luminosity LHC (HL-LHC)~\cite{Azzi:2019yne,Cepeda:2019klc,CidVidal:2018eel,Cerri:2018ypt,Atlas:2019qfx} and the Future Circular Collider (FCC)~\cite{Abada:2019lih,Abada:2019zxq,Benedikt:2018csr,Abada:2019ono}. For example, the $95\%$-CL limit on the branching ratio of $t \to cg$ at the $100$-$\mathrm{TeV}$ FCC in hadron-hadron mode (FCC-hh) with an integrated luminosity of $10~\text{ab}^{-1}$ is estimated to be of $\mathcal{O}(10^{-7}-10^{-8})$~\cite{Oyulmaz:2019jqr,Khanpour:2019qnw}, which is at least one order of magnitude higher than the maximum value predicted in the A2HDM. As a consequence, the predicted branching ratios of $t \to cg$ and $t \to cgg$ decays in the A2HDM are still out of the expected sensitivities of the future HL-LHC and FCC-hh.

This paper is organized as follows. In Section~\ref{sec:A2HDM}, we recapitulate the A2HDM, focusing only on the scalar and Yukawa sectors that are most relevant to us. In Section~\ref{sec:calculations}, we present our calculation of the decay rates of $t \to cg(g)$ decays, with the resulting $R$ functions present in the $t\to cgg$ decay supplemented in Appendix~\ref{sec:appendix}. Our numerical results are then given in Section~\ref{sec:numerical}, where we show the maximum branching ratios of $t\to cg(g)$ decays that can be reached in the A2HDM, and give also a simple analysis of these two decays in the four conventional 2HDMs with $\mathcal{Z}_2$ symmetries. Our conclusion is finally made in Section~\ref{sec:conclusions}. 

\section{Aligned two-Higgs-doublet model}
\label{sec:A2HDM} 

As a simple extension of the SM, the 2HDM is invariant under the SM gauge group and includes, besides the SM matter and gauge fields, two complex scalar $SU(2)_L$ doublets,
\begin{equation}
	\phi_{a}=\mathrm{e}^{i \theta_{a}}\left[\begin{array}{c}
		\phi_{a}^{+} \\[0.15cm]
		\frac{1}{\sqrt{2}}\left(v_{a}+\rho_{a}+i \eta_{a}\right)
	\end{array}\right]\,,
\end{equation}
with the weak hypercharge $Y=1/2$. Here we have assumed that the vacuum of the theory respects the electromagnetic gauge symmetry, which guarantees the vacuum expectation values (vevs) of the two scalar fields $\phi_{a}$ to be aligned in the $SU(2)_L$ space. After performing further a global $SU(2)_L$ transformation to put the non-zero vevs only in the lower (neutral) components of $\phi_{a}$, which guarantees the vacuum to be neutral and not charge-breaking, we can bring the vevs of the two scalar fields to the forms given by $\langle 0\vert \phi_a^T(x)\vert 0 \rangle=(0,v_a e^{i \theta_a}/\sqrt{2})$, where $v_1$ and $v_2$ are real and non-negative~\cite{Ginzburg:2004vp,Davidson:2005cw}. However, as one global phase can always be rotated away through an appropriate $U(1)_Y$ transformation, we can, without loss of generality, fix $\theta_1 = 0$ and leave the relative phase $\theta=\theta_2-\theta_1$~\cite{Ginzburg:2004vp,Davidson:2005cw,Pich:2009sp}.

Performing further a global $SU(2)$ transformation in the scalar space $(\phi_1,\phi_2)$, we can rotate the original scalar basis to the so-called Higgs basis~\cite{Davidson:2005cw,Haber:2006ue,Haber:2010bw,Botella:1994cs},
\begin{equation}
	\left( \begin{array}{c} \Phi_1 \\[0.1cm] -\Phi_2 \end{array} \right) \equiv
	\left( \begin{array}{cc} \cos\beta & \sin\beta \\[0.15cm] \sin\beta & -\cos\beta \end{array} \right)\,
	\left( \begin{array}{c} \phi_1 \\[0.1cm]  \mathrm{e}^{-i\theta}\phi_2 \end{array} \right)\; ,
\end{equation}
where the rotation angle $\beta$ is defined by the ratio of the absolute values of the two neutral scalar field vevs, $\tan\beta \equiv \frac{\left|\left\langle 0|\phi_{2}^{0}\right |0 \rangle\right|}{\left|\left\langle 0|\phi_{1}^{0}\right |0 \rangle\right|}=\frac{v_2}{v_1}$, and, by convention, its value is limited to the first quadrant due to $v_1,v_2\geq0$. In fact, both $\tan\beta$ and the relative phase $\theta$ are basis-dependent, and we can remove them entirely by transforming from a generic basis $(\phi_1,\phi_2)$ to the Higgs basis $(\Phi_1,\Phi_2)$~\cite{Davidson:2005cw,Haber:2006ue,Haber:2010bw,Botella:1994cs}. The true significance of $\tan\beta$ emerges only in specialized versions of the 2HDM, such as the four conventional 2HDMs with $\mathcal{Z}_2$ symmetries, where $\tan\beta$ is promoted to a physical parameter. In this case, we can remove the relative phase $\theta$ by making a phase rotation $\phi_2\to\mathrm{e}^{-i\theta}\phi_2$, such that the two neutral scalar vevs are real and the CP invariance is not spontaneously broken~\cite{Davidson:2005cw,Haber:2006ue,Haber:2010bw,Botella:1994cs}. In the Higgs basis, only the scalar doublet $\Phi_1$ gets a non-zero vev, $\langle 0|\Phi_1^T\left(x\right)|0\rangle = (0, v/\sqrt{2})$, with $v=\sqrt{v_1^2+v_2^2}=1/\sqrt{\sqrt{2}G_F}\simeq 246.22\,\gev$, and plays the role of the SM Higgs doublet. The two scalar doublets in the Higgs basis can now be parametrized, respectively, as~\cite{Pich:2009sp}
\begin{equation} \label{eq:Higgsbasis}
	\Phi_1=\left[ \begin{array}{c} G^+ \\[0.15cm] \frac{1}{\sqrt{2}}\, (v+S_1+iG^0) \end{array} \right] \; ,
	\qquad
	\Phi_2 = \left[ \begin{array}{c} H^+ \\[0.15cm] \frac{1}{\sqrt{2}}\, (S_2+iS_3) \end{array} \right] \; ,
\end{equation}
where $G^\pm$ and $G^0$ denote the electroweak Goldstone fields, and the remaining five physical degrees of freedom are given by the two charged fields $H^\pm(x)$ and the three neutral ones $\varphi^0_i(x) =\{h(x), H(x), A(x)\}=\mathcal{R}_{ij}S_j$, where $\mathcal{R}$ is the orthogonal matrix needed for diagonalizing the mass terms in the scalar potential~\cite{Celis:2013rcs}. Generally, none of these three neutral scalars can have a definite CP quantum number. Here, for simplicity, we shall assume a CP-conserving Higgs sector, as will be detailed in the next two subsections.

\subsection{Scalar sector}

In the Higgs basis, the most general scalar potential allowed by the electroweak gauge symmetry $SU(2)_L\otimes U(1)_Y$ takes the form~\cite{Davidson:2005cw,Haber:2006ue,Haber:2010bw}:
\begin{align} \label{eq:potential}
	V &= \mu_{1}\,\left(\Phi_{1}^{\dagger}\Phi_{1}\right)\,+\,\mu_{2}\,\left(\Phi_{2}^{\dagger}\Phi_{2}\right)\,
	+\,\left[\mu_{3}\,\left(\Phi_{1}^{\dagger}\Phi_{2}\right)\,+\,\mu_{3}^{*}\,\left(\Phi_{2}^{\dagger}\Phi_{1}
	\right)\right]\nonumber\\[0.15cm]
	& +\lambda_{1}\,\left(\Phi_{1}^{\dagger}\Phi_{1}\right)^{2}\,+\,\lambda_{2}\,\left(\Phi_{2}^{\dagger}\Phi_{2}\right)^{2}\,
	+\,\lambda_{3}\,\left(\Phi_{1}^{\dagger}\Phi_{1}\right)\left(\Phi_{2}^{\dagger}\Phi_{2}\right)\,
	+\,\lambda_{4}\,\left(\Phi_{1}^{\dagger}\Phi_{2}\right)\left(\Phi_{2}^{\dagger}\Phi_{1}\right)\nonumber\\[0.15cm]
	& +\left[\left(\lambda_{5}\,\Phi_{1}^{\dagger}\Phi_{2}\,+\,\lambda_{6}\,\Phi_{1}^{\dagger}\Phi_{1}\,+\,\lambda_{7}\,
	\Phi_{2}^{\dagger}\Phi_{2}\right)\left(\Phi_{1}^{\dagger}\Phi_{2}\right)\,+\,\mathrm{h.c.}\right]\,.
\end{align}
Due to Hermiticity of the scalar potential, the parameters $\mu_{1,2}$ and $\lambda_{1,2,3,4}$ are real, while $\mu_3$ and $\lambda_{5,6,7}$ could be generally complex. The minimization condition of the scalar potential imposes the relations 
\begin{align} \label{eq:minimizationcondition}
	\mu_1= -\lambda_1\,v^2,\qquad \mu_3 = -\frac{1}{2}\,\lambda_6\,v^2.
\end{align}

Inserting Eq.~\eqref{eq:Higgsbasis} into Eq.~\eqref{eq:potential} and imposing the minimization condition given by Eq.~\eqref{eq:minimizationcondition}, we can get the charged-Higgs mass
\begin{align}
	m_{H^{\pm}}^2= \mu_2+\frac{1}{2}\lambda_3\,v^2,
\end{align}
while the masses of the three neutral scalars can be obtained after diagonalizing the mass-squared matrix $\mathcal{M}$ of the three neutral-scalar fields $S_{1,2,3}$ by the orthogonal matrix $\mathcal{R}$, \textit{i.e.}, $\mathrm{diag}\left( m_h^2, m_H^2,m_A^2\right)=\mathcal{R}\,\mathcal{M}\,\mathcal{R}^T$~\cite{Celis:2013rcs}. In the CP-conserving limit of the scalar potential, the parameters $\lambda_{5,6,7}$ are all real and the three neutral scalars $h$, $H$, and $A$ are therefore CP eigenstates. In this case, the CP-odd scalar $A$ corresponds to the field $S_3$, while the two CP-even scalars $h$ and $H$ are the orthogonal combinations of $S_1$ and $S_2$:
\begin{align}
	\left(\begin{array}{c} h\\[0.1cm] H \end{array} \right)\; = \;
	\left(\begin{array}{cc} \cos{\tilde\alpha} & \sin{\tilde\alpha} \\[0.15cm] -\sin{\tilde\alpha} & \cos{\tilde\alpha} \end{array}\right)\;
	\left(\begin{array}{c} S_1\\[0.1cm] S_2 \end{array}\right) \,,
\end{align}
where the mixing angle $\tilde\alpha$ is determined by
\begin{equation}\label{eq:tanalphatilde}
	\tan{\tilde\alpha}\; =\; \frac{m_h^2 - 2\lambda_1 v^2}{v^2\lambda_6}
	\; =\; \frac{v^2\lambda_6}{2\lambda_1 v^2- m_H^2}\,.
\end{equation}
Note that one can always restrict $\tilde\alpha$ within the range $0 \leqslant \tilde \alpha < \pi$ by performing a phase redefinition of the two CP-even fields $S_1$ and $S_2$. In addition, the masses of $h$ and $H$ satisfy the relation $m_H^2\sin^2{\tilde\alpha}+m_h^2\cos^2{\tilde\alpha}=2\lambda_1 v^2$~\cite{Celis:2013ixa}, which, together with our convention $m_h \leq m_H$, implies that $2\lambda_1 v^2 - m_H^2\leq0$ and $m_h^2-2\lambda_1 v^2\leq0$. Then, from Eq.~\eqref{eq:tanalphatilde}, one can see that the sign of $\tan{\tilde\alpha}$ is uniquely determined by that of the parameter $\lambda_6$, since the vev $v=\sqrt{v_1^2+v_2^2}$ is, by definition, always positive. As the Higgs basis of a CP-conserving 2HDM is defined up to a global rephasing of the second scalar doublet $\Phi_2 \to \mathrm{e}^{i\eta} \Phi_2$, we can use this freedom to change the sign of $\lambda_6$~\cite{ONeil:2009fty}. Here, without loss of generality, we shall use the rephasing $\Phi_2 \to - \Phi_2$ to fix $\lambda_6 \leqslant 0$ so that $\tan{\tilde\alpha}\geq0$, which, together with $0 \leqslant \tilde \alpha < \pi$, results in our final choice $0 \leqslant \tilde \alpha \leqslant \pi/2$. The masses of the three neutral scalars are finally given, respectively, by
\begin{align}
	m_{h}^2= \frac{1}{2}(\Sigma-\Delta),\qquad m_{H}^2= \frac{1}{2}(\Sigma+\Delta),\qquad m_{A}^2= m_{H^{\pm}}^2+v^2 \left(\frac{\lambda_4}{2}-\lambda_5\right),
\end{align}
with
\begin{eqnarray}
	\Sigma=m_{H^{\pm}}^2+v^2\,\left(2\lambda_2+\frac{\lambda_4}{2}-\lambda_5\right),\\[0.2cm]
	\Delta=\sqrt{\left[m_{A}^2+2v^2\,(\lambda_5-\lambda_1)\right]^2+4v^4\,\lambda_6^{2}}.
\end{eqnarray}
Here the convention $m_h \leqslant m_H$ has been chosen, and the SM limit is recovered when $\tilde\alpha=0$.

\subsection{Yukawa sector}

The Yukawa Lagrangian of a 2HDM in the Higgs basis can be written as~\cite{Pich:2009sp,Branco:2011iw}
\begin{align}\label{eq:Yukawa_interaction}
	\mathcal{L}_{Y,\text{weak}} = & -\frac{\sqrt{2}}{v}\,\Big[\bar{Q}^\prime_L (M^\prime_d \Phi_1 + Y^\prime_d \Phi_2) d^\prime_R + \bar{Q}^\prime_L (M^\prime_u \tilde{\Phi}_1 + Y^\prime_u \tilde{\Phi}_2) u^\prime_R \nn\\[0.1cm]
	& \quad \quad\,\, + \bar{L}^\prime_L (M^\prime_\ell \Phi_1 + Y^\prime_\ell \Phi_2) \ell^\prime_R \Big] + \mathrm{h.c.}
\end{align}
Here, $Q_L^\prime$ and $L_L^\prime$ are the left-handed quark and lepton doublets, while $u^\prime_R$, $d^\prime_R$, $\ell^\prime_R$ denote the right-handed fermion singlets, all being given in the weak-interaction basis; $\tilde{\Phi}_a(x)\equiv i\sigma_2\Phi_a^{\ast}(x)$, with $\sigma_2$ the Pauli matrix, are the charge-conjugated fields with hypercharge $Y=-1/2$; $M^{\prime}_f$ ($f=u,d,\ell$) denote the non-diagonal matrices that encode both the fermion masses and the Yukawa couplings of the scalar doublet $\Phi_1$ to fermions, while $Y^{\prime}_f$ characterize only the Yukawa couplings of the second doublet $\Phi_2$ to fermions.

In general, the two matrices $M^{\prime}_f$ and $Y^{\prime}_f$ associated with the same type of right-handed fermion $u^\prime_R$, $d^\prime_R$, or $\ell^\prime_R$ in Eq.~\eqref{eq:Yukawa_interaction} cannot be simultaneously diagonalized in flavour space. As a consequence, in the fermion mass-eigenstate basis with diagonal mass matrices $M_f=U_L^{f\dagger}M^{\prime}_fU_R^f$, the corresponding Yukawa matrices $Y_f=U_L^{f\dagger}Y^{\prime}_fU_R^f$ remain still non-diagonal and hence give rise to non-vanishing tree-level FCNC interactions. These unwanted tree-level FCNC couplings can, however, be eliminated by requiring that the two matrices $M^{\prime}_f$ and $Y^{\prime}_f$ are aligned in flavour space~\cite{Pich:2009sp}. This results in the following alignment relations between the mass and the Yukawa matrices~\cite{Pich:2009sp}:
\begin{equation}\label{eq:alignment}
	Y_{d,\ell}=\varsigma_{d,\ell} \, M_{d,\ell},\qquad  Y_u=\varsigma_u^* \, M_u,
\end{equation}
where $\varsigma_f$ are arbitrary complex parameters and could introduce new sources of CP violation beyond that of the CKM quark-mixing matrix. They are also scalar-basis independent and satisfy universality among the three different generations~\cite{Pich:2009sp}.

In terms of the fermion mass eigenstates, $f_{L,R}=U_{L,R}^{f\dagger}f_{L,R}^{\prime}$, and making use of the alignment conditions specified by Eq.~\eqref{eq:alignment}, we can finally rewrite the Yukawa Lagrangian of the A2HDM as~\cite{Pich:2009sp}
\begin{align}
	\mathcal{L}_{Y,\text{mass}} =&-i \frac{G^0}{v} \Big\{\bar{d}_{L} M_{d} d_{R}+\bar{u}_{R} M_{u}^{\dagger} u_{L}+\bar{\ell}_{L} M_{\ell} \ell_{R}\Big\} \nn \\[0.1cm]
	& -\left(1+\frac{S_{1}}{v}\right) \Big\{\bar{u}_{L} M_{u} u_{R}+\bar{d}_{L} M_{d} d_{R}+\bar{\ell}_{L} M_{\ell} \ell_{R}\Big\} \nn \\[0.1cm]
	& -\frac{1}{v} \left(S_2+i S_3\right) \Big\{\bar{d}_{L} Y_{d} d_{R}+\bar{u}_{R} Y_{u}^{\dagger} u_{L}+\bar{\ell}_{L} Y_{\ell} \ell_{R}\Big\}\nn \\[0.1cm]
	& - \frac{\sqrt{2}}{v} G^{+} \Big\{\bar{u}_{L} V_{\mathrm{CKM}} M_{d} d_{R} -\bar{u}_{R} M_{u}^{\dagger} V_{\mathrm{CKM}} d_{L} 
	+\bar{\nu}_{L} M_{\ell} \ell_{R} \Big\}\nn \\[0.1cm]
	& - \frac{\sqrt{2}}{v} H^{+} \Big\{\varsigma_d\,\bar{u}_{L} V_{\mathrm{CKM}} M_{d} d_{R} - \varsigma_u\,\bar{u}_{R} M_{u}^{\dagger} V_{\mathrm{CKM}} d_{L} 
	+ \varsigma_\ell\,\bar{\nu}_{L} M_{\ell} \ell_{R}\Big\}+\mathrm{h.c.}\,,
\end{align}
where $V_{\text{CKM}}=U_L^{u\dagger}U_L^d$ is the usual CKM matrix~\cite{Cabibbo:1963yz,Kobayashi:1973fv}, and the diagonal fermion mass matrices are now given, respectively, by
\begin{align}
    M_{u}=\mathrm{diag}\left(m_{u}, m_{c}, m_{t}\right), \quad
    M_{d}=\mathrm{diag}\left(m_{d}, m_{s}, m_{b}\right), \quad M_{\ell}=\mathrm{diag}\left(m_{e}, m_{\mu}, m_{\tau}\right)\,.
\end{align}
For the rare FCNC decays $t\to cg(g)$ considered here, besides the SM part, only the interactions of the charged scalars $H^{\pm}$ with the quarks are involved, which means that only the alignment parameters $\varsigma_{u,d}$ and the charged-Higgs mass $m_{H^{\pm}}$ will be involved throughout this work.

\begin{table}[t]
	\let\oldarraystretch=\arraystretch
	\tabcolsep 0.2in
	\renewcommand*{\arraystretch}{1.3}
	\begin{center}
		\begin{tabular}{|c|c|c|c|}
			\hline \hline  \rowcolor{Gray}
			Model & $\varsigma_d$ & $\varsigma_u$ & $\varsigma_\ell$ \\
			\hline
			Type~I & $\cot{\beta}$ &$\cot{\beta}$ & $\cot{\beta}$ \\ \rowcolor{Gray}
			Type~II & $-\tan{\beta}$ & $\cot{\beta}$ & $-\tan{\beta}$ \\ 
			Type~X & $\cot{\beta}$ & $\cot{\beta}$ & $-\tan{\beta}$ \\ \rowcolor{Gray}
			Type~Y & $-\tan{\beta}$ & $\cot{\beta}$ & $\cot{\beta}$ \\ 
			\hline \hline
		\end{tabular}
		\caption{The one-to-one correspondence between the different choices of the alignment parameters $\varsigma_f$ and the four conventional 2HDMs based on discrete $\mathcal{Z}_2$ symmetries.}
		\label{tab:Z2symmetries}
	\end{center}
\end{table}

As shown in Table~\ref{tab:Z2symmetries}, the four conventional 2HDMs~\cite{Branco:2011iw,Gunion:1989we} based on discrete $\mathcal{Z}_2$ symmetries can be easily recovered from the A2HDM by choosing particular values of the alignment parameters $\varsigma_f$. As the parameter $\varsigma_\ell$ is not involved in the rare decays $t \to cg(g)$, we are actually left with only two cases, type-I~(type-X) 2HDM with $\varsigma_d=\varsigma_u=\cot{\beta}$, and type-II~(type-Y) 2HDM with $\varsigma_d=-1/\varsigma_u=-\tan{\beta}$. Numerical results of the branching ratios of $t \to cg(g)$ decays in these two specific cases as well as their comparisons with the SM predictions and the results obtained in the A2HDM will also be discussed in Section~\ref{sec:numerical}.

\section{Calculation of the decay rates}
\label{sec:calculations} 

\subsection{General remarks}
\label{sec:generalremarks}

The rare top-quark decays $t \to qg(g)$, with $q=u,c$, occur firstly at one-loop level within the SM. As such, the decay rates of these processes are not only suppressed by the loop factor, but also receive a strong CKM and GIM suppression relative to the dominant tree-level decay $t \to b W^+$~\cite{Eilam:1990zc,AguilarSaavedra:2002ns,Eilam:2006uh}. Due to the additional charged-Higgs contributions at the loop level, however, these decay rates could be enhanced in the A2HDM as well as in the conventional 2HDMs with $\mathcal{Z}_2$ symmetries~\cite{Eilam:1990zc,Deshpande:1991pn,Bejar:2000ub}. Here we shall focus on the decays into a charm instead of an up quark, because the decay rates of $t \to ug(g)$ are further suppressed by the ratio $|V_{ub}/V_{cb}|^2 \simeq 0.0087$ compared to that of $t \to cg(g)$, both within the SM as well as in the 2HDMs considered. 

Within the SM, due to the hierarchy of the CKM matrix elements, $|V_{tb}|\gg |V_{td}|, |V_{ts}|$, the total width of the top quark is dominated by the tree-level two-body decay $t \to b W^+$, and we can, therefore, take the approximation $\Gamma_\mathrm{tot}(t) \simeq \Gamma(t \to b W^+)$. Keeping only the dominant next-to-leading-order (NLO) QCD correction~\cite{Jezabek:1988iv,Czarnecki:1990kv,Li:1990qf} while neglecting the $W$-boson width effect~\cite{Jezabek:1988iv}, the NLO electroweak~\cite{Denner:1990ns,Eilam:1991iz}, as well as the next-to-next-to-leading-order QCD correction~\cite{Brucherseifer:2013iv,Gao:2012ja,Blokland:2004ye,Chetyrkin:1999ju,Czarnecki:1998qc}, all being only of a few percentages of the Born term~\cite{Gao:2012ja}, we can write the partial decay width of the channel $t \to b W^+$ as~\cite{Li:1990qf,Eilam:1991iz} 
\begin{eqnarray}\label{eq:t2bWNLO}
\Gamma(t \to b W^+) & = & \Gamma_0(t \to b W^+) \Biggl\{ 1+\frac{C_{F}\alpha_s}{2\pi}\left[\,2\frac{(1-\beta_W^2)(2\beta_W^2-1)(\beta_W^2-2)}
{\beta_W^4(3-2\beta_W^2)}\ln(1-\beta_W^2)\right.\notag\\[0.15cm]
&& \hspace{-1.2cm} \left. - \frac{9-4\beta_W^2}{3-2\beta_W^2}\ln\beta_W^2+2\mathrm{Li}_2(\beta_W^2) -2\mathrm{Li}_2(1-\beta_W^2)-\frac{6\beta_W^4-3\beta_W^2-8}{2\beta_W^2(3-2\beta_W^2)}-\pi^2\,\right]\Biggr\},
\end{eqnarray}
where $C_F = 4/3$ is the colour factor, and $\alpha_s=g_s^2/(4\pi)$ denotes the strong coupling constant. The parameter $\beta_W = \sqrt{1-m_W^2/m_t^2}$, with $m_W$ and $m_t$ the $W$-boson and the top-quark mass respectively, is the velocity of $W^+$ in the top-quark rest frame, and the dilogarithm function is defined by $\mbox{Li}_2(x)=-\int_0^x \text{ln}(1-t)/t\,dt$. The leading-order (LO) decay width $\Gamma_0(t \to b W^+)$ in Eq.~\eqref{eq:t2bWNLO} is given by~\cite{Denner:1990ns}
\begin{align}\label{eq:t2bWLO}
\Gamma_0(t \to b W^+) =\frac{G_F |V_{tb}|^2 \sqrt{\lambda(m_t,m_b,m_W)}}{8\pi\sqrt{2}\, m_t^3}\Big[(m_t^2-m_b^2)^2+(m_t^2+m_b^2)m_W^2-2m_W^4\Big],
\end{align}
where $G_F$ is the Fermi constant, $m_b$ the bottom-quark mass, and $\lambda(x,y,z)=(x^2-y^2-z^2)^2-4y^2z^2$ the usual triangle (or K\"{a}ll\'{e}n) function for a two-body decay.  

To a very good approximation, $\Gamma_\mathrm{tot}(t) \simeq \Gamma(t \to b W^+)$ holds also in the A2HDM as well as in the four conventional 2HDMs with $\mathcal{Z}_2$ symmetries, except when $m_{H^{\pm}} < m_t - m_b$; in this regime, the top quark can have an appreciable chance to decay into the final state $bH^+$, which must be taken into account for the total top width. In this case, we shall have $\Gamma_\mathrm{tot}(t) \simeq \Gamma(t \to b W^+) + \Gamma(t \to b H^+)$, with the LO partial width of the decay $t \to b H^+$ in the A2HDM given by~\cite{Abbas:2015cua}
\begin{eqnarray}\label{eq:t2bHLO}
	\Gamma(t \to bH^{+}) &=& \frac{G_F|V_{tb}|^2\sqrt{\lambda(m_t,m_b,m_{H^{\pm}})}}{8\pi\sqrt{2}\, m_t^3} \Big[\left(m_t^2 + m_b^2 - m_{H^{\pm}}^2\right)\left(m_b^2|\varsigma_d|^2 + m_t^2|\varsigma_u|^2\right)\notag\\
	&& \hspace{5.0cm}  
	- 4 m_b^2 m_t^2 \,\mathrm{Re}(\varsigma_d \varsigma_u^*)\,\Big].
\end{eqnarray}
As we are primarily interested in checking if the ``higher-order dominance'' phenomenon observed in $t\to cg(g)$ decays within the SM~\cite{Eilam:2006uh} is still valid for the A2HDM and the four conventional 2HDMs with $\mathcal{Z}_2$ symmetries, the charged-Higgs mass $m_{H^{\pm}}$ will be assumed to be larger than the top-quark mass $m_t$, and the decay mode $t \to b H^+$ is therefore kinematically forbidden. Thus, the branching ratios of $t \to cg(g)$ decays can be defined as
\begin{equation}\label{eq:br}
	\mathcal{B}(t \to c g(g)) = \frac{\Gamma(t \to c g(g))}{\Gamma (t \to b W^+)}\,.	
\end{equation}

In the next two subsections, we shall detail the calculation of the decay rates $\Gamma(t \to c g(g))$ in the A2HDM. To this end, we shall follow the following procedure: Firstly, we implement the A2HDM into the Mathematica-based package  \texttt{FeynRules}~\cite{Christensen:2008py,Alloul:2013bka} to generate the model file together with a complete set of Feynman rules, which is then fed into the package \texttt{FeynArts}~\cite{Kublbeck:1990xc,Hahn:2000kx} to generate the one-loop Feynman diagrams and the corresponding amplitudes for the $t\to cg(g)$ decays. Then, we resort to the packages \texttt{FormCalc}~\cite{Hahn:1998yk,Hahn:1999mt} and \texttt{LoopTools}~\cite{Hahn:1998yk} to manipulate the decay amplitudes and evaluate numerically the scalar and tensor one-loop Feynman integrals. Meanwhile, some partial cross-checks have been done by using the package \texttt{FeynCalc}~\cite{Mertig:1990an,Shtabovenko:2016sxi,Shtabovenko:2020gxv}. Finally, we can obtain the decay rates from the amplitudes squared by performing the phase space integration. Throughout this work, we carry out the one-loop calculation in $D=4-2\epsilon$ space-time dimensions using the dimensional regularization scheme~\cite{tHooft:1972tcz,Bollini:1972ui,Cicuta:1972jf,Ashmore:1972uj} for the ultraviolet divergence in loop integrals, and make the GIM mechanism manifest by dropping the terms independent of the internal down-type quark masses. For the collinear and infrared singularities present in the $t\to cgg$ decay, we take a non-zero charm-quark mass to avoid the collinear divergence generated when one of the gluons in the final state travels parallelly to the charm quark, and introduce phase-space cuts to eliminate the collinear (generated when the two gluons travel parallelly to each other) and infrared (generated when one of the gluons is soft) divergences~\cite{Simma:1990nr,Eilam:2006uh}.\footnote{A more precise approach to obtain a finite result requires both performing a complete NLO QCD calculation of the virtually corrected decay width $\Gamma(t\to cg)$ and achieving a dimensionally regularized version of the decay width $\Gamma(t\to cgg)$ in which both infrared and collinear singularities become manifest~\cite{Greub:2000sy}. The finiteness of the decay rate is then guaranteed by the Kinoshita-Lee-Nauenberg theorem~\cite{Kinoshita:1962ur,Lee:1964is}.} 

As the two gluons are both on-shell and hence transversely polarized, care must be taken regarding the gluon polarization sums when squaring the amplitude of $t\to cgg$ decay. One can maintain their transversality either by introducing contributions from the ghost fields while using the simple polarization sums~\cite{Eilam:2006uh,Eilam:2006rb}\footnote{Throughout this work, the colour indices in the gluon polarization vectors will be suppressed.} 
\begin{equation}
	\sum_{\text{polarizations}} \epsilon_{\mu}^{*}(k_3) \epsilon_{\nu}(k_3)
   =\sum_{\text{polarizations}} \epsilon_{\mu}^{*}(k_4) \epsilon_{\nu}(k_4)=-g_{\mu \nu},
\end{equation}  
or via the explicit construction~\cite{Hou:1988wt,Simma:1990nr}
\begin{equation}\label{eq:polarizationsum}
	\sum_{\text{polarizations}} \epsilon_{\mu}^{*}(k_3) \epsilon_{\nu}(k_3)
   =\sum_{\text{polarizations}} \epsilon_{\mu}^{*}(k_4) \epsilon_{\nu}(k_4)=-g_{\mu \nu}+\frac{k_{3 \mu} k_{4 \nu}+k_{4 \mu} k_{3 \nu}}{k_{3} \cdot k_{4}},
\end{equation} 
which are now simultaneously transverse with respect to the two gluon momenta $k_3$ and $k_4$. Here, $\epsilon_{\mu}(k_3)$ and $ \epsilon_{\mu}(k_4)$ denote the polarization vectors of the final-state gluons with momenta $k_3$ and $k_4$ respectively, while $g_{\mu\nu}$ is the metric tensor in Minkowski space, with the Greek indices running over $0,1,2,3$. In this paper, we shall follow the second method. In practice, however, one needs only keep one of the two gluon polarization sums as in Eq.~\eqref{eq:polarizationsum} while replacing the other one by $-g_{\mu \nu}$~\cite{Hou:1988wt}. Conveniently, the package \texttt{FormCalc} provides already a function \textit{PolarizationSum} to achieve such a task. 

In our calculation, we shall use the 't Hooft-Feynman gauge and take the external top and charm quarks to be on-shell. However, the gauge independence of our final results has been checked by performing additionally all the calculations in the unitary gauge. Due to the smallness of the down- and strange-quark masses $m_{d,s}$ and the unitarity of the CKM matrix $\sum_{q=d,s,b}V_{tq}^*V_{cq}=0$, the decay rates turn out to be only sensitive to the bottom-quark mass $m_b$~\cite{AguilarSaavedra:2004wm,AguilarSaavedra:2002ns}. For these internal quark masses, the most adequate choice is the $\overline{\rm MS}$ running mass evaluated at a typical scale of $\mathcal{O}(m_t)$~\cite{AguilarSaavedra:2004wm,AguilarSaavedra:2002ns}, and such a prescription will be followed by us. The external top- and charm-quark masses are, on the other hand, taken as the on-shell pole masses. For the sake of simplicity for our presentation, however, we shall set the charm-quark mass $m_c$ to zero and keep only the contributions from the internal bottom quark to these decays, while taking into account all these effects in our numerical analyses presented in Section~\ref{sec:numerical}.

\subsection{Two-body decay $t \to cg$}

\begin{figure}[t]
	\vspace{-0.8cm}
	\centering 
	\centerline{\hspace{0.10cm}
		\includegraphics[width=0.99\textwidth]{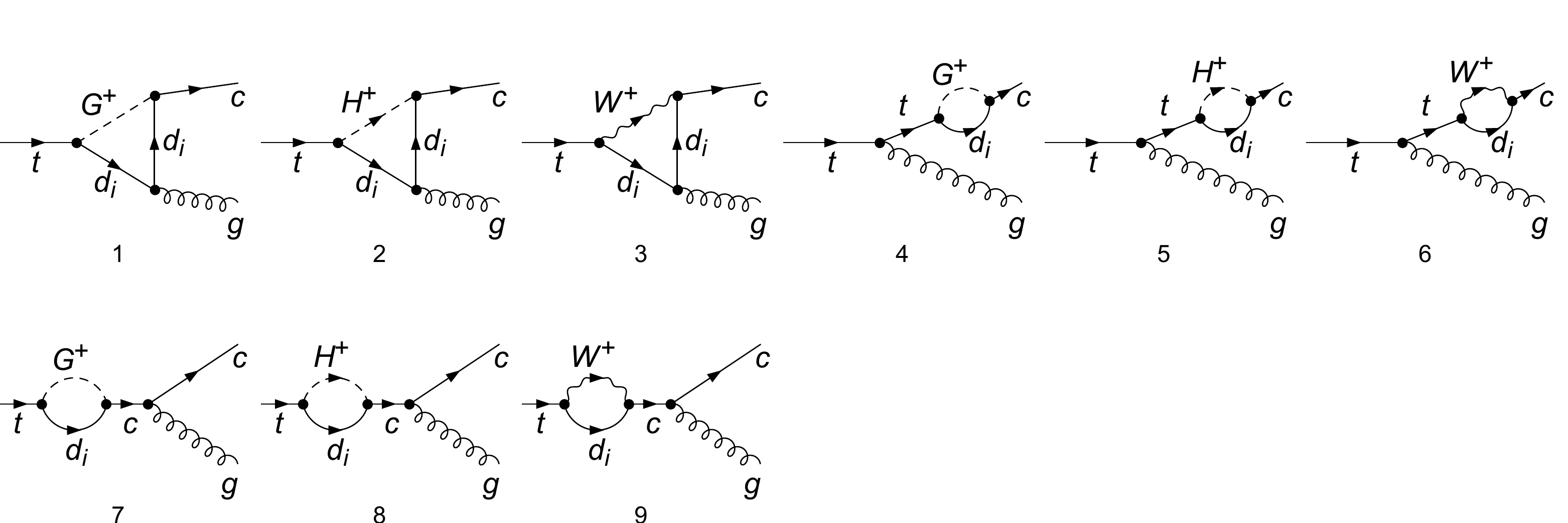}}
	\caption{One-loop Feynman diagrams contributing to the $t \to cg$ decay in the 't Hooft-Feynman gauge, with $d_i=\{d,s,b\}$, in the A2HDM. The first 1-3 diagrams are the vertex diagrams, and the rest 4-9 are the $t-c$ fermion self-energy diagrams.}
	\label{Fig:tcg}
\end{figure}

Let us firstly calculate the decay rate of the two-body decay $t \to cg$ in the A2HDM. The one-loop Feynman diagrams responsible for this decay in the 't Hooft-Feynman gauge are shown in Fig.~\ref{Fig:tcg}, where the first three ones correspond to the vertex diagrams while the remaining ones to the flavour-changing ($t-c$) fermion self-energy diagrams. During the calculation of these Feynman diagrams, we have kept all the external and internal quark masses non-vanishing~\cite{Eilam:1990zc,Abbas:2015cua}. Especially for the flavour-changing fermion self-energy diagrams, we have followed the treatment proposed in Ref.~\cite{Logan:2000iv}: It is most straightforward to avoid the issue of counterterms~\cite{Simma:1990nr,Liu:1989pc} altogether by simply calculating these diagrams, because the internal top (charm) quark line is off-shell and hence does not contribute to the one-particle pole of the charm (top) quark and needs not be truncated. Furthermore, due to the fact that the flavour-changing fermion self-energies are part of the non-truncated Green's function, one must include their contributions with a factor of $1$ rather than a factor of $1/2$, as the latter applies only in the standard flavour-conserving case. 

The resulting transition amplitude for the two-body decay $t \to cg$ can be written as\footnote{When the charm-quark mass is kept, there are two more form factors in Eq.~\eqref{eq:Amptcg}, with the associated Dirac strings obtained by replacing $P_R$ with $P_L$. Applying the on-shell conditions and the Gordon identities, we can also recast Eq.~\eqref{eq:Amptcg} into the standard dipole structure form~\cite{Deshpande:1991pn}, as will be discussed in Section~\ref{sec:numerical}.}
\begin{align}\label{eq:Amptcg}
\mathcal{A}_{t \to cg} &= \frac{G_F}{4\sqrt{2}\,\pi^2}\, g_s T^a_{ij} \sum_{q=d,s,b}V_{tq}^*V_{cq} \left[R_1^{(q)}(\epsilon^*(k_3)\cdot k_1)\,\bar{c}_i(k_2,0)\, P_R\, t_j(k_1,m_t)\,\right.\nn\\
& \left. \hspace{4.6cm} + R_2^{(q)}\,\bar{c}_i(k_2,0)\, P_R\, \epsilon\!\!/^*(k_3)\,t_j(k_1,m_t)\,\right],
\end{align}
where $g_s=\sqrt{4\pi\alpha_s}$ is the strong coupling constant, $T^a_{ij}$ the generators of the $SU(3)$ gauge group with the indexes $a=1,\cdots,8$ and $i,j=1,2,3$, and $P_{R,L}=(1\pm\gamma_5)/2$ the chirality projectors. Here, $\epsilon^a_\mu(k_3)$ is the polarization vector of the gluon, while $k_1$, $k_2$, and $k_3=k_1-k_2$ are the momenta of the top, the charm, and the gluon, respectively. The two form factors $R_{1,2}^{(q)}$ in Eq.~\eqref{eq:Amptcg} are the sums of the contribution from each of the nine diagrams shown in Fig.~\ref{Fig:tcg}. They can be expressed in terms of the masses of the external and internal quarks and bosons, and contain the $D$-dimensional loop integrals resulting from dimensional regularization. Specific to the internal bottom-quark contribution, we have explicitly 
\begin{align}
R_1^{(b)} &= 2m_t\Bigl\{2 m_W^2 C_{12}(0,m_t^2,0,m_b^2,m_b^2,m_W^2)
-(m_b^2-2m_W^2)\,C_{2}(0,m_t^2,0,m_b^2,m_b^2,m_W^2) \nn\\[0.1cm]
& \hspace{1.2cm} -m_b^2\bigl[C_0(0,m_t^2,0,m_b^2,m_b^2,m_W^2) -C_{12}(0,m_t^2,0,m_b^2,m_b^2,m_W^2) \nn\\[0.1cm]
&\hspace{2.2cm}+\text{X}_{du}\left(C_0(0,m_t^2,0,m_b^2,m_b^2,m_{H^\pm}^2) +C_{2}(0,m_t^2,0,m_b^2,m_b^2,m_{H^\pm}^2)\right) \nn\\[0.1cm]
&\hspace{2.2cm} -\text{X}_{dd}\,C_{12}(0,m_t^2,0,m_b^2,m_b^2,m_{H^\pm}^2)\bigr] \Bigr\},
\end{align}
and
\begin{align}
R_2^{(b)} &= m_W^2\Bigl\{2 m_t^2\big[C_{1}(0,m_t^2,0,m_b^2,m_b^2,m_W^2)- m_b^2\,C_0(0,m_t^2,0,m_b^2,m_b^2,m_W^2)\nn\\[0.1cm]
&\hspace{0.8cm} -C_{2}(0,m_t^2,0,m_b^2,m_b^2,m_W^2)\big] +2 B_{1}(m_t^2,m_b^2,m_W^2)-4 C_{00}(0,m_t^2,0,m_b^2,m_b^2,m_W^2)\Bigr\} \nn\\[0.1cm]
&+ m_b^2\,\text{X}_{dd}\Bigl[(m_{H^\pm}^2-m_b^2)C_0(0,m_t^2,0,m_b^2,m_b^2,m_{H^\pm}^2)-2 C_{00}(0,m_t^2,0,m_b^2,m_b^2,m_{H^\pm}^2)\nn\\[0.1cm]
&\hspace{0.8cm} + B_{1}(m_t^2,m_b^2,m_{H^\pm}^2)\Bigr] -\left(m_b^4-m_t^2 m_b^2-2 m_W^4\right) C_0(0,m_t^2,0,m_b^2,m_b^2,m_W^2) \nn\\[0.1cm]
&+ m_b^2\Bigl\{m_t^2\big[C_{1}(0,m_t^2,0,m_b^2,m_b^2,m_W^2) +\text{X}_{dd}\,C_{1}(0,m_t^2,0,m_b^2,m_b^2,m_{H^\pm}^2) \nn\\[0.1cm]
&\hspace{0.8cm}  +\text{X}_{du}\,C_0(0,m_t^2,0,m_b^2,m_b^2,m_{H^\pm}^2)\big] -\text{X}_{du}\big[B_{0}(0,m_b^2,m_{H^\pm}^2)+B_{0}(m_t^2,m_b^2,m_{H^\pm}^2)\big] \nn\\[0.1cm]
&\hspace{0.8cm}  -B_{0}(0,m_b^2,m_W^2)+B_{0}(m_t^2,m_b^2,m_W^2)+B_{1}(m_t^2,m_b^2,m_W^2) \nn\\[0.1cm]
&\hspace{0.8cm} -2C_{00}(0,m_t^2,0,m_b^2,m_b^2,m_W^2)\Bigr\} + \left[(\text{X}_{dd}+1) m_b^2+2 m_W^2\right] B_{0}(0,m_b^2,m_b^2).
\end{align}
Here, following the standard procedure and notation of Refs.~\cite{Denner:2019vbn,Denner:1991kt,tHooft:1978jhc,Passarino:1978jh}, we have reduced the vector and tensor integrals via the Passarino-Veltman method~\cite{Passarino:1978jh} to the scalar one-loop integrals $B_{l}$ and $C_{l,lm}$, which can be further decomposed into the basic scalar functions $A_0$, $B_0$, and $C_0$. However, as the package \texttt{FormCalc} does not require such a further reduction, we have expressed our results in terms of $B_{l}$ and $C_{l,lm}$ instead of the more basic ones to save CPU time. This is especially preferable for the three-body decay $t \to c gg$, in which the more complicated four-point one-loop integrals $D_{l,lm,lmn}$ will be involved. In addition, we have introduced the symbols $\text{X}_{qq^{\prime}}=\varsigma_q\varsigma_{q^{\prime}}^*$, with $q,q^{\prime}=d,u$, to represent the possible combinations of the alignment parameters in the A2HDM, and the SM result~\cite{Eilam:1990zc,AguilarSaavedra:2002ns} is reproduced by setting $\text{X}_{qq^{\prime}}=0$. It should be noted that, although the individual contribution from each internal quark $d_i\in\{d,s,b\}$ is separately divergent, the total amplitude is of course ultraviolet finite after summing over all the three internal quark flavours and taking into account the unitarity of the CKM matrix, $\sum_{q=d,s,b}V_{tq}^*V_{cq}=0$.

With the transition amplitude at hand, it is straightforward to calculate the decay rate $\Gamma(t \to cg)$, with the result given in the top-quark rest frame and in the limit $m_c=0$ by 
\begin{align}
\Gamma(t\to cg) &= \frac{\sqrt{\lambda(m_t,m_c,0)}}{16 \pi m_t^3} \sum_{\rm{polarizations}}\frac{1}{3}\sum_{\rm{colours}}\frac{1}{2}\sum_{\rm{spins}}\left|\mathcal{A}_{t \to cg}\right|^2.
\end{align}
For our later numerical analyses, however, we always take into account the finite charm-quark mass effect. 

\subsection{Three-body decay $t \to cgg$}

\begin{figure}[ht]
	\vspace{-0.8cm}
	\centering 
	\centerline{\hspace{0.10cm}
    \includegraphics[width=0.99\textwidth]{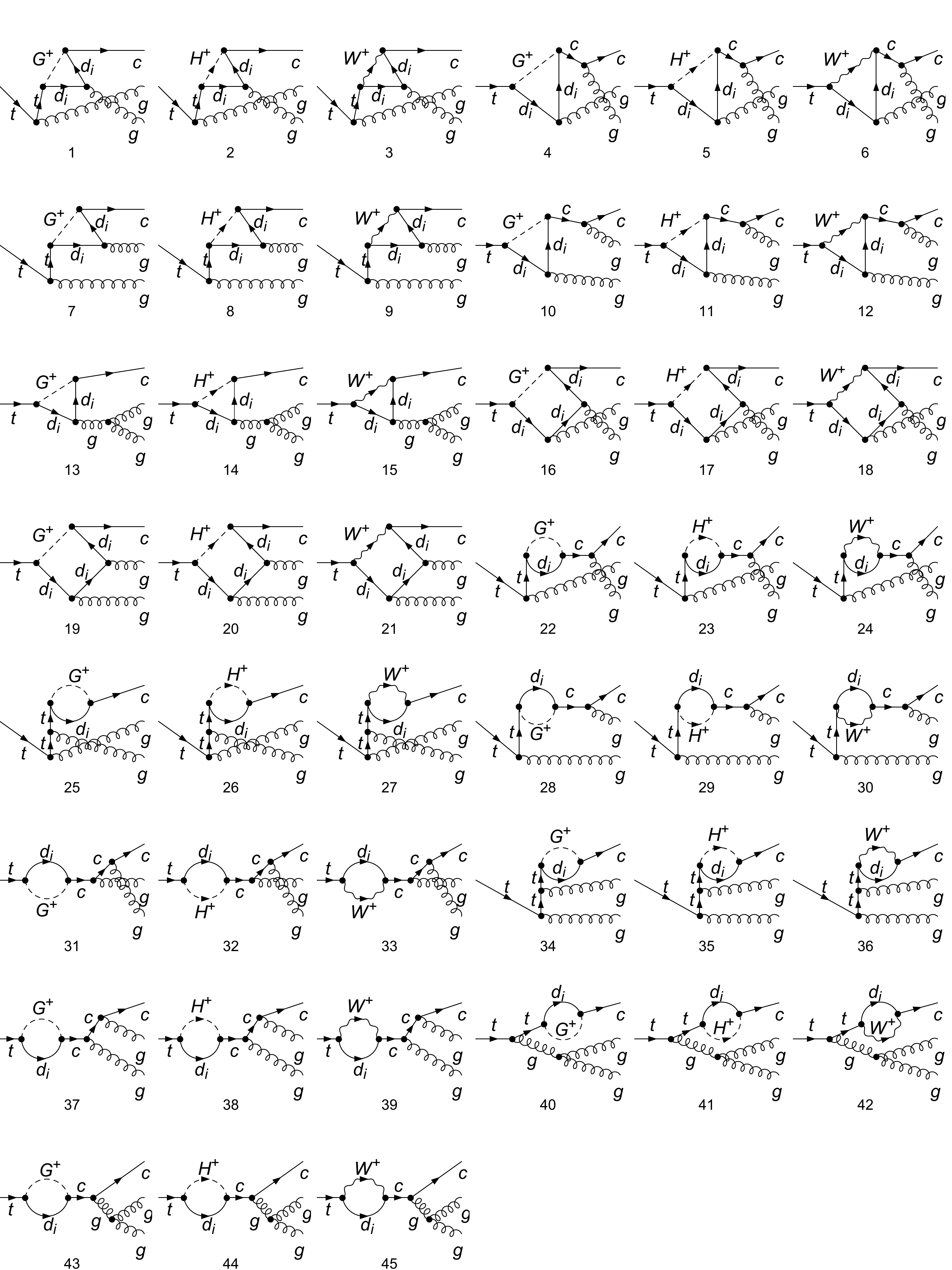}}
    \caption{One-loop Feynman diagrams contributing to the $t \to cgg$ decay in the 't Hooft-Feynman gauge, with $d_i=\{d,s,b\}$, in the A2HDM. The first 1-15 diagrams are the vertex diagrams, the diagrams 16-21 are the box diagrams, and the rest 22-45 are the $t-c$ fermion self-energy diagrams.}
    \label{Fig:tcgg}
\end{figure}

In the A2HDM, the three-body decay $t \to cgg$ receives contributions from the one-loop Feynman diagrams shown in Fig.~\ref{Fig:tcgg}, where $d_i=\{d,s,b\}$ and the 't Hooft-Feynman gauge is adopted. The transition amplitude of the decay can be written as
\begin{equation} \label{eq:Amptcgg}
\mathcal{A}_{t \to cgg} \!=\! \sum_{q=d,s,b} \Big[\mathcal{A}_{(q)}^{\text{vertex}} + \mathcal{A}_{(q)}^{\text{box}} + \mathcal{A}_{(q)}^{\text{self}}\,\Big],
\end{equation}
where $\mathcal{A}_{(q)}^{\text{vertex}}$, $\mathcal{A}_{(q)}^{\text{box}}$, and $\mathcal{A}_{(q)}^{\text{self}}$ represent the contributions from the vertex, the box, and the self-energy diagrams, respectively. Specific again to the internal bottom-quark contribution and in the limit $m_c=0$, these separate terms can be written in a compact form as
\begin{align}
\mathcal{A}_{(b)}^{\text{vertex}}&=\frac{G_F\,\alpha_s V_{tb}^{*}V_{cb}}{\sqrt{2}\,\pi\, t_{12}\, s_{23}\, t\, R_1 R_2}\Big[\,T_d\,t_{12}\,t\,R_2\big(R_1R_1^{\text{vertex}}+s_{23}R_2^{\text{vertex}}\big) \nn\\[0.15cm]
& \qquad + T_e\,t_{12}\,s_{23}\,R_1\big(R_2R_3^{\text{vertex}} + t R_4^{\text{vertex}}\big) + \big(T_d-T_e\big) s_{23}\,t\,R_1R_2R_5^{\text{vertex}}\,\Big],\label{eq:Amptcggvertex}\\[0.2cm]
\mathcal{A}_{(b)}^{\text{box}}&=\frac{G_F\,\alpha_s V_{tb}^{*}V_{cb}}{\sqrt{2}\,\pi\, t_{12}\, s_{23}\, t\, R_1 R_2}\Big[\,m_t\big(F_1 R_1^{\text{box}}+ F_{12}R_5^{\text{box}}+2F_{13}R_6^{\text{box}}+2F_{14}R_7^{\text{box}}\big) \nn\\[0.15cm]
& \qquad - \big(F_{3}R_2^{\text{box}}+F_{4}R_3^{\text{box}}+F_{5}R_4^{\text{box}} -F_{16}R_8^{\text{box}}\big)\,\Big],\label{eq:Amptcggbox}\\[0.2cm]
\mathcal{A}_{(b)}^{\text{self}}&=\frac{G_F\,\alpha_s V_{tb}^{*}V_{cb}}{\sqrt{2}\,\pi\, t_{12}\, s_{23}\, t\, R_1 R_2}\Big[\,T_d\,t_{12}\,t\,R_2\big(R_1R_1^{\text{self}}+s_{23}\,m_b^2R_2^{\text{self}}+R_6^{\text{self}}\big) \nn\\[0.15cm]
& \quad + T_e\,t_{12}\,s_{23}\,R_1\big(R_2R_3^{\text{self}} +t\,m_b^2R_4^{\text{self}}+R_7^{\text{self}}\big)+ \big(T_d-T_e\big) s_{23}\,t\,R_1 R_2 R_5^{\text{self}}\,\Big].\label{eq:Amptcggself}
\end{align}
Here the kinematics of the process is specified in the parentheses of the initial- and final-state particles as $t(k_1) \to c(k_2)+g(k_3)+g(k_4)$.
For convenience, we have also introduced in Eqs.~\eqref{eq:Amptcggvertex}--\eqref{eq:Amptcggself} two products of the colour generators, $T_d=(T^aT^b)_{ij}$ and $T_e=(T^bT^a)_{ij}$, as well as the (generalized) Mandelstam variables $t=(k_{1}-k_{3})^2$, $t_{12}=(k_{1}-k_{2})^2$, and $s_{23}=(k_{2}+k_{3})^2$.\footnote{It should be noted that we can also organize Eqs.~\eqref{eq:Amptcggvertex}--\eqref{eq:Amptcggself} according to whether the two on-shell final-state gluons are in a colour-symmetric ($\{T^a,T^b\}$) or in a colour-antisymmetric ($[T^a,T^b]$) configuration~\cite{Simma:1990nr,Greub:2000sy}. Although taking the same symbol, the Mandelstam variable $t$ can be clearly distinguished from the top-quark spinor through the context.} In addition, the independent fermion chains involved are defined, respectively, by
\begin{equation}
\begin{aligned}
F_1 &= \bar{c}_i(k_2,0)\, P_R \,t_j(k_1,m_t), && \;\,
F_2 = \bar{c}_i(k_2,0)\, P_L \,t_j(k_1,m_t), \\[0.2cm]
F_3 &= \bar{c}_i(k_2,0)\, P_R \, \epsilon\!\!/^*(k_3)\,t_j(k_1,m_t), && \;\,
F_4 = \bar{c}_i(k_2,0)\, P_R\, \epsilon\!\!/^*(k_4)\, t_j(k_1,m_t), \\[0.2cm]
F_5 &= \bar{c}_i(k_2,0)\, P_R\, k\!\!\!/_3 \,t_j(k_1,m_t), && \;\,
F_6 = \bar{c}_i(k_2,0)\, P_L\, \epsilon\!\!/^*(k_3) \,t_j(k_1,m_t), \\[0.2cm]
F_7 &=  \bar{c}_i(k_2,0)\, P_L\, \epsilon\!\!/^*(k_4) \,t_j(k_1,m_t), && \;\,
F_8 = \bar{c}_i(k_2,0)\, P_L\, k\!\!\!/_3 \,t_j(k_1,m_t), \\[0.2cm]
F_9 &= \bar{c}_i(k_2,0)\, P_L\, \epsilon\!\!/^*(k_3)\,
\epsilon\!\!/^*(k_4)\,t_j(k_1,m_t), && 
F_{10} = \bar{c}_i(k_2,0)\, P_L\, \epsilon\!\!/^*(k_3)\, k\!\!\!/_3 \,t_j(k_1,m_t), \\[0.2cm]
F_{11} &= \bar{c}_i(k_2,0)\, P_L\, \epsilon\!\!/^*(k_4)\, k\!\!\!/_3 \,t_j(k_1,m_t), &&
F_{12} = \bar{c}_i(k_2,0)\, P_R\, \epsilon\!\!/^*(k_3)\, \epsilon\!\!/^*(k_4)
\,t_j(k_1,m_t), \\[0.2cm]
F_{13} &= \bar{c}_i(k_2,0)\, P_R\, \epsilon\!\!/^*(k_3)\, k\!\!\!/_3 \,t_j(k_1,m_t), &&
F_{14} = \bar{c}_i(k_2,0)\, P_R\, \epsilon\!\!/^*(k_4)\, k\!\!\!/_3 \,t_j(k_1,m_t), \\[0.2cm]
F_{15} &= \bar{c}_i(k_2,0)\, P_L\, \epsilon\!\!/^*(k_3)\, \epsilon\!\!/^*(k_4)\,
k\!\!\!/_3 \,t_j(k_1,m_t), &&
F_{16} = \bar{c}_i(k_2,0)\, P_R\, \epsilon\!\!/^*(k_3)\, \epsilon\!\!/^*(k_4)\,
k\!\!\!/_3 \,t_j(k_1,m_t).
\end{aligned}
\end{equation}
For the $R$ functions present in Eqs.~\eqref{eq:Amptcggvertex}--\eqref{eq:Amptcggself}, we have $R_1=s_{23}-m_t^2$ and $R_2=t-m_t^2$, while the explicit expressions of all the remaining ones are collected in Appendix~\ref{sec:appendix}. 

In the top-quark rest frame, the differential decay rate of $t \to cgg$ can be written as
\begin{align}\label{eq:dGammat2cgg}
d\Gamma(t\to cgg) &= \frac{1}{2m_t}\left(\prod_{i=2}^{4} \int \frac{d^{3} k_{i}}{(2 \pi)^{3}}\frac{1}{2 E_{i}}\right)(2 \pi)^{4} \delta^{(4)}\big(k_1-k_2-k_3-k_4\big)\nn\\[0.10cm] 
&\qquad \times \sum_\text{polarizations}\frac{1}{3}\sum_\text{colours}\frac{1}{2}\sum_\text{spins}
\frac{1}{2} \left|\mathcal{A}_{t \to cgg}\right|^2,
\end{align}
where $E_i=k_i^0$ represents the energy of the ($i$-1)-th final-state particle in the top-quark rest frame, and the last factor $1/2$ accounts for the statistical factor due to the two gluons in the final state. In this frame, we have $\vec{k}_2+\vec{k}_3+\vec{k}_4=0$, and thus the three final-state momentum vectors lie in a plane. Then, the three-body phase space in Eq.~\eqref{eq:dGammat2cgg} can be integrated over the orientation of this plane by using the momentum delta functions and the azimuthal symmetry. In this way, we are left with only two non-trivial integration variables, which can be chosen as the energy of the charm quark ($k_2^0$) and the energy of one of the gluons (say $k_3^0$), and the decay width of $t \to cgg$ can be finally written as
\begin{align}\label{eq:Gammat2cgg}
	\Gamma(t\to cgg) &= \frac{1}{64\,\pi^3\,m_t} \sum_\text{polarizations}\frac{1}{3}\sum_\text{colours}\frac{1}{2}\sum_\text{spins}
	\int dk_3^0\int dk_2^0 \times \frac{1}{2}
	\left|\mathcal{A}_{t \to cgg}\right|^2.
\end{align}

Because of the infrared and collinear divergences mentioned in Section~\ref{sec:generalremarks}, which are also evident from Eqs.~\eqref{eq:Amptcggvertex}--\eqref{eq:Amptcggself} and appear at the vertices and along the boundary lines of the Dalitz plot in the $(k_2^0, k_3^0)$ space respectively, we have to impose proper cuts on both $k_2^0$ and $k_3^0$ when performing the integration of Eq.~\eqref{eq:Gammat2cgg} over these two variables. Following the treatment of Refs.~\cite{Simma:1990nr,Eilam:2006uh}, we shall use the kinematic constraints
\begin{equation}
\begin{split}
& t=(k_1-k_3)^2=m_{t}^{2} - 2 m_{t} k_{3}^0 \geqslant C \cdot 2 m_{t}^{2}, \\[0.15cm]
& t_{12}=(k_1-k_2)^2=m_{t}^{2} - 2 m_{t} k_{2}^0 \geqslant C \cdot 2 m_{t}^{2}, \\[0.15cm]
& s_{23}=(k_2+k_3)^2=(k_1-k_4)^2=-m_{t}^{2} + 2 m_{t} (k_{2}^0 + k_{3}^0) \geqslant C \cdot 2 m_{t}^{2},
\end{split}
\end{equation}
along with the fact that $k_{2,\text{max}}^0 \geqslant k_{2,\text{min}}^0$, to get the integration ranges of the two variables:\footnote{To ensure that the divergent point at $k_2^0+k_3^0=m_t/2$ resulting from $s_{23}=0$ in Eqs.~\eqref{eq:Amptcggvertex}--\eqref{eq:Amptcggself} is also excluded, our choice is  slightly different from that of Ref.~\cite{Eilam:2006uh}. Recall that $m_c=0$ is assumed here.}
\begin{equation}\label{eq:interalranges}
	k_2^0 \in \left[\frac{m_t}{2}-k_3^0+Cm_t,\frac{m_t}{2}(1-2C)\right], \qquad k_3^0 \in \left[2Cm_t,\frac{m_t}{2}(1-2C)\right],
\end{equation}
where $C$ is the cutoff parameter, and its value must be taken within the range of jet energy resolution of a detector. Note that, when a non-zero charm-quark mass is considered, Eq.~\eqref{eq:interalranges} must be modified as
\begin{equation}\label{eq:interalranges_mc}
	k_2^0 \in \left[\frac{m_t}{2}-k_3^0+Cm_t+\frac{m_c^2}{2m_t},\frac{m_t}{2}(1-2C)+\frac{m_c^2}{2m_t}\right], \quad k_3^0 \in \left[2Cm_t,\frac{m_t}{2}(1-2C)-\frac{m_c^2}{2m_t}\right].
\end{equation}
Here, for simplicity, we shall take $C\simeq 0.01$ as the default value (\textit{i.e.}, requiring the energy of each decay product to be larger than $2.0\,\gev$ in the top-quark rest frame), while varying $C$ within the range $[0.001,0.05]$ to study its numerical impacts on the branching ratio of $t \to cgg$ decay. Note that the choice $C=0.01$ is already large enough to reach the jet energy resolution sensitivity of the HL-LHC~\cite{Atlas:2019qfx} and FCC~\cite{Abada:2019lih,Abada:2019zxq,Benedikt:2018csr,Abada:2019ono} facilities.  

\section{Numerical results and discussions}
\label{sec:numerical}

\subsection{Input parameters}
\label{sec:inputs}

\begin{table}[t]
	\centering
	\let\oldarraystretch=\arraystretch
	\renewcommand*{\arraystretch}{1.3}
	{\tabcolsep=0.60cm \begin{tabular}{|cccc|}
			\hline\hline
			\multicolumn{4}{|l|}{\textbf{QCD and electroweak parameters~\cite{Zyla:2020zbs}}}
			\\
			\hline
			$G_{F}[10^{-5}\,\gev^{-2}]$
			& $m_{W}[\gev]$
			& $m_{Z}[\gev]$
			& $\alpha_s^{(5)}(m_Z)$
			\\
			$1.1663787$
			& $80.379$
			& $91.1876$
			& $0.1179$
			\\
			\hline
			& 
			& 
			& $\alpha_s^{(6)}(m_t^{\rm pole})$
			\\
			& 
			& 
			& $0.1076$
			\\
			\hline
	\end{tabular}}
	{\tabcolsep=0.385cm \begin{tabular}{|cccccc|}
			\multicolumn{6}{|l|}{\hspace{0.10cm} \textbf{Quark masses~[GeV]~\cite{Zyla:2020zbs}}}
			\\
			\hline
			$m_t^{\rm pole}$
			& $m_b^{\rm pole}$
			& $m_c^{\rm pole}$
			& $\bar{m}_b(\bar{m}_b)$
			& $\bar{m}_s(2\,\gev)$
			& $\bar{m}_d(2\,\gev)$
			\\
			  $172.5$
			& $4.78$
			& $1.67$
			& $4.18$
			& $0.093$
			& $0.00467$
			\\
			\hline
			&
			&
			& $\bar{m}_b(m_t^{\rm pole})$
			& $\bar{m}_s(m_t^{\rm pole})$
			& $\bar{m}_d(m_t^{\rm pole})$
			\\
			&
			&
			& $2.60$
			& $0.048$
			& $0.00241$
			\\
			\hline
	\end{tabular}}
	{\tabcolsep=1.115cm \begin{tabular}{|cccc|}
			\multicolumn{4}{|l|}{\hspace{-0.5cm} \textbf{CKM parameters~\cite{Eberhardt:2020dat}}}
			\\
			\hline
			$\lambda$
			& $A$
			& $\bar{\rho}$
			& $\bar{\eta}$
			\\
			$0.2256$
			& $0.829$
			& $0.182$
			& $0.360$
			\\
			\hline\hline
	\end{tabular}}
	\caption{Summary of the input parameters used throughout this work. 
	For convenience, the derived values of the $\overline{\rm MS}$ running parameters evaluated at the scale $\mu_t=m_t^{\rm pole}$ are also given. In the Wolfenstein parametrization~\cite{Wolfenstein:1983yz} of the CKM matrix, the four independent Wolfenstein parameters can be chosen as $\lambda$, $A$, $\bar{\rho}=\rho(1-\lambda^2/2)$, and $\bar{\eta}=\eta(1-\lambda^2/2)$~\cite{Buras:1994ec}.}
	\label{tab:inputs}
\end{table}

The relevant input parameters needed for evaluating the branching ratios of $t \to cg(g)$ decays are listed in Table~\ref{tab:inputs}. For the QCD and electroweak parameters as well as the quark masses, their values are taken from the \textsc{Particle Data Group}~\cite{Zyla:2020zbs}. Here, for the external on-shell quarks, we take their pole masses as input, and evolve the internal quark masses from their respective initial scales to the scale $\mu_t=m_t^\mathrm{pole}$ characteristic for the top-quark decay, using the two-loop renormalization group equations for the $\overline{\rm MS}$ running quark masses.\footnote{Note that the down- and strange-quark running masses $\bar{m}_{d,s}$ are quoted in the $\overline{\rm MS}$ scheme at a scale $\mu=2\,\gev$, while the bottom-quark running mass $\bar{m}_b$ is evaluated at a scale equal to its mass, \textit{i.e.}, $\bar{m}_b(\bar{m}_b)$.} The $\overline{\rm MS}$ coupling constant $\alpha_s$ is given at the $Z$-boson mass scale with the top quark decoupled (\textit{i.e.}, with $n_f=5$ active quark flavours). To obtain its values at the scale $\mu_t$ and also at other scales when evolving the $\overline{\rm MS}$ running quark masses, we use the two-loop running of $\alpha_s$ by taking into account the decoupling effect when crossing a flavour threshold~\cite{Chetyrkin:2000yt,Herren:2017osy}. 
For convenience, we compile also in Table~\ref{tab:inputs} the derived values of the $\overline{\rm MS}$ running parameters evaluated at the scale $\mu_t=m_t^{\rm pole}$ following the above prescription.

Concerning the needed entries of the CKM matrix, we adopt the Wolfenstein parametrization~\cite{Wolfenstein:1983yz}, generalized to include the higher-order terms of $\lambda=|V_{us}|$~\cite{Buras:1994ec}. To be consistent with the study of NP, the four Wolfenstein parameters $\lambda$, $A$, $\bar{\rho}=\rho(1-\lambda^2/2)$, and $\bar{\eta}=\eta(1-\lambda^2/2)$ should be fitted to the tree-level observables as well as the ratio of the mass differences between the neutral $B$-meson eigenstates, $\Delta M_d/\Delta M_s$, which are all expected to have only a minimal sensitivity to the charged-Higgs contributions~\cite{Jung:2010ik}. To this end, we shall use as input the central values of the fitting results obtained in Ref.~\cite{Eberhardt:2020dat}. For any further details, the readers are referred to Ref.~\cite{Eberhardt:2020dat}.

As pointed out already in Section~\ref{sec:A2HDM}, only the parameters $\varsigma_u$, $\varsigma_d$, and $m_{H^\pm}$ are involved for the rare decays $t \to cg(g)$ in the A2HDM. Let us now fix the allowed ranges of these model parameters. For simplicity, we shall entertain ourselves with the CP-conserving limit and assume that the observed $125\,\gev$ Higgs is the lightest CP-even scalar of the model, which corresponds exactly to the \textit{light scenario} considered in Ref.~\cite{Eberhardt:2020dat}. For the two alignment parameters $\varsigma_{u,d}$, we choose the ranges 
\begin{equation}\label{eq:zetauzetadrange}
	|\varsigma_{u}| \in [0,1.5], \qquad |\varsigma_{d}| \in [0,50],
\end{equation}
as constrained mostly by the precision flavour physics, such as $Z\to b\bar{b}$, $\bar{B}\to X_s\gamma$, as well as $B_{s,d}^0-\bar{B}_{s,d}^0$ and $K^0-\bar{K}^0$ mixings~\cite{Jung:2012vu,Jung:2010ab,Jung:2010ik,Chang:2015rva,Cho:2017jym}. It is especially interesting to note that, for large values of $|\varsigma_{u} \varsigma_{d}|$, important correlated constraints on $\varsigma_{u}$ and $\varsigma_{d}$ can be derived from the weak radiative decay $\bar{B}\to X_s\gamma$, with the region $\varsigma_{d} \varsigma_{u}<0$ almost excluded by the current data on the branching ratio of $\bar{B}\to X_s\gamma$~\cite{Jung:2010ik}. The ranges in Eq.~\eqref{eq:zetauzetadrange} are also roughly consistent with the perturbativity requirement, $|\sqrt{2}\varsigma_{u,d} m_{u,d}/v| \lesssim 1$. For the charged-Higgs mass, we focus only on the case with $m_{H^\pm}$ varied within the ranges $m_{H^\pm}\in [200, 600]\,\gev$ for the $t \to cg$ and $m_{H^\pm}\in [200, 400]\,\gev$ for the $t \to cgg$ decay. Here the lower limit is chosen to be larger than the top-quark mass such that the decay mode $t \to b H^+$ is kinematically forbidden, while the upper limit is motivated by the observation that the maximum branching ratio of $t \to cg(g)$ for $m_{H^\pm}\gtrsim 600\,(400)\,\gev$ approaches already the corresponding SM prediction, as will be demonstrated in the next subsection.\footnote{Direct constraints on the charged-Higgs mass can also be obtained from collider searches for the production and decay of on-shell charged-Higgs bosons~\cite{ALEPH:2013htx,ATLAS:2018gfm,ATLAS:2021upq,CMS:2019bfg,CMS:2019rlz}; see, \textit{e.g.}, Refs.~\cite{Akeroyd:2016ymd,Arbey:2017gmh} for recent reviews. Our choices of the model parameters comply with these direct constraints.} Recently, a detailed global fit has been performed to the A2HDM by considering a CP-conserving scalar potential and real alignment parameters, and including the most constraining flavour observables, electroweak precision measurements, and the available data on Higgs signal strengths and collider searches for heavy scalars, together with the theoretical requirements of perturbativity and positivity of the scalar potential~\cite{Eberhardt:2020dat}. We shall use the relevant constraints on the alignment parameters $\varsigma_{u,d}$ resulting from such a global fit obtained at the $95.5\%$ CL (cf. the brown region in the left plot of Fig.~9) in Ref.~\cite{Eberhardt:2020dat}; for further details, the readers are also referred to Ref.~\cite{Eberhardt:2020dat}.  

\subsection{$\mathcal{B}(t \to cg(g))$ in the A2HDM}
\label{sec:t2cggina2HDM}

Firstly, with the input parameters collected in Table~\ref{tab:inputs}, we give for reference the partial width of the dominant tree-level $t \to b W^+$ decay,
\begin{align}
	\Gamma(t \to b W^+)=1.35\,(1.47)\,\gev,
\end{align}
at the NLO (LO) in QCD, which is found to be well consistent with the current data on the top-quark total width, $\Gamma_\mathrm{tot}^\mathrm{exp}(t) = 1.42^{+0.19}_{-0.15}\,\gev$~\cite{Zyla:2020zbs}. This in turn implies that the extra contribution from the partial width $\Gamma(t \to b H^+)$ in the $m_{H^{\pm}} < m_t - m_b$ case should be small, supporting therefore our choice of the lower limit of the charged-Higgs mass, $m_{H^{\pm}} \geq 200\,\gev$, as required to forbid the decay $t \to b H^+$ kinematically. Otherwise, for $m_{H^{\pm}}<150\,\gev$, the alignment parameter $\varsigma_{d}$ is constrained to be small by the direct charged-Higgs searches at the LHC via the decay $t \to b H^+$~\cite{ATLAS:2018gfm,CMS:2019bfg}, implying a very strong suppression on the decay rates of $t \to c V$~($V = \gamma, Z$) and $t \to c h$ processes~\cite{Abbas:2015cua}. 

To display the dependence of the branching ratios of $t\to cg$ and $t\to cgg$ decays on the alignment parameters $\varsigma_{u,d}$, we fix the charged-Higgs mass to three benchmark values, $m_{H^{\pm}}=200$, $300$, and $500\,(400)\,\gev$. In this way, we obtain numerically
\begin{eqnarray}\label{eq:tcgnum}
	\mathcal{B}^{\text{A2HDM}}(t \to cg) &=& 4.50\times 10^{-12} \nn \\[0.1cm] 
	&& \hspace{-0.9cm} \times
	\begin{cases}
	\begin{aligned}
		\Bigl[1 &+ 2.46\times 10^{-5}\, \varsigma_u^4 + 7.19\times 10^{-5}\, \varsigma_d \varsigma_u^3 + 0.261\, \varsigma_d^2 \varsigma_u^2 \\[0.15cm]
		& - 1.30\times 10^{-2}\, \varsigma_d^3 \varsigma_u + 1.63\times 10^{-4}\, \varsigma_d^4 + 4.91\times 10^{-6}\, \varsigma_u^2 \\[0.15cm]
		& + 0.848\, \varsigma_d \varsigma_u - 2.12\times 10^{-2}\, \varsigma_d^2\,\Bigr], \qquad \qquad \text{for $m_{H^\pm}=200\,\gev$}, \\[0.2cm]
		\Bigl[1 &+ 1.96\times 10^{-7}\, \varsigma_u^4 + 2.88\times 10^{-6}\, \varsigma_d \varsigma_u^3 + 3.48\times 10^{-2}\, \varsigma_d^2 \varsigma_u^2 \\[0.15cm]
		& - 1.49\times 10^{-3}\, \varsigma_d^3 \varsigma_u + 1.60\times 10^{-5}\, \varsigma_d^4 + 2.92\times 10^{-6}\, \varsigma_u^2 \\[0.15cm]
		& + 0.310\, \varsigma_d \varsigma_u - 6.63\times 10^{-3}\, \varsigma_d^2\,\Bigr], \qquad \qquad \text{for $m_{H^\pm}=300\,\gev$}, \\[0.2cm]
		\Bigl[1 &+ 3.42\times10^{-9}\, \varsigma_u^4 + 1.49\times 10^{-7}\, \varsigma_d \varsigma_u^3 + 4.64\times 10^{-3}\, \varsigma_d^2 \varsigma_u^2 \\[0.15cm]
		& - 1.73\times 10^{-4}\, \varsigma_d^3 \varsigma_u + 1.61\times 10^{-6}\, \varsigma_d^4 + 5.41\times 10^{-7}\, \varsigma_u^2 \\[0.15cm]
		& + 0.113\, \varsigma_d \varsigma_u - 2.11\times 10^{-3}\, \varsigma_d^2\,\Bigr], \qquad \qquad \text{for $m_{H^\pm}=500\,\gev$},
	 \end{aligned}
     \end{cases}
\end{eqnarray} 
for the two-body decay $t \to cg$, and
\begin{eqnarray}\label{eq:tcggnum}
	\mathcal{B}^{\text{A2HDM}}(t \to cgg) &=& 8.31\times 10^{-10} \nn \\[0.1cm] 
	&& \hspace{-0.9cm} \times
	\begin{cases}
	\begin{aligned}
		\Bigl[1 &+ 6.31\times 10^{-7}\, \varsigma_u^4 + 2.47\times 10^{-7}\, \varsigma_d \varsigma_u^3 + 1.06\times 10^{-3}\, \varsigma_d^2 \varsigma_u^2 \\[0.15cm]
		& - 4.67\times 10^{-5}\, \varsigma_d^3 \varsigma_u + 5.67\times 10^{-7}\, \varsigma_d^4 - 1.19\times 10^{-6}\, \varsigma_u^2 \\[0.15cm]
		& + 2.72\times 10^{-3}\, \varsigma_d \varsigma_u - 9.63\times 10^{-5}\, \varsigma_d^2\,\Bigr], \qquad \text{for $m_{H^\pm}=200\,\gev$}, \\[0.2cm]
		\Bigl[1 &+ 4.99\times 10^{-8}\, \varsigma_u^4 + 1.40\times 10^{-8}\, \varsigma_d \varsigma_u^3 + 1.35\times 10^{-4}\, \varsigma_d^2 \varsigma_u^2 \\[0.15cm]
		& - 5.20\times 10^{-6}\, \varsigma_d^3 \varsigma_u + 5.47\times 10^{-8}\, \varsigma_d^4 - 3.11\times 10^{-7}\, \varsigma_u^2 \\[0.15cm]
		& + 1.02\times 10^{-3}\, \varsigma_d \varsigma_u - 3.00\times 10^{-5}\, \varsigma_d^2\,\Bigr], \qquad \text{for $m_{H^\pm}=300\,\gev$}, \\[0.2cm]
		\Bigl[1 &+ 1.32\times 10^{-8}\, \varsigma_u^4 + 3.17\times 10^{-9}\, \varsigma_d \varsigma_u^3 + 4.12\times 10^{-5}\, \varsigma_d^2 \varsigma_u^2 \\[0.15cm]
		& - 1.49\times 10^{-6}\, \varsigma_d^3 \varsigma_u + 1.45\times 10^{-8}\, \varsigma_d^4 - 1.52\times 10^{-7}\, \varsigma_u^2 \\[0.15cm]
		& + 5.78\times 10^{-4}\, \varsigma_d \varsigma_u - 1.54\times 10^{-5}\, \varsigma_d^2\,\Bigr], \qquad \text{for $m_{H^\pm}=400\,\gev$},
	\end{aligned}
	\end{cases}
\end{eqnarray}
for the three-body decay $t \to cgg$. Here, the SM results are obtained by taking the limits $\varsigma_{u,d}\to 0$ and read
\begin{align}
	\mathcal{B}^{\text{SM}}(t \to cg) &= 4.50\times 10^{-12}, \label{eq:NbrSMtcg} \\[0.2cm]
	\mathcal{B}^{\text{SM}}(t \to cgg) &= 8.31\times 10^{-10}. \label{eq:NbrSMtcgg}
\end{align}
It is found that our SM predictions are generally consistent with that made in Refs.~\cite{Balaji:2020qjg,AguilarSaavedra:2002ns} for the two-body decay $t \to cg$, as well as in Ref.~\cite{Eilam:2006uh} for the three-body decay $t \to cgg$. Especially, the branching ratio of the three-body decay $t \to cgg$ is predicted to be about two orders of magnitude larger than that of the two-body decay $t \to cg$, which is a clear demonstration of the ``higher-order dominance'' phenomenon observed already in Ref.~\cite{Eilam:2006uh}. It should be noted that the branching ratio of $t \to cgg$ depends on the cutoff parameter $C$, and the results given by Eqs.~\eqref{eq:tcggnum} and \eqref{eq:NbrSMtcgg} are obtained with the default value $C=0.01$. The cutoff dependence of this branching ratio within the SM is shown in Table~\ref{tab:cutdepen} for six different values of $C$, together with the comparison with that obtained in Ref.~\cite{Eilam:2006uh}. It can be seen that, although being increased by about $7\%$ compared to that obtained in Ref.~\cite{Eilam:2006uh} for the same choice of $C$,\footnote{Note that the $\overline{\rm MS}$ running masses of both the top and charm quarks as well as the $\alpha$ scheme for the electroweak parameters were used in Ref.~\cite{Eilam:2006uh}, while we take the pole masses for the external on-shell quarks as input and adopt the $G_F$ scheme for the electroweak parameters~\cite{Eilam:1991iz}.} our updated results depend only slightly on the cutoff parameter $C$, with the variation being only about two times when $C$ varies within the range $[0.001,0.05]$. From Eqs.~\eqref{eq:tcgnum} and \eqref{eq:tcggnum}, one can also see that the branching ratios of $t \to cg(g)$ decays display a different dependence on the alignment parameters $\varsigma_{u,d}$: Firstly, due to the much smaller coefficients, together with the allowed ranges of $\varsigma_{u}$ and $\varsigma_{d}$ specified by Eq.~\eqref{eq:zetauzetadrange}, the terms proportional to $\varsigma_u^4$, $\varsigma_d \varsigma_u^3$, and $\varsigma_u^2$ can be safely neglected. Secondly, for moderate values of $|\varsigma_d|$ and large values of $|\varsigma_u|$, the branching ratios will be dominated by the product $\varsigma_{d} \varsigma_{u}$ and, to a less extent, also by $\varsigma_{d}^2 \varsigma_{u}^2$. Finally, the significance of $\varsigma_{d}$ occurs only when $|\varsigma_d|$ and $|\varsigma_u|$ take much larger and smaller values, respectively. 

\begin{table}[t]
	\begin{center}	
		\let\oldarraystretch=\arraystretch
		\renewcommand*{\arraystretch}{1.3}
		\tabcolsep=0.315cm\begin{tabular}{|cccccccc|}
			\hline\hline
			$\textit{C}$
			& $0.001$
			& $0.003$
			& $0.005$
			& $0.01$
			& $0.03$
			& $0.05$
			& 
			\\\hline
			$\mathcal{B}^{\text{SM}}(t \to cgg)\,[10^{-10}]$
			& $10.7$
			& $9.80$
			& $9.33$
			& $8.31$ 
			& $6.06$
			& $4.68$ 
			& This work
			\\
			$\mathcal{B}^{\text{SM}}(t \to cgg)\,[10^{-10}]$
			& $10.2$
			& $9.04$
			& $8.76$
			& $7.78$ 
			& ---
			& ---
			& Ref.~\cite{Eilam:2006uh}
			\\
			\hline\hline
		\end{tabular}
		\caption{The cutoff dependence of the branching ratio of $t \to cgg$ decay within the SM, as well as the comparison with that obtained in Ref.~\cite{Eilam:2006uh}. Note that the $\overline{\rm MS}$ running masses of both the top and charm quarks as well as the $\alpha$ scheme for the electroweak parameters were used in Ref.~\cite{Eilam:2006uh}.} \label{tab:cutdepen}
	\end{center}
    \vspace{-0.2cm}
\end{table}

\begin{figure}[t]
	\centering
	\includegraphics[width=0.99\textwidth]{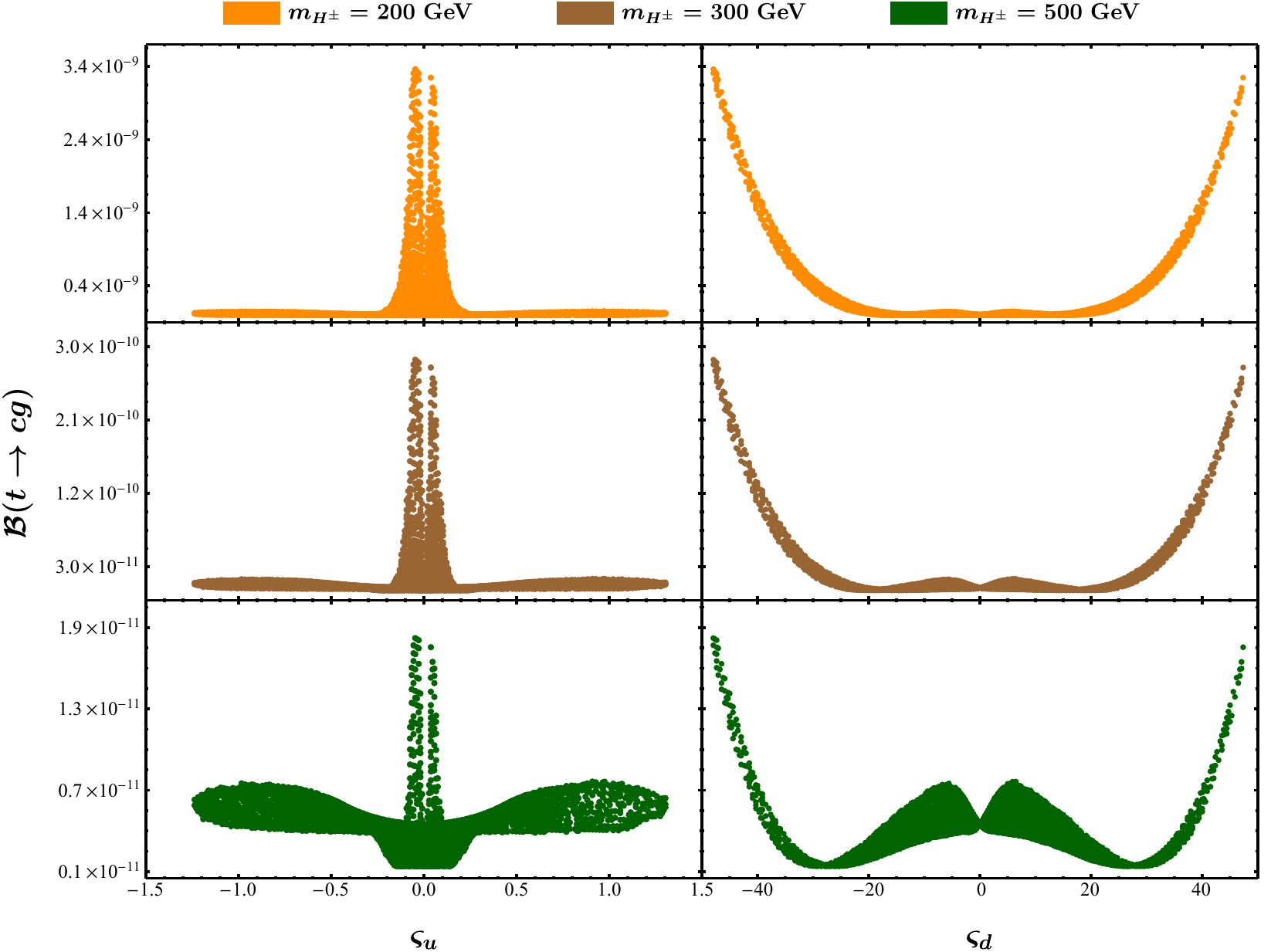}
	\caption{The branching ratio $\mathcal{B}(t\to cg)$ as a function of the alignment parameters $\varsigma_u$ (left) and $\varsigma_d$ (right), for three benchmark values of the charged-Higgs mass, $m_{H^{\pm}}=200$, $300$, and $500\,\gev$, represented by the orange (top), brown (middle), and dark green (bottom) scatter plots, respectively. Here we have used the relevant constraints on the alignment parameters $\varsigma_{u,d}$ from a global fit obtained at the $95.5\%$ CL in Ref.~\cite{Eberhardt:2020dat}. The thickness of the shaded regions reflects the corresponding allowed range for $\varsigma_{d(u)}$ for a given value of $\varsigma_{u(d)}$, as can be read from the brown region of the left plot of Figure~9 in Ref.~\cite{Eberhardt:2020dat}. } \label{tcgdu}
\end{figure}

\begin{figure}[t]
	\centering
	\includegraphics[width=0.99\textwidth]{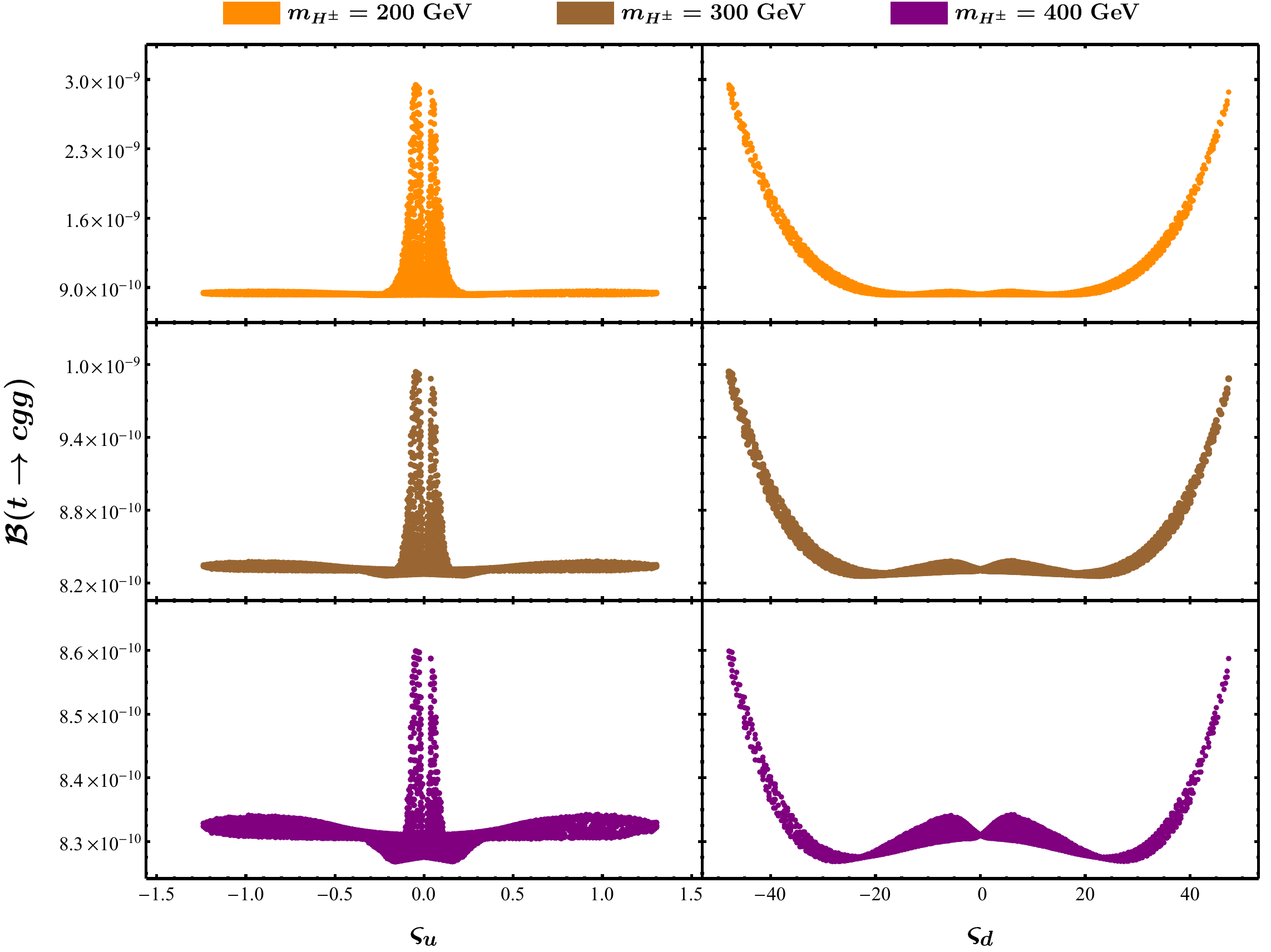}
	\caption{The branching ratio $\mathcal{B}(t\to cgg)$ as a function of the alignment parameters $\varsigma_u$ (left) and $\varsigma_d$ (right), for three benchmark values of the charged-Higgs mass, $m_{H^{\pm}}=200$, $300$, and $400\,\gev$, represented by the orange (top), brown (middle), and purple (bottom) scatter plots, respectively. The other captions are the same as in Fig.~\ref{tcgdu}.} \label{tcggdu}
\end{figure}

\begin{figure}[t]
	\centering
	\includegraphics[width=0.49\textwidth]{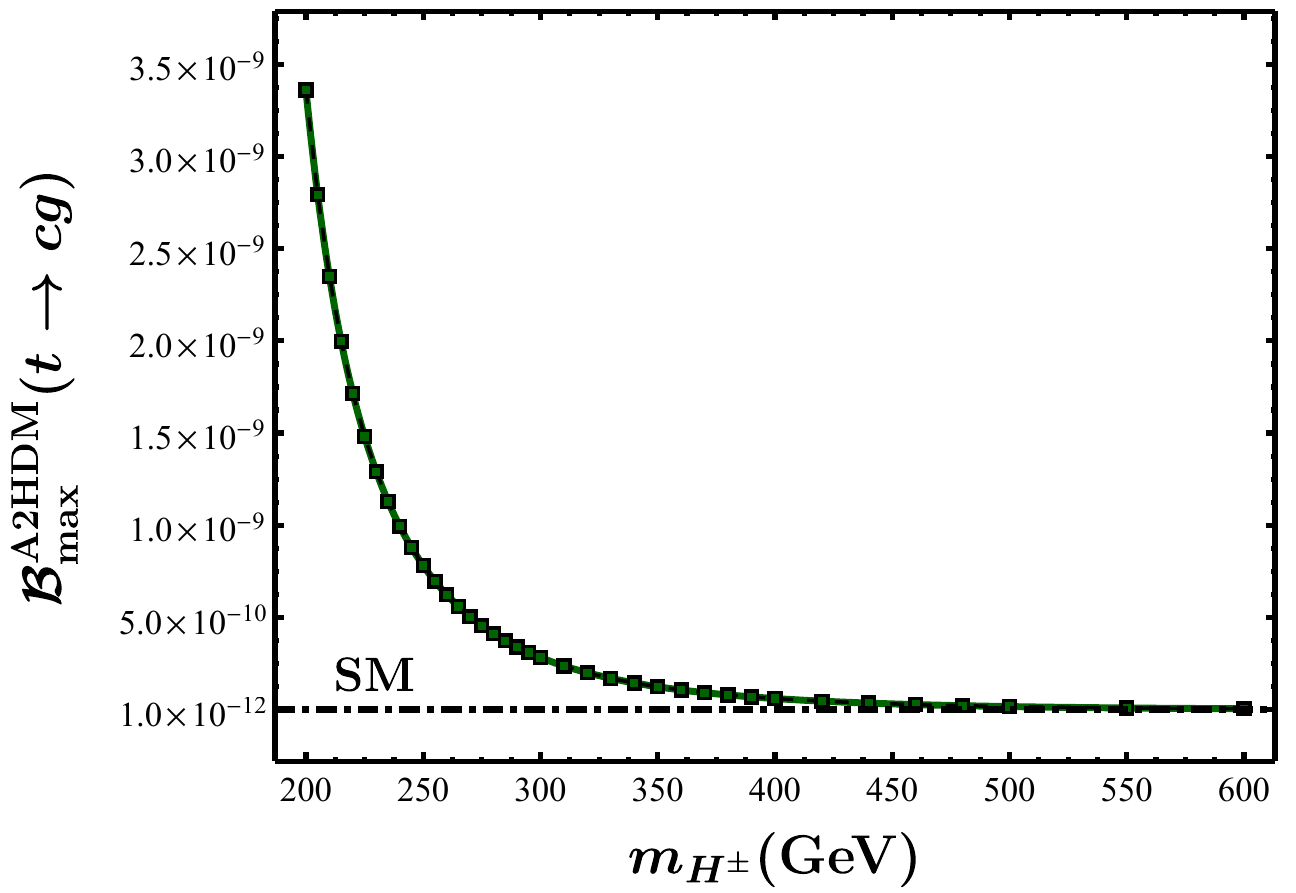}\hspace{0.12cm}
	\includegraphics[width=0.49\textwidth]{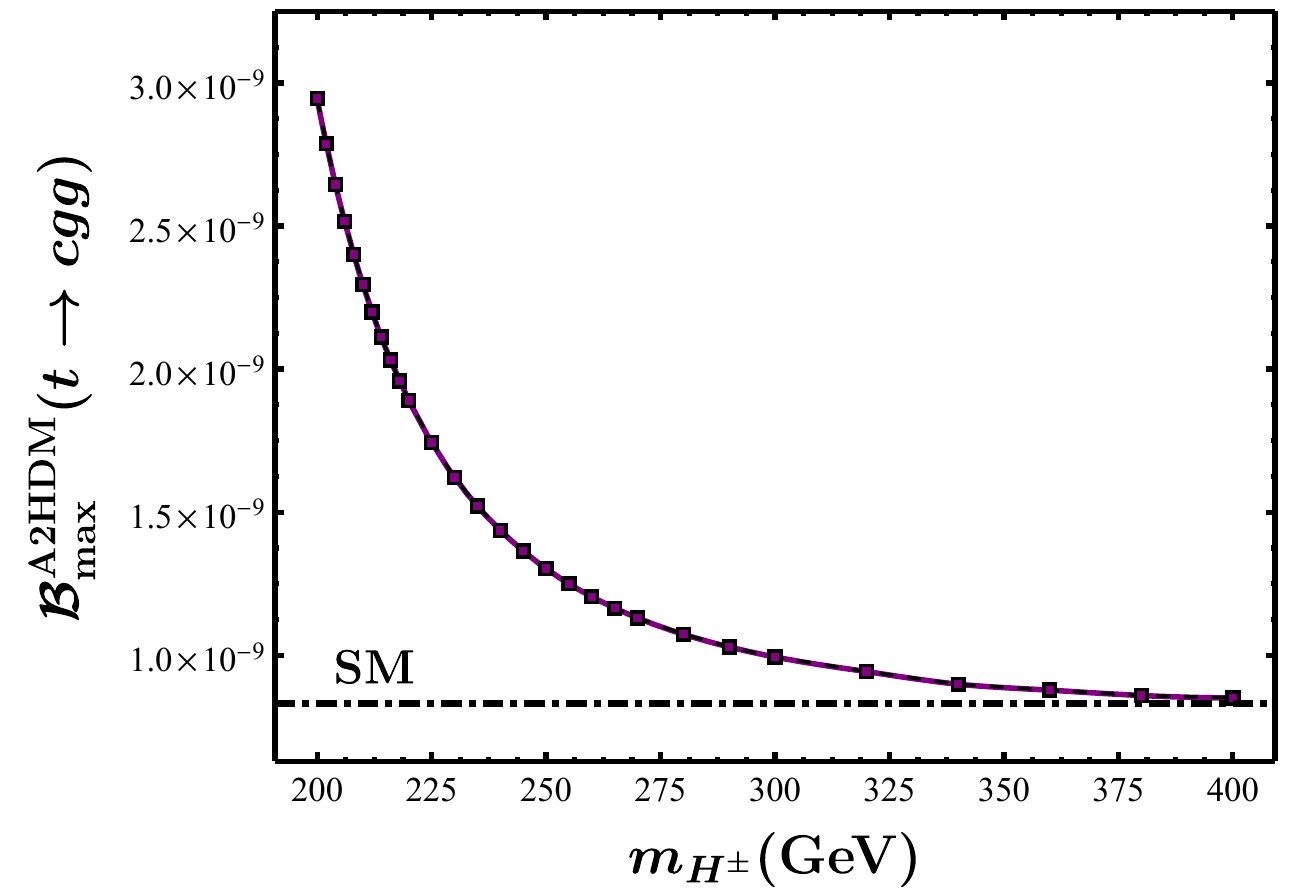}
	\caption{The dependence of the maximum branching ratios of $t\to cg$ (left) and $t\to cgg$ (right) on the charged-Higgs mass $m_{H^{\pm}}$, with the alignment parameters fixed at $(\varsigma_u,\varsigma_d) \simeq (-0.047,-48)$. The horizontal dash-dotted lines represent the SM predictions given by Eqs.~\eqref{eq:NbrSMtcg} and \eqref{eq:NbrSMtcgg}, respectively. } \label{BrmH}
\end{figure}

Picking up randomly the values of the alignment parameters $\varsigma_{u,d}$ within the brown region (corresponding to the global fit obtained at the $95.5\%$ CL) shown by the left plot of Figure~9 in Ref.~\cite{Eberhardt:2020dat}, we show in Figs.~\ref{tcgdu} and~\ref{tcggdu} the branching ratios $\mathcal{B}(t\to cg)$ and $\mathcal{B}(t\to cgg)$ as a function of $\varsigma_u$ (left) and $\varsigma_d$ (right), for three benchmark values of the charged-Higgs mass, $m_{H^{\pm}}=200$, $300$, and $500\,(400)\,\gev$, represented by the orange (top), brown (middle), and dark green (purple) (bottom) scatter plots, respectively. It can be seen that the maximum branching ratios of both $t \to cg$ and $t \to cgg$ decays are reached in the limits $|\varsigma_u|\to 0$ and $|\varsigma_d|\to 48$, which correspond to the minimum $|\varsigma_u|$ and maximum $|\varsigma_d|$ allowed by the combined constraints from the most constraining observables and the theoretical requirements~\cite{Eberhardt:2020dat}. Furthermore, almost irrespective of the values of the charged-Higgs mass, the maxima of the branching ratios are always reached with the same set of alignment parameters, $(\varsigma_u,\varsigma_d) \simeq (-0.047,-48)$, when $m_{H^\pm}$ varies within the ranges $[200, 600]\,\gev$ for the $t \to cg$ and $[200, 400]\,\gev$ for the $t \to cgg$ decay. Numerically, we obtain
\begin{align}
	\mathcal{B}^{\text{A2HDM}}_\text{max}(t \to cg)&=
	\begin{cases}
	3.36\times 10^{-9},  & \text{for $m_{H^\pm}=200\,\gev$},\\[0.1cm]
	2.84\times 10^{-10}, & \text{for $m_{H^\pm}=300\,\gev$},\\[0.1cm]
	1.82\times 10^{-11}, & \text{for $m_{H^\pm}=500\,\gev$},
	\end{cases}
\end{align}
and 
\begin{align}
	\mathcal{B}^{\text{A2HDM}}_\text{max}(t \to cgg)&=
	\begin{cases}
		2.95\times 10^{-9}, & \text{for $m_{H^\pm}=200\,\gev$},\\[0.1cm]
		9.94\times 10^{-10}, & \text{for $m_{H^\pm}=300\,\gev$},\\[0.1cm]
		8.60\times 10^{-10}, & \text{for $m_{H^\pm}=400\,\gev$}.
	\end{cases}
\end{align}
It can be seen that, with the $95.5\%$-CL constraints on the alignment parameters from the global fit taken into account~\cite{Eberhardt:2020dat}, the maximum branching ratios of $t \to cg$ and $t \to cgg$ decays are both reached at $m_{H^\pm}=200\,\gev$ and given by $3.36\times 10^{-9}$ and $2.95\times 10^{-9}$ respectively, being therefore of the same order, in the A2HDM. This is obviously different from the SM case, where the predicted branching ratio of $t \to cgg$ is about two orders of magnitude larger than that of $t \to cg$ (cf. Eqs.~\eqref{eq:NbrSMtcg} and \eqref{eq:NbrSMtcgg})~\cite{Eilam:2006uh}. For convenience, we show in Fig.~\ref{BrmH} the dependence of the maximum branching ratios of $t\to cg$ (left) and $t\to cgg$ (right) on the charged-Higgs mass $m_{H^{\pm}}$, with the alignment parameters fixed at $(\varsigma_u,\varsigma_d) \simeq (-0.047,-48)$.\footnote{Here the plots are obtained by firstly calculating the maximum branching ratios of $t \to cg$ and $t \to cgg$ decays with different values of $m_{H^{\pm}}$ and the same set of alignment parameters $(\varsigma_u,\varsigma_d) \simeq (-0.047,-48)$, and then constructing an interpolation of the maximum branching ratios as a function of $m_{H^{\pm}}$.} One can see that the maximum branching ratios of $t \to cg$ and $t \to cgg$ decays decrease sharply as the charged-Higgs mass $m_{H^{\pm}}$ increases, and approach the corresponding SM predictions for $m_{H^\pm}\gtrsim 600\,\gev$ and $m_{H^\pm}\gtrsim 400\,\gev$, respectively.\footnote{Numerically, we find that the maximum branching ratios of $t \to cg$ and $t \to cgg$ decays in the A2HDM are given by $6.63\times 10^{-12}$ for $m_{H^\pm}=600\,\gev$ and $8.60\times 10^{-10}$ for $m_{H^\pm}=400\,\gev$ respectively, which indeed approach the corresponding SM predictions given by Eqs.~\eqref{eq:NbrSMtcg} and \eqref{eq:NbrSMtcgg}.} In addition, the maximum branching ratio $\mathcal{B}^{\text{A2HDM}}_\text{max}(t \to cg(g))$ will change only slightly when $m_{H^\pm}\gtrsim 600~(400)\,\gev$. These observations motivate our choice of the upper limits for the charged-Higgs mass mentioned before. On the other hand, although being enhanced by about three orders of magnitude relative to the SM prediction, the maximum branching ratio of $t\to cg$ decay that can be reached in the A2HDM, $\mathcal{B}^{\text{A2HDM}}_\text{max}(t \to cg)=3.36\times 10^{-9}$, is still at least one order of magnitude lower than the projected $95\%$-CL limits at the future colliders, which are estimated to be $3.21 \times 10^{-5}$ at the HL-LHC with $3$-$\text{ab}^{-1}$ luminosity data~\cite{Cerri:2018ypt,Atlas:2019qfx}, and of $\mathcal{O}(10^{-7}-10^{-8})$ at the $100$-$\mathrm{TeV}$ FCC-hh with an integrated luminosity of $10~\text{ab}^{-1}$~\cite{Oyulmaz:2019jqr,Khanpour:2019qnw}, respectively. As a consequence, the predicted branching ratios of $t \to cg$ and $t \to cgg$ decays in the A2HDM are still out of the expected sensitivities of the future HL-LHC and FCC-hh.

\subsection{$\mathcal{B}(t \to cg(g))$ in the conventional 2HDMs with $\mathcal{Z}_2$ symmetries}
\label{sec:t2cggin2HDM}

\begin{figure}[t]
	\centering
	\includegraphics[width=0.49\textwidth]{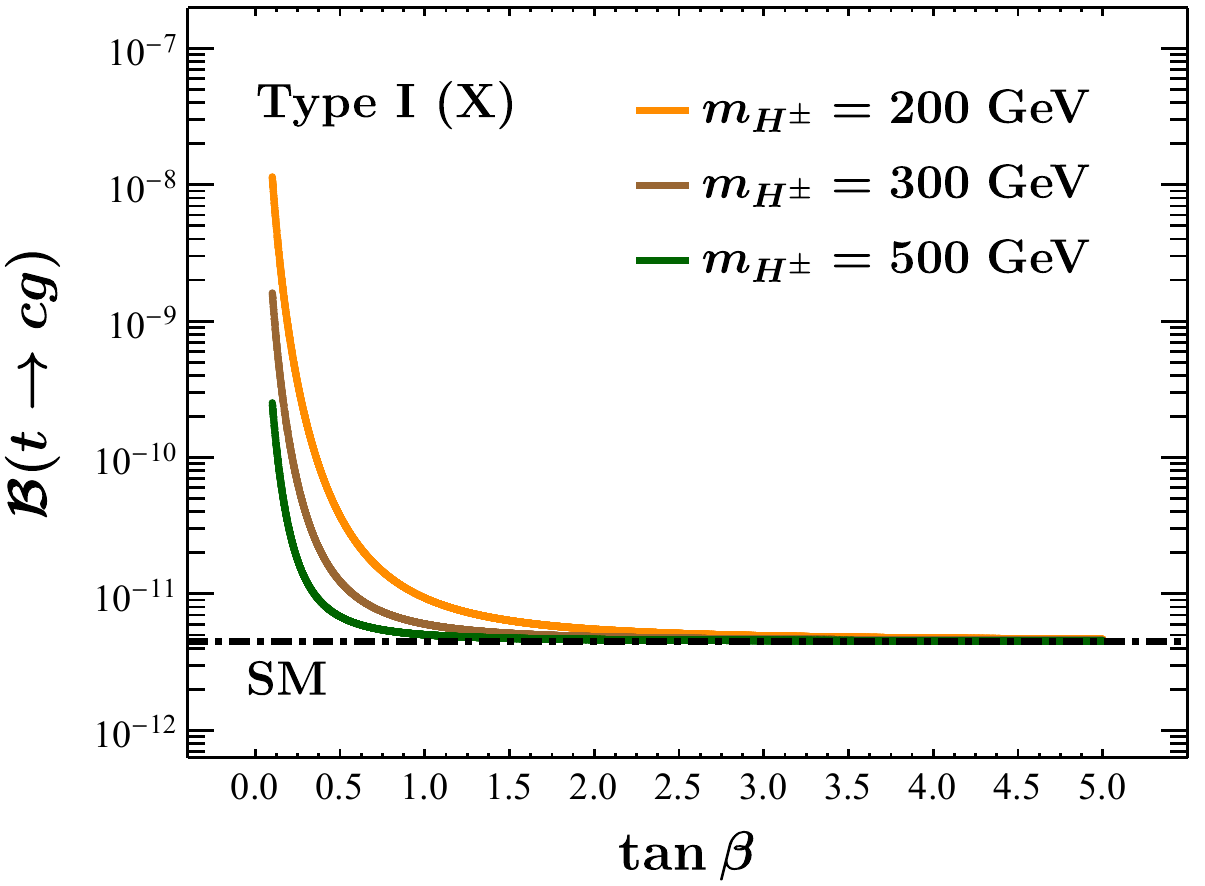}\hspace{0.12cm}
	\includegraphics[width=0.49\textwidth]{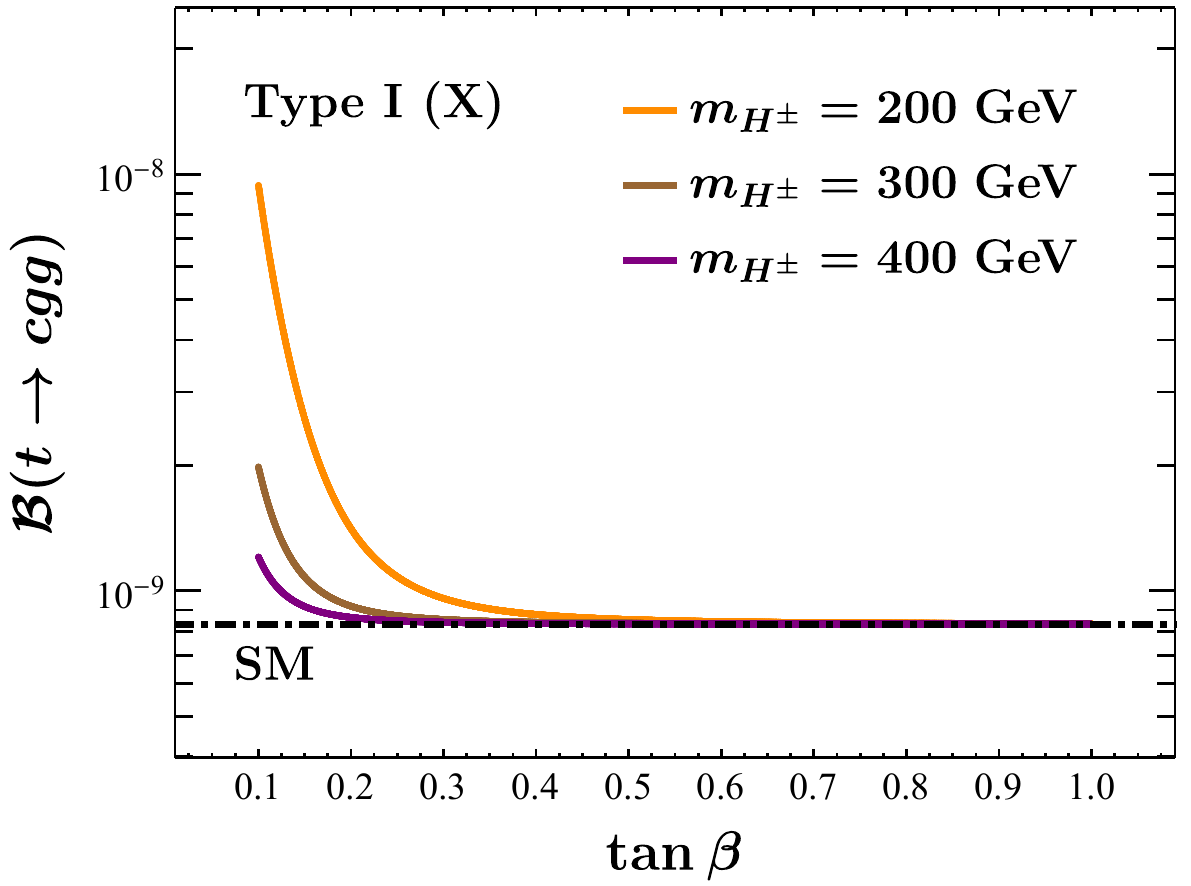}\vspace{0.2cm}
	\includegraphics[width=0.49\textwidth]{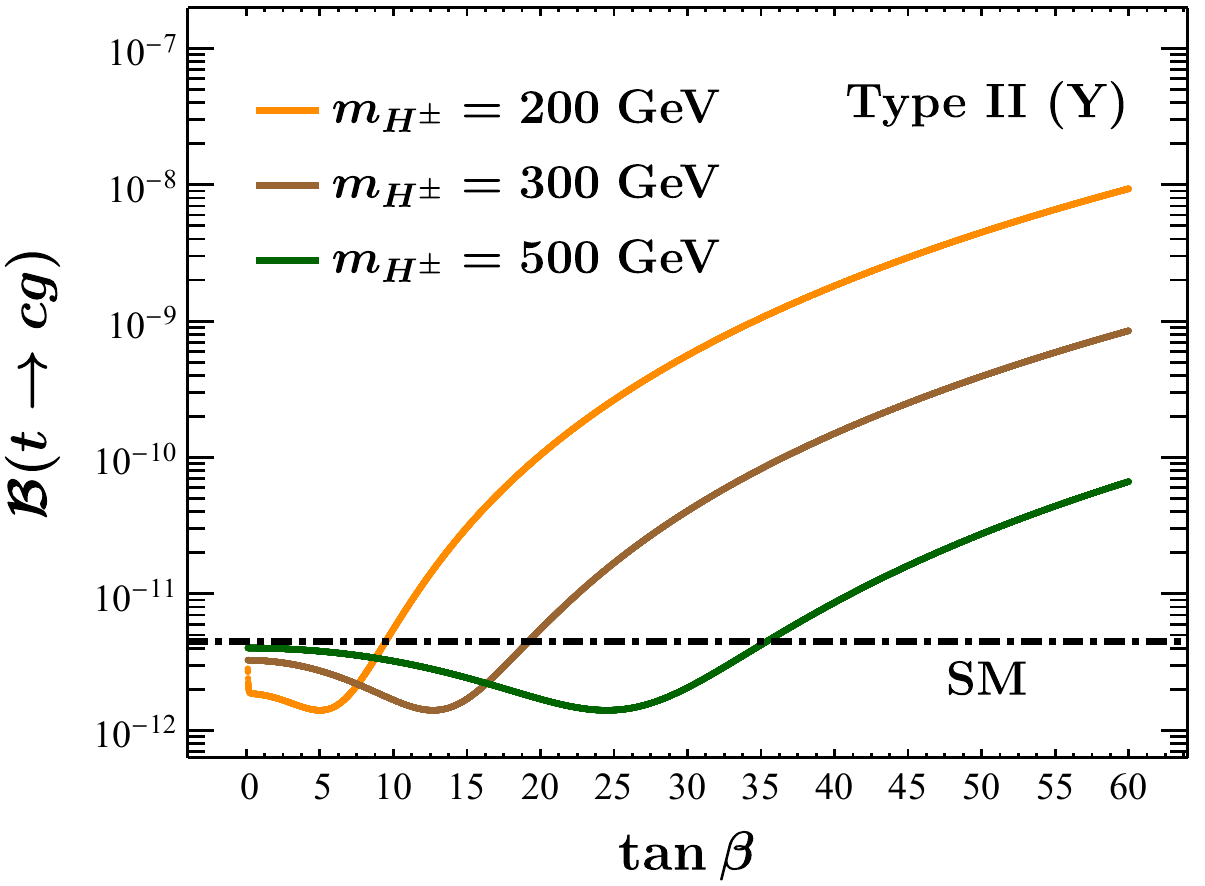}\hspace{0.12cm}
	\includegraphics[width=0.49\textwidth]{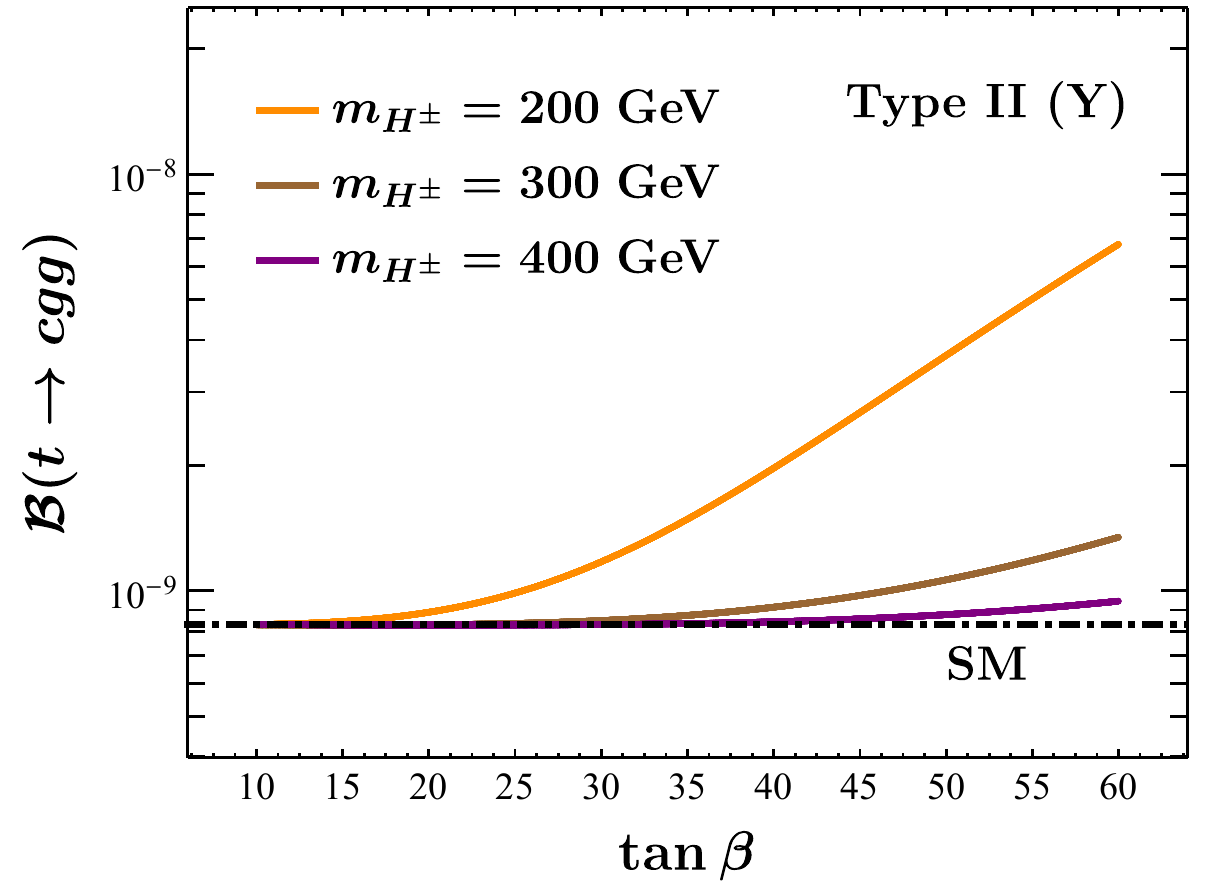}
	\caption{The branching ratios of $t\to cg$ (left) and $t\to cgg$ (right) as a function of the parameter $\tan{\beta}=[0.1,60]$, for three different choices of the charged-Higgs mass $m_{H^{\pm}}$, in the four conventional 2HDMs with $\mathcal{Z}_2$ symmetries. The horizontal dash-dotted lines represent the SM predictions. We show only the ranges of $\tan{\beta}$ where the resulting branching ratios display significant enhancements compared with the corresponding SM predictions. Note that the charged-Higgs mass $m_{H^{\pm}}$ in type-II~(type-Y) 2HDM is already bounded to be larger than $600\,\gev$~\cite{Misiak:2020vlo,Atkinson:2021eox,Hermann:2012fc}, while the parameter $\tan\beta$ in type-I~(type-X) 2HDM is constrained to be larger than one~\cite{Arbey:2017gmh}, both of which imply that no significant enhancements of the branching ratios are possible in these conventional 2HDMs.} \label{fig:t2cgandt2cgg2HDMwithZ2}
\end{figure}

Let us now discuss the branching ratios of $t \to cg$ and $t \to cgg$ decays in the four conventional 2HDMs with $\mathcal{Z}_2$ symmetries, the Yukawa types of which have been given explicitly in Table~\ref{tab:Z2symmetries}. As only the charged-Higgs interactions with the quark sector are involved, we are actually left with only two different cases, type-I~(type-X) 2HDM with $\varsigma_d=\varsigma_u=1/\tan{\beta}$, and type-II~(type-Y) 2HDM with $\varsigma_d=-1/\varsigma_u=-\tan{\beta}$. Here, to predict the branching ratios of $t \to cg$ and $t \to cgg$ decays, we need only the knowledge of the charged-Higgs mass $m_{H^{\pm}}$ and the parameter $\tan{\beta}$, both of which receive severe constraints from many flavour processes and theoretical considerations~\cite{Akeroyd:2016ymd,Arbey:2017gmh,Chowdhury:2017aav,Haller:2018nnx}. For convenience, we show in Fig.~\ref{fig:t2cgandt2cgg2HDMwithZ2} the dependence of the two branching ratios $\mathcal{B}(t \to cg)$~(left) and $\mathcal{B}(t \to cgg)$~(right) on $\tan{\beta}$, for three different choices of $m_{H^{\pm}}$, in the four conventional 2HDMs with $\mathcal{Z}_2$ symmetries. It should be noted that the branching ratios of $t\to cg$ and $t\to cgg$ decays are calculated in these four models by varying $\tan{\beta}=[0.1,60]$, while the ranges where the resulting branching ratios are not significantly enhanced compared with the corresponding SM predictions are not shown in Fig.~\ref{fig:t2cgandt2cgg2HDMwithZ2}. It can be seen that the sensitivity of the two branching ratios to $m_{H^{\pm}}$ occurs at low and high $\tan\beta$ for the type-I~(type-X) and the type-II~(type-Y) 2HDM respectively, and such a sensitivity becomes, however, less significant for larger $m_{H^{\pm}}$. 

For the type-II 2HDM, the branching ratio of the weak radiative decay $\bar{B}\to X_s\gamma$ plays a significant role in setting a lower bound on the charged-Higgs mass $m_{H^{\pm}}$, with the latest updated results given by $m_{H^{\pm}} \gtrsim 800\,\gev$ at $95\%$ CL~\cite{Misiak:2020vlo} and $m_{H^{\pm}} \gtrsim 790\,(1510)\,\gev$ at $2\sigma\,(1\sigma)$~\cite{Atkinson:2021eox}. These bounds are nearly $\tan\beta$-independent for $\tan\beta > 2$ and show a strong dependence on $\tan\beta$ only for $\tan\beta < 2$~\cite{Hermann:2012fc}. Quite recently, a comprehensive study of the allowed parameter space of the type-II 2HDM has been performed by combining the most recent flavour, collider, and electroweak precision observables with theoretical constraints, with the best fit point lying around $m_{H^{\pm}} \simeq 2\,\mathrm{TeV}$ and $\tan\beta \simeq 4$~\cite{Atkinson:2021eox}. The experimental lower bound on $m_{H^{\pm}}$ from direct collider searches is, however, only $m_{H^{\pm}} \gtrsim 160\,\gev$~\cite{ATLAS:2018gfm}, and hence well below the indirect constraints from flavour physics. With these bounds taken into account, it is then found that the branching ratios of both $t \to cg$ and $t \to cgg$ decays remain almost identical with the corresponding SM predictions, as given by Eqs.~\eqref{eq:NbrSMtcg} and \eqref{eq:NbrSMtcgg} respectively. The same conclusion can also be applied to the type-Y 2HDM, because its Yukawa type shares the same coupling pattern for the quarks as of the type-II 2HDM~\cite{Arbey:2017gmh}. In fact, for these two types of 2HDMs, we have $\varsigma_d=-1/\varsigma_u=-\tan{\beta}$ and hence $\varsigma_d \varsigma_u=-1$, while the region $\varsigma_{d} \varsigma_{u}<0$ is almost excluded by the current data on the branching ratio of the weak radiative decay $\bar{B}\to X_s\gamma$~\cite{Jung:2010ik}. This also implies that significant enhancements of the branching ratios of $t \to cg$ and $t \to cgg$ decays relative to the SM predictions are impossible in these two types of 2HDMs.

For the type-I~(type-X) 2HDM, on the other hand, a $95\%$-CL lower limit on $m_{H^{\pm}}$ from the branching ratio of $\bar{B}\to X_s\gamma$ can be derived for low $\tan\beta$ only~\cite{Hermann:2012fc}. As an illustration, a bound $\tan\beta \lesssim 2.5$ can be set when the robust lower charged-Higgs mass limit $m_{H^{\pm}} \gtrsim 80\,\gev$ from the combined charged-Higgs searches at LEP~\cite{ALEPH:2013htx} is considered. A detailed analysis of the phenomenological status of the charged-Higgs sector in the conventional 2HDMs with $\mathcal{Z}_2$ symmetries has been performed by combing all the constraints from direct collider searches for charged and neutral Higgs bosons, as well as the most relevant constraints from flavour physics~\cite{Arbey:2017gmh}. Using the combined flavour constraints at $95\%$ CL obtained there, we extract the values $\tan\beta \simeq 1.8$ for $m_{H^\pm}=200\,\gev$, $\tan\beta \simeq 1.5$ for $m_{H^\pm}=300\,\gev$, and $\tan\beta \simeq 1.2$ for $m_{H^\pm}=500\,\gev$~\cite{Arbey:2017gmh}. With these typical values taken as inputs, the corresponding branching ratios of $t \to cg$ and $t \to cgg$ decays are given in Table~\ref{tab:typeIX}. It can be seen that, in these two types of 2HDMs, the branching ratio of $t\to cg$ decay is predicted to be of the same order as in the SM (cf. Eq.~\eqref{eq:NbrSMtcg}), while the branching ratio of $t\to cgg$ decay remains almost identical with the SM expectation (cf. Eq.~\eqref{eq:NbrSMtcgg}). 

\begin{table}[t]
	\begin{center}	
		\let\oldarraystretch=\arraystretch
		\renewcommand*{\arraystretch}{1.3}
		\tabcolsep=0.67cm
		\begin{tabular}{|cccc|}
			\hline\hline
			$(m_{H^\pm}\,[\gev],\,\tan\beta)$
			& $(200,\, 1.8)$
			& $(300,\, 1.5)$
			& $(500,\, 1.2)$
			\\\hline
			$\mathcal{B}^{\text{Type I\,(X)}}(t \to cg)$
			& $5.75\times 10^{-12}$
			& $5.13\times 10^{-12}$
			& $4.85\times 10^{-12}$
			\\
			$\mathcal{B}^{\text{Type I\,(X)}}(t \to cgg)$
			& $8.32\times 10^{-10}$
			& $8.31\times 10^{-10}$
			& $8.31\times 10^{-10}$
			\\
			\hline\hline
		\end{tabular}
		\caption{The branching ratios of $t \to cg$ and $t \to cgg$ decays in the type-I~(type-X) 2HDM, with three sets of input parameters of $(m_{H^\pm},\tan\beta)$ extracted from Ref.~\cite{Arbey:2017gmh}.} \label{tab:typeIX}
	\end{center}
\end{table}

In fact, at low $\tan\beta$, the flavour constraints are similar among all the four conventional 2HDMs with $\mathcal{Z}_2$ symmetries, because the contributions in this case are now dominated by the universal charged-Higgs couplings to the up-type quarks (cf. Table~\ref{tab:Z2symmetries}). Then, it is found that values of $\tan\beta <1$ are ruled out for all values of $m_{H^\pm} <650\,\gev$, irrespective of the Yukawa types considered~\cite{Arbey:2017gmh}. As a summary, we can therefore conclude that, compared with the SM predictions, no significant enhancements are observed in the four conventional 2HDMs with $\mathcal{Z}_2$ symmetries for the branching ratios of both $t \to cg$ and $t \to cgg$ decays, once the flavour constraints on the model parameters are taken into account.

\subsection{Further discussions and comments}
\label{sec:discussion}

Let us now explain why the ``higher-order dominance'' phenomenon is more obvious within the SM than in the A2HDM as well as the four conventional 2HDMs with $\mathcal{Z}_2$ symmetries. Naively, one would expect that the rate of the two-body decay $t \to cg$ is larger than that of the three-body decay $t \to cgg$, due to the coupling-constant and phase-space suppression of the latter. This is, however, not true within the SM; instead, we find that the predicted branching ratio of $t \to cgg$ is about two orders of magnitude larger than that of $t \to cg$ (cf. Eqs.~\eqref{eq:NbrSMtcg} and \eqref{eq:NbrSMtcgg})~\cite{Eilam:2006uh}. To understand such a ``higher-order dominance'' phenomenon, one should firstly note that both the $t \to cg$ and $t \to cgg$ decays are dominated by the one-loop effective vertex $V_{tcg^{\ast}}^\mu$, which can be generically written as~\cite{Deshpande:1991pn}
\begin{align}
	V_{tcg^{\ast}}^{\mu} = \left(k^{2} \gamma^{\mu} - k^{\mu} k\!\!\!/\right)\left(F_{1}^{L} P_L+F_{1}^{R} P_R\right) + i \sigma^{\mu \nu} k_{\nu}\left(F_{2}^{L} m_{c} P_L+F_{2}^{R} m_{t} P_R\right)\,,
\end{align}
as required by Lorentz and gauge invariance. Here we have suppressed both the colour labels
and a sum over the internal quark flavours together with the CKM matrix elements. The form factors $F_{1}^{L,R}$ and $F_{2}^{L,R}$ are a function of the gluon virtuality $k^2$, and are called the monopolar (or charge radius) and the dipolar (or dipole moment) form factors, respectively. Note that when $k^2 \to 0$, \textit{i.e.}, when the gluon goes on-shell, the $F_{1}^{L,R}$ contributions vanish. Therefore, the two-body decay $t \to cg$ is entirely determined by the dipole structure, while both the dipole and the monopole structure contribute to the three-body decay $t \to cgg$. It also turns out that the contribution from the monopolar form factors can be considerably larger than from the dipolar ones within the SM~\cite{Hou:1988wt,Simma:1990nr,Liu:1989pc,Eilam:2006uh}. This explains the ``higher-order dominance'' phenomenon observed within the SM. For the A2HDM, however, the charged-Higgs enhancements are mostly through the dipolar form factors, and hence show up only through a greatly enhanced branching ratio of the two-body decay $t \to cg$. This helps us to understand why the branching ratios of $t \to cg$ and $t \to cgg$ decays are predicted to be of the same order in the A2HDM. For the four conventional 2HDMs with $\mathcal{Z}_2$ symmetries, on the other hand, after taking into account the relevant flavour constraints on the model parameters, significant enhancements are no longer allowed for the branching ratios of both $t \to cg$ and $t \to cgg$ decays.     

Finally, we would like to make a brief comment on the observability of $t \to cgg$ decay. Note that this process can generally be treated in twofold way: One can either treat it inclusively with $t \to cg$ or consider it as a separate channel~\cite{Eilam:2006rb}. The former means that the process $t \to cgg$ can be considered as a QCD correction to the $t \to cg$ decay by assuring that two of the three final-state jets are collinear so that only two can be resolved by the detector. The latter is especially promising if the branching ratio of $t \to cgg$ is considerably larger than that of $t \to cg$, as observed within the SM~\cite{Eilam:2006uh}. However, to detect $t \to cgg$ before $t \to cg$, one must avoid collinearity by applying certain cuts. To this end, the cutoff parameter $C$ introduced in the phase space plays an essential role in distinguishing the $t \to cgg$ from the $t \to cg$ decay. Here, $C$ must be taken in the range of jet-energy resolution of the detector, and a better jet resolution will be definitely welcome.

\section{Conclusion}
\label{sec:conclusions}

In this paper, we have firstly updated the SM predictions for the branching ratios of both $t \to cg$ and $t \to cgg$ decays, with the results given respectively by $\mathcal{B}^{\text{SM}}(t \to cg) = 4.50\times 10^{-12}$ and $\mathcal{B}^{\text{SM}}(t \to cgg) = 8.31\times 10^{-10}$, finding that the latter is about two orders of magnitude larger than the former, which is a clear demonstration of the ``higher-order dominance'' phenomenon observed in Ref.~\cite{Eilam:2006uh}. This is due to the fact that the contribution from the monopolar form factors can be considerably larger than from the dipolar ones within the SM~\cite{Hou:1988wt,Simma:1990nr,Liu:1989pc,Eilam:2006uh}. Note that the branching ratio of the three-body decay $t \to cgg$ is obtained with the cutoff parameter $C=0.01$. 

To check if such a ``higher-order dominance'' phenomenon is also applied in the A2HDM as well as in the four conventional 2HDMs with $\mathcal{Z}_2$ symmetries, we have also evaluated the maximum branching ratios of $t \to cg$ and $t \to cgg$ decays that can be reached in these different models. After taking into account the relevant constraints on the model parameters resulting from a global fit obtained at the $95.5\%$ CL~\cite{Eberhardt:2020dat}, we find that the branching ratios of $t \to cg$ and $t \to cgg$ decays can reach up to $3.36\times 10^{-9}$ and $2.95\times 10^{-9}$ respectively, being therefore of the same order, in the A2HDM. This is obviously different from the SM case, and can be understood by the fact that the charged-Higgs enhancements are mostly through the dipolar form factors, and hence show up only through a greatly enhanced branching ratio of the two-body decay $t \to cg$. On the other hand, compared with the SM predictions, significant enhancements are no longer possible in the four conventional 2HDMs with $\mathcal{Z}_2$ symmetries for the branching ratios of these two decays, once the flavour constraints on the model parameters~\cite{Atkinson:2021eox,Arbey:2017gmh} are properly taken into account. 

Nevertheless, the predicted branching ratios of $t \to cg$ and $t \to cgg$ decays in the A2HDM are still out of the expected sensitivities of the future HL-LHC and FCC-hh. To have an opportunity to detect $t \to cgg$ before $t \to cg$, we have to improve the jet resolution of the future detector. Note that, due to the imaginary components of the loop functions, the CKM matrix elements, as well as the complex alignment parameters $\varsigma_{u,d}$ in the A2HDM, it would also be interesting to investigate the CP asymmetries in these rare top-quark decays~\cite{Balaji:2020qjg,AguilarSaavedra:2002ns}. 

\section*{Acknowledgements}
We are very grateful to Thomas Hahn for useful communications on the function \textit{PolarizationSum} defined in the package \texttt{FormCalc}. This work is supported by the National Natural Science Foundation of China under Grant Nos.~12135006, 12075097, 11947213, 11675061 and 11775092, as well as by the Fundamental Research Funds for the Central Universities under Grant Nos.~CCNU20TS007, CCNU19TD012 and CCNU22LJ004. S.F. is also supported by the International Postdoctoral Exchange Fellowship Program. 

\appendix

\section{\boldmath The $R$ functions relevant for $t\to cgg$ decay}
\label{sec:appendix}

In this appendix, we give the explicit expressions of the remaining $R$ functions present in Eqs.~\eqref{eq:Amptcggvertex}--\eqref{eq:Amptcggself}. For convenience, the scalar products involved are defined in the same way as in Ref.~\cite{Eilam:2006uh}, with
\begin{align}
	\text{SP1} &= \epsilon^*(k_3)\cdot \epsilon^*(k_4),  &&
	\text{SP2} = \epsilon^*(k_3)\cdot k_1, &&
	\text{SP3} = \epsilon^*(k_3)\cdot k_2, \nn \\[0.15cm]
	\text{SP4} &= \epsilon^*(k_4)\cdot k_1, &&
	\text{SP5} = \epsilon^*(k_4)\cdot k_2.
\end{align}
For the vertex part (cf. Eqs.~\eqref{eq:Amptcggvertex}), we have
\begin{eqnarray}
R_1^{\text{vertex}}&=&-(\text{SP1}\,F_5-F_{16}) (m_W^2(1-2 (B_{0}^{(3)}(m_W^2)-2 C_{00}^{(3)}(m_W^2)-s_{23} C_{2}^{(3)}(m_W^2))) \nn\\[0.1cm]
&-&m_b^2 (m_t^2(C_{0}^{(3)}(m_W^2)+C_{2}^{(3)}(m_W^2)+ \text{X}_{dd}C_{2}^{(3)}(m_{H^\pm}^2)+\text{X}_{du}C_{0}^{(3)}(m_{H^\pm}^2) ) +B_{0}^{(3)}(m_W^2)\nn\\[0.1cm]
&+&\text{X}_{dd}B_{0}^{(3)}(m_{H^\pm}^2)-2 (C_{00}^{(3)}(m_W^2)+\text{X}_{dd}C_{00}^{(3)}(m_{H^\pm}^2) )))+m_t (s_{23} (\text{SP1}\,F_1-F_{12})\nn\\[0.1cm]
&~& (2m_W^2 C_{2}^{(3)}(m_W^2) -m_b^2(C_{0}^{(3)}(m_W^2)+C_{2}^{(3)}(m_W^2)+\text{X}_{dd}C_{2}^{(3)}(m_{H^\pm}^2)\nn\\[0.1cm] &+&\text{X}_{du}C_{0}^{(3)}(m_{H^\pm}^2) ))
+2 \text{SP4} (2 \text{SP3}\,F_1-F_{13}) ( m_b^2(C_{0}^{(3)}(m_W^2)+2 C_{2}^{(3)}(m_W^2)\nn\\[0.1cm]
&+&C_{22}^{(3)}(m_W^2)+\text{X}_{dd}(C_{12}^{(3)}(m_{H^\pm}^2)+C_{2}^{(3)}(m_{H^\pm}^2)
+C_{22}^{(3)}(m_{H^\pm}^2)) +\text{X}_{du}(C_{0}^{(3)}(m_{H^\pm}^2)\nn\\[0.1cm]
&+&C_{2}^{(3)}(m_{H^\pm}^2)) )+2m_W^2C_{22}^{(3)}(m_W^2)+(m_b^2+2 m_W^2)C_{12}^{(3)}(m_W^2) )) \nn\\[0.1cm]
&+&F_3 (-( m_t^2(\text{SP4} C_{2}^{(3)}(m_W^2)-\text{SP5}\,\text{X}_{dd}C_{2}^{(3)}(m_{H^\pm}^2) )+\text{SP4} (B_{0}^{(3)}(m_W^2)-2 C_{00}^{(3)}(m_W^2))\nn\\[0.1cm]
&+&(\text{SP4}-\text{SP5}) (m_t^2(C_{0}^{(3)}(m_W^2)+\text{X}_{du}C_{0}^{(3)}(m_{H^\pm}^2) ) +(B_{0}^{(3)}(m_{H^\pm}^2)\nn\\[0.1cm]
&-&2 C_{00}^{(3)}(m_{H^\pm}^2)) \text{X}_{dd})) m_b^2+(\text{SP4}-\text{SP5}) m_W^2+(B_{0}^{(3)}(m_W^2)-2 C_{00}^{(3)}(m_W^2))(\text{SP5}m_b^2 \nn\\[0.1cm]
&-&2 (\text{SP4}-\text{SP5}) m_W^2)+C_{2}^{(3)}(m_W^2) (\text{SP5} m_b^2 m_t^2-2 s_{23} (\text{SP4}+\text{SP5}) m_W^2)\nn\\[0.1cm]
&-&\text{SP4} ((2 s_{23} C_{12}^{(3)}(m_{H^\pm}^2)+m_t^2 C_{2}^{(3)}(m_{H^\pm}^2)) \text{X}_{dd} m_b^2+2 s_{23} C_{12}^{(3)}(m_W^2) (m_b^2+2 m_W^2)))\nn\\[0.1cm]
&+&\text{SP3}\,F_4 ((4 (C_{00}^{(3)}(m_W^2)+C_{00}^{(3)}(m_{H^\pm}^2) \text{X}_{dd})-2 ((C_{0}^{(3)}(m_W^2)+C_{2}^{(3)}(m_W^2)\nn\\[0.1cm]
&+&C_{2}^{(3)}(m_{H^\pm}^2) \text{X}_{dd}+C_{0}^{(3)}(m_{H^\pm}^2) \text{X}_{du}) m_t^2+B_{0}^{(3)}(m_W^2)+B_{0}^{(3)}(m_{H^\pm}^2) \text{X}_{dd})) m_b^2\nn\\[0.1cm]
&+&(-4 B_{0}^{(3)}(m_W^2)+8 C_{00}^{(3)}(m_W^2)+4 s_{23} C_{2}^{(3)}(m_W^2)+2) m_W^2),\\[0.3cm]
R_2^{\text{vertex}}&=& (2 (s_{23} (\text{SP4}-\text{SP5}) C_{2}^{(2)}(m_W^2) F_3+\text{SP3}\,F_4) m_W^2-(\text{SP1}\,F_5-F_{16}) ((-2 B_{0}^{(3)}(m_W^2)\nn\\[0.1cm]
&+&4 C_{00}^{(2)}(m_W^2)+2 s_{23} C_{2}^{(2)}(m_W^2)+1) m_W^2-m_b^2 ((C_{0}^{(2)}(m_W^2)+C_{0}^{(2)}(m_{H^\pm}^2) \text{X}_{du}) m_t^2\nn\\[0.1cm]
&+&B_{0}^{(3)}(m_W^2)+B_{0}^{(3)}(m_{H^\pm}^2) \text{X}_{dd}-2 (C_{00}^{(2)}(m_W^2)+C_{00}^{(2)}(m_{H^\pm}^2) \text{X}_{dd})))+(\text{SP4}-\text{SP5}) F_3 \nn\\[0.1cm]
&~&(-B_{0}^{(3)}(m_{H^\pm}^2) \text{X}_{dd} m_b^2+(2 (C_{00}^{(2)}(m_W^2)+C_{00}^{(2)}(m_{H^\pm}^2) \text{X}_{dd})-m_t^2 (C_{0}^{(2)}(m_W^2)\nn\\[0.1cm]
&+&C_{0}^{(2)}(m_{H^\pm}^2) \text{X}_{du})) m_b^2+(4 C_{00}^{(2)}(m_W^2)+1) m_W^2-B_{0}^{(3)}(m_W^2) (m_b^2+2 m_W^2))\nn\\[0.1cm]
&-&\text{SP3}\,F_4 (4 (B_{0}^{(3)}(m_W^2)-2 C_{00}^{(2)}(m_W^2)) m_W^2+2 ((-(C_{2}^{(2)}(m_W^2)+C_{2}^{(2)}(m_{H^\pm}^2) \text{X}_{du}) m_t^2\nn\\[0.1cm]
&+&B_{0}^{(3)}(m_W^2)+B_{0}^{(3)}(m_{H^\pm}^2) \text{X}_{dd}-2 (C_{00}^{(2)}(m_W^2)+C_{00}^{(2)}(m_{H^\pm}^2) \text{X}_{dd})) m_b^2\nn\\[0.1cm]
&+&s_{23} (C_{12}^{(2)}(m_{H^\pm}^2) \text{X}_{dd} m_b^2+C_{12}^{(2)}(m_W^2) (m_b^2+2 m_W^2))))+m_t (-F_{12} ((1-2 (B_{0}^{(3)}(m_W^2)\nn\\[0.1cm]
&-&2 C_{00}^{(2)}(m_W^2)-s_{23} C_{2}^{(2)}(m_W^2))) m_W^2-m_b^2 (B_{0}^{(3)}(m_W^2)+B_{0}^{(3)}(m_{H^\pm}^2) \text{X}_{dd}\nn\\[0.1cm]
&-&2 (C_{00}^{(2)}(m_W^2)+C_{00}^{(2)}(m_{H^\pm}^2) \text{X}_{dd})+s_{23} (C_{0}^{(2)}(m_W^2)+C_{0}^{(2)}(m_{H^\pm}^2) \text{X}_{du})))\nn\\[0.1cm]
&-&\text{SP3}\,F_{14} ((2 (C_{12}^{(2)}(m_W^2)+C_{12}^{(2)}(m_{H^\pm}^2) \text{X}_{dd})-2(C_{0}^{(2)}(m_W^2)+C_{2}^{(2)}(m_W^2)\nn\\[0.1cm]
&+&(C_{0}^{(2)}(m_{H^\pm}^2)+C_{2}^{(2)}(m_{H^\pm}^2)) \text{X}_{du})) m_b^2+4 (C_{12}^{(2)}(m_W^2)+C_{2}^{(2)}(m_W^2)) m_W^2)\nn\\[0.1cm]
&+&F_1 (-((s_{23} \text{SP1}-2 \text{SP3} (\text{SP4}-\text{SP5})) C_{0}^{(2)}(m_W^2)-2 (\text{SP1} (C_{00}^{(2)}(m_W^2)\nn\\[0.1cm]
&+&C_{00}^{(2)}(m_{H^\pm}^2) \text{X}_{dd})+\text{SP3} (\text{SP4}-\text{SP5}) C_{0}^{(2)}(m_{H^\pm}^2) \text{X}_{du})) m_b^2+2 C_{2}^{(2)}(m_W^2) (s_{23} \text{SP1} m_W^2\nn\\[0.1cm]
&+&\text{SP3} (\text{SP4}-\text{SP5}) (m_b^2-2 m_W^2))+\text{SP1} (-(B_{0}^{(3)}(m_{H^\pm}^2) \text{X}_{dd}+s_{23} C_{0}^{(2)}(m_{H^\pm}^2) \text{X}_{du}) m_b^2\nn\\[0.1cm]
&+&(4 C_{00}^{(2)}(m_W^2)+1) m_W^2-B_{0}^{(3)}(m_W^2) (m_b^2+2 m_W^2))-2 \text{SP3} (\text{SP4}-\text{SP5})\nn\\[0.1cm]
&~& ((C_{12}^{(2)}(m_{H^\pm}^2) \text{X}_{dd}-C_{2}^{(2)}(m_{H^\pm}^2) \text{X}_{du}) m_b^2+C_{12}^{(2)}(m_W^2) (m_b^2+2 m_W^2)))),\\[0.3cm]
R_3^{\text{vertex}}&=&-((2 (\text{SP4}-\text{SP5}) t C_{2}^{(5)}(m_W^2) F_3-2 \text{SP2}\,F_4) m_W^2-\text{SP1}\,F_5 ((-2 B_{0}^{(4)}(m_W^2)\nn\\[0.1cm]
&+&4 C_{00}^{(5)}(m_W^2)+2 t C_{2}^{(5)}(m_W^2)+1) m_W^2-m_b^2 ( m_t^2(C_{0}^{(5)}(m_W^2)+C_{2}^{(5)}(m_W^2)\nn\\[0.1cm]
&+&C_{2}^{(5)}(m_{H^\pm}^2) \text{X}_{dd}+C_{0}^{(5)}(m_{H^\pm}^2) \text{X}_{du})+B_{0}^{(4)}(m_W^2)+B_{0}^{(4)}(m_{H^\pm}^2) \text{X}_{dd}-2 (C_{00}^{(5)}(m_W^2)\nn\\[0.1cm]
&+&C_{00}^{(5)}(m_{H^\pm}^2) \text{X}_{dd})))-F_{16} ((1-2 (B_{0}^{(4)}(m_W^2)-2 C_{00}^{(5)}(m_W^2)-t C_{2}^{(5)}(m_W^2))) m_W^2\nn\\[0.1cm]
&-&m_b^2 ( m_t^2(C_{0}^{(5)}(m_W^2)+C_{2}^{(5)}(m_W^2)+C_{2}^{(5)}(m_{H^\pm}^2) \text{X}_{dd}+C_{0}^{(5)}(m_{H^\pm}^2) \text{X}_{du})+B_{0}^{(4)}(m_W^2)\nn\\[0.1cm]
&+&B_{0}^{(4)}(m_{H^\pm}^2) \text{X}_{dd}-2 (C_{00}^{(5)}(m_W^2)+C_{00}^{(5)}(m_{H^\pm}^2) \text{X}_{dd})))\nn\\[0.1cm]
&+&(\text{SP4}-\text{SP5}) F_3 (-B_{0}^{(4)}(m_{H^\pm}^2) \text{X}_{dd} m_b^2+m_b^2(2 (C_{00}^{(5)}(m_W^2)+C_{00}^{(5)}(m_{H^\pm}^2) \text{X}_{dd})\nn\\[0.1cm]
&-&m_t^2 (C_{0}^{(5)}(m_W^2)+C_{2}^{(5)}(m_W^2)+C_{2}^{(5)}(m_{H^\pm}^2) \text{X}_{dd}+C_{0}^{(5)}(m_{H^\pm}^2) \text{X}_{du})) \nn\\[0.1cm]
&+&(4 C_{00}^{(5)}(m_W^2)+1) m_W^2-B_{0}^{(4)}(m_W^2) (m_b^2+2 m_W^2))+\text{SP2}\,F_4 (-(2 m_b^2(C_{2}^{(5)}(m_W^2)\nn\\[0.1cm]
&+&C_{22}^{(5)}(m_W^2)+C_{22}^{(5)}(m_{H^\pm}^2) \text{X}_{dd}+C_{2}^{(5)}(m_{H^\pm}^2) \text{X}_{du}) +4 C_{22}^{(5)}(m_W^2) m_W^2) m_t^2\nn\\[0.1cm]
&+&4m_W^2(B_{0}^{(4)}(m_W^2)-2 C_{00}^{(5)}(m_W^2)) +2 ((-C_{12}^{(5)}(m_W^2) m_t^2+B_{0}^{(4)}(m_W^2)+(B_{0}^{(4)}(m_{H^\pm}^2)\nn\\[0.1cm]
&-&(s_{23}+t_{12}) C_{12}^{(5)}(m_{H^\pm}^2)) \text{X}_{dd}-2 (C_{00}^{(5)}(m_W^2)+C_{00}^{(5)}(m_{H^\pm}^2) \text{X}_{dd})) m_b^2+C_{12}^{(5)}(m_W^2) (t m_b^2\nn\\[0.1cm]
&-&2 (s_{23}+t_{12}) m_W^2)))+m_t (-\text{SP2}\,F_{14} (2  m_b^2(C_{0}^{(5)}(m_W^2)+C_{12}^{(5)}(m_W^2)+C_{22}^{(5)}(m_W^2)\nn\\[0.1cm]
&+&(C_{12}^{(5)}(m_{H^\pm}^2)+C_{22}^{(5)}(m_{H^\pm}^2)) \text{X}_{dd}+C_{0}^{(5)}(m_{H^\pm}^2) \text{X}_{du}+C_{2}^{(5)}(m_{H^\pm}^2) (\text{X}_{dd}+\text{X}_{du}))\nn\\[0.1cm]
&+&4 (C_{2}^{(5)}(m_W^2) m_b^2+(C_{12}^{(5)}(m_W^2)+C_{22}^{(5)}(m_W^2)) m_W^2))+F_{12} (m_W^2(-2 B_{0}^{(4)}(m_W^2)\nn\\[0.1cm]
&+&4 C_{00}^{(5)}(m_W^2)+1) -m_b^2 (B_{0}^{(4)}(m_W^2)+B_{0}^{(4)}(m_{H^\pm}^2) \text{X}_{dd}-2 (C_{00}^{(5)}(m_W^2)\nn\\[0.1cm]
&+&C_{00}^{(5)}(m_{H^\pm}^2) \text{X}_{dd})+(s_{23}+t_{12}) (C_{0}^{(5)}(m_W^2)+C_{2}^{(5)}(m_W^2)+C_{2}^{(5)}(m_{H^\pm}^2) \text{X}_{dd}\nn\\[0.1cm]
&+&C_{0}^{(5)}(m_{H^\pm}^2) \text{X}_{du})))-F_1 (m_b^2((s_{23} \text{SP1}+t_{12} \text{SP1}+\text{SP2} (2 \text{SP5}-2 \text{SP4})) C_{0}^{(5)}(m_W^2)\nn\\[0.1cm]
&-&(4 \text{SP2} (\text{SP4}-\text{SP5})+\text{SP1} t) C_{2}^{(5)}(m_W^2)-\text{SP1} (-C_{2}^{(5)}(m_W^2) m_t^2+2 C_{00}^{(5)}(m_W^2)\nn\\[0.1cm]
&+&(2 C_{00}^{(5)}(m_{H^\pm}^2)+t C_{2}^{(5)}(m_{H^\pm}^2)) \text{X}_{dd})-2 \text{SP2} (\text{SP4} C_{22}^{(5)}(m_W^2)\nn\\[0.1cm]
&+&(\text{SP4}-\text{SP5}) C_{22}^{(5)}(m_{H^\pm}^2) \text{X}_{dd})-(2 \text{SP2} (\text{SP4}-\text{SP5})+\text{SP1} t) C_{0}^{(5)}(m_{H^\pm}^2) \text{X}_{du}\nn\\[0.1cm]
&+&C_{2}^{(5)}(m_{H^\pm}^2) (\text{SP1} m_t^2 \text{X}_{dd}-2 \text{SP2} (\text{SP4}-\text{SP5}) (\text{X}_{dd}+\text{X}_{du}))) +2 \text{SP2} ((C_{12}^{(5)}(m_W^2)\nn\\[0.1cm]
&+&C_{22}^{(5)}(m_W^2)) (\text{SP5} m_b^2-2 (\text{SP4}-\text{SP5}) m_W^2)-m_b^2 (\text{SP4} C_{12}^{(5)}(m_W^2)\nn\\[0.1cm]
&+&(\text{SP4}-\text{SP5}) C_{12}^{(5)}(m_{H^\pm}^2) \text{X}_{dd}))-\text{SP1} (-m_b^2(C_{0}^{(5)}(m_{H^\pm}^2) \text{X}_{du} m_t^2+B_{0}^{(4)}(m_{H^\pm}^2) \text{X}_{dd}) \nn\\[0.1cm]
&+&(4 C_{00}^{(5)}(m_W^2)+1) m_W^2-B_{0}^{(4)}(m_W^2) (m_b^2+2 m_W^2))))),\\[0.3cm]
R_4^{\text{vertex}}&=& F_{16} ((-2 B_{0}^{(4)}(m_W^2)+4 C_{00}^{(4)}(m_W^2)+2 t C_{2}^{(4)}(m_W^2)+1) m_W^2-m_b^2 ((C_{0}^{(4)}(m_W^2)\nn\\[0.1cm]
&+&C_{0}^{(4)}(m_{H^\pm}^2) \text{X}_{du}) m_t^2+B_{0}^{(4)}(m_W^2)+B_{0}^{(4)}(m_{H^\pm}^2) \text{X}_{dd}-2 (C_{00}^{(4)}(m_W^2)\nn\\[0.1cm]
&+&C_{00}^{(4)}(m_{H^\pm}^2) \text{X}_{dd})))+\text{SP1}\,F_5 (m_W^2(1-2 (B_{0}^{(4)}(m_W^2)-2 C_{00}^{(4)}(m_W^2)-t C_{2}^{(4)}(m_W^2))) \nn\\[0.1cm]
&-&m_b^2 ((C_{0}^{(4)}(m_W^2)+C_{0}^{(4)}(m_{H^\pm}^2) \text{X}_{du}) m_t^2+B_{0}^{(4)}(m_W^2)+B_{0}^{(4)}(m_{H^\pm}^2) \text{X}_{dd}\nn\\[0.1cm]
&-&2 (C_{00}^{(4)}(m_W^2)+C_{00}^{(4)}(m_{H^\pm}^2) \text{X}_{dd})))+m_t (((s_{23} \text{SP1}+t_{12} \text{SP1}+4 \text{SP2}\,\text{SP5}) F_1\nn\\[0.1cm]
&+&(s_{23}+t_{12}) F_{12}-2 \text{SP5}\,F_{13}) (C_{0}^{(4)}(m_W^2)+C_{0}^{(4)}(m_{H^\pm}^2) \text{X}_{du}) m_b^2+2 \text{SP5} (2 \text{SP2}\,F_1\nn\\[0.1cm]
&-&F_{13}) (-m_b^2(C_{12}^{(4)}(m_{H^\pm}^2) \text{X}_{dd}-C_{2}^{(4)}(m_{H^\pm}^2) \text{X}_{du}) -C_{12}^{(4)}(m_W^2) (m_b^2+2 m_W^2)\nn\\[0.1cm]
&+&C_{2}^{(4)}(m_W^2) (m_b^2-2 m_W^2)))+\text{SP2}\,F_4 (m_b^2(4 (C_{00}^{(4)}(m_W^2)+C_{00}^{(4)}(m_{H^\pm}^2) \text{X}_{dd})\nn\\[0.1cm]
&-&2 ((C_{0}^{(4)}(m_W^2)+C_{0}^{(4)}(m_{H^\pm}^2) \text{X}_{du}) m_t^2+B_{0}^{(4)}(m_W^2)+B_{0}^{(4)}(m_{H^\pm}^2) \text{X}_{dd})) \nn\\[0.1cm]
&+&(-4 B_{0}^{(4)}(m_W^2)+8 C_{00}^{(4)}(m_W^2)+4 t C_{2}^{(4)}(m_W^2)+2) m_W^2)-F_3 (-(\text{SP4} (B_{0}^{(4)}(m_W^2)\nn\\[0.1cm]
&-&2 C_{00}^{(4)}(m_W^2))+(\text{SP4}-\text{SP5}) ((C_{0}^{(4)}(m_W^2)+C_{0}^{(4)}(m_{H^\pm}^2) \text{X}_{du}) m_t^2+(B_{0}^{(4)}(m_{H^\pm}^2)\nn\\[0.1cm]
&-&2 C_{00}^{(4)}(m_{H^\pm}^2)) \text{X}_{dd})) m_b^2+(\text{SP4}-\text{SP5}+2 (\text{SP4}+\text{SP5}) t C_{2}^{(4)}(m_W^2)) m_W^2\nn\\[0.1cm]
&+&(B_{0}^{(4)}(m_W^2)-2 C_{00}^{(4)}(m_W^2)) (\text{SP5} m_b^2-2 (\text{SP4}-\text{SP5}) m_W^2)-\text{SP5} (4 C_{2}^{(4)}(m_W^2) m_t^2 m_W^2\nn\\[0.1cm]
&+&2 (s_{23}+t_{12}) (C_{12}^{(4)}(m_{H^\pm}^2) \text{X}_{dd} m_b^2+C_{12}^{(4)}(m_W^2) (m_b^2+2 m_W^2)))),\\[0.3cm]
R_5^{\text{vertex}}&=&X_{\text{dd}}m_b^2(2 C_{00}^{(6)}(m_{H^\pm}^2) (2 (\text{SP4}-\text{SP5}) F_3+2 (\text{SP3}-\text{SP2}) F_4-2 \text{SP1}\,F_5+\text{SP1}\,F_1 m_t)\nn\\[0.1cm]
&+&C_{0}^{(6)}(m_{H^\pm}^2) (m_b^2-m_{H^\pm}^2) (2 (\text{SP4}-\text{SP5}) F_3+2 (\text{SP3}-\text{SP2}) F_4-2 \text{SP1}\,F_5\nn\\[0.1cm]
&+&\text{SP1}\,F_1 m_t)+m_t (C_{12}^{(6)}(m_{H^\pm}^2) F_1 (\text{SP1} m_t^2-4 \text{SP3}\,\text{SP4}+4 \text{SP2}\,\text{SP5}-\text{SP1} (2 t+t_{12}))\nn\\[0.1cm]
&+&C_{1}^{(6)}(m_{H^\pm}^2) m_t (2 (\text{SP5}-\text{SP4}) F_3+2 (\text{SP2}-\text{SP3}) F_4+\text{SP1} (2 F_5-F_1 m_t))))  \nn\\[0.1cm]
&+&m_t ( F_1 (\text{SP1} m_t^2-4 \text{SP3}\,\text{SP4}+4 \text{SP2}\,\text{SP5}-\text{SP1} (2 t+t_{12}))
(C_{12}^{(6)}(m_W^2)-C_{2}^{(6)}(m_W^2))\nn\\[0.1cm]
&+&m_t (2 C_{0}^{(6)}(m_W^2) ((\text{SP2}-\text{SP3}) F_4+\text{SP1} (F_5-F_1 m_t))+C_{1}^{(6)}(m_W^2) (2 (\text{SP5}-\text{SP4}) F_3\nn\\[0.1cm]
&+&2 (\text{SP2}-\text{SP3}) F_4+\text{SP1} (2 F_5-F_1 m_t)))+(-\text{SP1} (2 C_{0}^{(6)}(m_{H^\pm}^2)+C_{2}^{(6)}(m_{H^\pm}^2)) F_1 m_t^2\nn\\[0.1cm]
&+&2 C_{0}^{(6)}(m_{H^\pm}^2) ((\text{SP5}-\text{SP4}) F_3+(\text{SP2}-\text{SP3}) F_4+\text{SP1}\,F_5) m_t+(4 \text{SP3}\,\text{SP4}\nn\\[0.1cm]
&-&4 \text{SP2}\,\text{SP5}+2 \text{SP1} t+\text{SP1} t_{12}) (C_{0}^{(6)}(m_{H^\pm}^2)+C_{2}^{(6)}(m_{H^\pm}^2)) F_1) X_{\text{du}}) m_b^2\nn\\[0.1cm]
&+&(2 \text{SP1} m_t^3(-C_{1}^{(6)}(m_W^2)+C_{12}^{(6)}(m_W^2)+2 C_{2}^{(6)}(m_W^2)) F_1 -4 (C_{2}^{(6)}(m_W^2)-C_{1}^{(6)}(m_W^2))\nn\\[0.1cm]
&~& ((\text{SP5}-\text{SP4}) F_3+(\text{SP2}-\text{SP3}) F_4+\text{SP1}\,F_5) m_t^2+F_1 m_t(\text{SP1} C_{0}^{(6)}(m_W^2) m_b^2+\text{SP1}\nn\\[0.1cm]
&-&2 (4 \text{SP3}\,\text{SP4}-4 \text{SP2}\,\text{SP5}+2 \text{SP1} t+\text{SP1} t_{12}) C_{12}^{(6)}(m_W^2)+(-4 m_t^2 \text{SP1}+4 s_{23} \text{SP1}\nn\\[0.1cm]
&-&8 \text{SP3}\,\text{SP4}+8 \text{SP2}\,\text{SP5}) C_{2}^{(6)}(m_W^2)) -2 ((\text{SP4}-\text{SP5}) F_3+(\text{SP3}-\text{SP2}) F_4\nn\\[0.1cm]
&-&\text{SP1}\,F_5) (-C_{0}^{(6)}(m_W^2) m_b^2+2 t_{12} C_{2}^{(6)}(m_W^2)-1)) m_W^2+C_{0}^{(6)}(m_W^2) (-2 m_b^4((\text{SP2}\nn\\[0.1cm]
&-&\text{SP3}) F_4+\text{SP1}\,F_5) +F_1 (\text{SP1} m_b^2+4 \text{SP3}\,\text{SP4}-4 \text{SP2}\,\text{SP5}+2 \text{SP1} t+\text{SP1} t_{12}) m_t m_b^2\nn\\[0.1cm]
&+&2 (2 (\text{SP2}-\text{SP3}) F_4+\text{SP1} (2 F_5-F_1 m_t)) m_W^4+2 (\text{SP4}-\text{SP5}) F_3 (m_b^4-m_t^2 m_b^2\nn\\[0.1cm]
&-&2 m_W^4))+(2 (\text{SP4}-\text{SP5}) F_3+2 (\text{SP3}-\text{SP2}) F_4-2 \text{SP1}\,F_5+\text{SP1}\,F_1 m_t)\nn\\[0.1cm]
&~& (2 C_{00}^{(6)}(m_W^2) (m_b^2+2 m_W^2)-B_0^{(5)}(m_b^2) ((X_{\text{dd}}+1) m_b^2+2 m_W^2)).
\end{eqnarray}
For the box part (cf. Eqs.~\eqref{eq:Amptcggbox}), we obtain
\begin{eqnarray}
R_1^{\text{box}}&=&m_b^2(2 (\text{SP3} (\text{SP4}-\text{SP5}) T_dD_{0}^{(1)}(m_W^2)+(2 \text{SP2}\,\text{SP5}-\text{SP3} (\text{SP4}+\text{SP5})) T_eD_{0}^{(2)}(m_W^2)\nn\\[0.1cm]
&-&\text{SP1} T_dD_{00}^{(1)}(m_W^2))+((\text{SP4}-\text{SP5}) (T_d(2 (\text{SP2}-\text{SP3}) (D_{12}^{(1)}(m_{H^\pm}^2)+2 D_{122}^{(1)}(m_{H^\pm}^2))\nn\\[0.1cm]
&-&2 \text{SP3} D_{13}^{(1)}(m_{H^\pm}^2))+4 (\text{SP2}-\text{SP3}) (T_dD_{222}^{(1)}(m_{H^\pm}^2)+T_eD_{222}^{(2)}(m_{H^\pm}^2)))\nn\\[0.1cm]
&-&4 (\text{SP3}\,\text{SP4}+\text{SP2}\,\text{SP5}) T_dD_{223}^{(1)}(m_{H^\pm}^2)-(4 \text{SP2}-6 \text{SP3}) \text{SP5} T_eD_{23}^{(2)}(m_{H^\pm}^2)\nn\\[0.1cm]
&-&\text{SP3} (T_d(4 (\text{SP4}-\text{SP5}) D_{123}^{(1)}(m_{H^\pm}^2)-8 \text{SP5} D_{223}^{(1)}(m_{H^\pm}^2)+6 (\text{SP4}-\text{SP5}) D_{23}^{(1)}(m_{H^\pm}^2))\nn\\[0.1cm]
&+&2 \text{SP4} T_e(2 (D_{123}^{(2)}(m_{H^\pm}^2)+D_{223}^{(2)}(m_{H^\pm}^2))+D_{23}^{(2)}(m_{H^\pm}^2)))+4 \text{SP3}\,\text{SP5} (T_dD_{233}^{(1)}(m_{H^\pm}^2)\nn\\[0.1cm]
&+&T_eD_{233}^{(2)}(m_{H^\pm}^2))+(T_dD_{22}^{(1)}(m_{H^\pm}^2)+T_eD_{22}^{(2)}(m_{H^\pm}^2)) (6 (\text{SP2}-\text{SP3}) (\text{SP4}-\text{SP5})\nn\\[0.1cm]
&-&\text{SP1} m_t^2)+T_e((\text{SP2}-\text{SP3}) (4 \text{SP4} D_{112}^{(2)}(m_{H^\pm}^2)+2 (\text{SP4}-\text{SP5}) (4 D_{122}^{(2)}(m_{H^\pm}^2)\nn\\[0.1cm]
&+&D_{2}^{(2)}(m_{H^\pm}^2)))-4 \text{SP5} ((2 \text{SP2}-3 \text{SP3}) D_{123}^{(2)}(m_{H^\pm}^2)+(\text{SP2}-\text{SP3}) (D_{112}^{(2)}(m_{H^\pm}^2)\nn\\[0.1cm]
&+&D_{113}^{(2)}(m_{H^\pm}^2)+D_{13}^{(2)}(m_{H^\pm}^2))-\text{SP3} D_{133}^{(2)}(m_{H^\pm}^2)+(\text{SP2}-2 \text{SP3}) D_{223}^{(2)}(m_{H^\pm}^2))\nn\\[0.1cm]
&-&\text{SP1} t (D_{11}^{(2)}(m_{H^\pm}^2)+D_{12}^{(2)}(m_{H^\pm}^2)-D_{13}^{(2)}(m_{H^\pm}^2)-D_{23}^{(2)}(m_{H^\pm}^2))\nn\\[0.1cm]
&+&D_{12}^{(2)}(m_{H^\pm}^2) (6 (\text{SP2}-\text{SP3}) (\text{SP4}-\text{SP5})-\text{SP1} m_t^2))) \text{X}_{dd}-(4 (\text{SP1} (T_dD_{00}^{(1)}(m_{H^\pm}^2)\nn\\[0.1cm]
&+&T_eD_{00}^{(2)}(m_{H^\pm}^2))+(\text{SP2}-\text{SP3}) (\text{SP4}-\text{SP5}) (T_d(D_{12}^{(1)}(m_{H^\pm}^2)+D_{22}^{(1)}(m_{H^\pm}^2))\nn\\[0.1cm]
&+&T_e(D_{12}^{(2)}(m_{H^\pm}^2)+D_{22}^{(2)}(m_{H^\pm}^2))))-4 (\text{SP3}\,\text{SP4}+\text{SP2}\,\text{SP5}) T_dD_{23}^{(1)}(m_{H^\pm}^2)\nn\\[0.1cm]
&-&2 T_e((2 \text{SP2}\,\text{SP5}-\text{SP3} (\text{SP4}+\text{SP5})) D_{0}^{(2)}(m_{H^\pm}^2)+(\text{SP2}-\text{SP3}) (\text{SP4}+\text{SP5})\nn\\[0.1cm]
&~& D_{1}^{(2)}(m_{H^\pm}^2)+2 \text{SP5} ((\text{SP2}-\text{SP3}) D_{13}^{(2)}(m_{H^\pm}^2)+(\text{SP2}-2 \text{SP3}) D_{23}^{(2)}(m_{H^\pm}^2)))\nn\\[0.1cm]
&-&(4 \text{SP2}-6 \text{SP3}) \text{SP5} T_eD_{3}^{(2)}(m_{H^\pm}^2)+\text{SP1} (s_{23} T_d(D_{0}^{(1)}(m_{H^\pm}^2)-D_{1}^{(1)}(m_{H^\pm}^2)\nn\\[0.1cm]
&+&D_{3}^{(1)}(m_{H^\pm}^2))+t T_e(D_{0}^{(2)}(m_{H^\pm}^2)-D_{1}^{(2)}(m_{H^\pm}^2)+D_{3}^{(2)}(m_{H^\pm}^2)))\nn\\[0.1cm]
&-&\text{SP3} (T_d(2 (\text{SP4}-\text{SP5}) (D_{0}^{(1)}(m_{H^\pm}^2)+2 D_{13}^{(1)}(m_{H^\pm}^2))-8 \text{SP5} D_{23}^{(1)}(m_{H^\pm}^2)\nn\\[0.1cm]
&+&2 (\text{SP4}-3 \text{SP5}) D_{3}^{(1)}(m_{H^\pm}^2))+\text{SP4} T_e(4 D_{23}^{(2)}(m_{H^\pm}^2)-2 D_{3}^{(2)}(m_{H^\pm}^2)))\nn\\[0.1cm]
&+&4 \text{SP3}\,\text{SP5} (T_dD_{33}^{(1)}(m_{H^\pm}^2)+T_eD_{33}^{(2)}(m_{H^\pm}^2))+(T_dD_{2}^{(1)}(m_{H^\pm}^2)\nn\\[0.1cm]
&+&T_eD_{2}^{(2)}(m_{H^\pm}^2)) (2 (\text{SP2}-\text{SP3}) (\text{SP4}-3 \text{SP5})-\text{SP1} m_t^2)) \text{X}_{du}) -(T_dD_{22}^{(1)}(m_W^2)\nn\\[0.1cm]
&+&T_eD_{22}^{(2)}(m_W^2)) (\text{SP1} m_t^2 (m_b^2+2 m_W^2)-2 (\text{SP2}-\text{SP3}) (\text{SP4}-\text{SP5}) (m_b^2+6 m_W^2))\nn\\[0.1cm]
&-&\text{SP5} (2 T_eD_{23}^{(2)}(m_W^2) (\text{SP3} m_b^2+2 (4 \text{SP2}-5 \text{SP3}) m_W^2)-4 (\text{SP3} (T_dD_{233}^{(1)}(m_W^2)\nn\\[0.1cm]
&+&T_eD_{233}^{(2)}(m_W^2)) (m_b^2+2 m_W^2)+\text{SP2} T_eD_{3}^{(2)}(m_W^2) (m_b^2-2 m_W^2))\nn\\[0.1cm]
&+&\text{SP3} (4 (T_dD_{33}^{(1)}(m_W^2)+T_eD_{33}^{(2)}(m_W^2)) (m_b^2-2 m_W^2)+T_eD_{3}^{(2)}(m_W^2) (6 m_b^2-4 m_W^2)))\nn\\[0.1cm]
&+&T_e((2 (\text{SP2}-\text{SP3}) (\text{SP4}+\text{SP5}) D_{1}^{(2)}(m_W^2)+4 \text{SP2}\,\text{SP4} D_{112}^{(2)}(m_W^2)) m_b^2-8 (\text{SP3}\,\text{SP4}\nn\\[0.1cm]
&-&\text{SP2} (\text{SP4}-\text{SP5})) D_{122}^{(2)}(m_W^2) (m_b^2+2 m_W^2)+4 (\text{SP2}-\text{SP3}) D_{2}^{(2)}(m_W^2) (\text{SP5} m_b^2\nn\\[0.1cm]
&+&(\text{SP4}-\text{SP5}) m_W^2)-4 \text{SP4} D_{112}^{(2)}(m_W^2) (\text{SP3} m_b^2-2 (\text{SP2}-\text{SP3}) m_W^2)\nn\\[0.1cm]
&-&D_{12}^{(2)}(m_W^2) (\text{SP1} m_t^2 (m_b^2+2 m_W^2)-2 (\text{SP2}-\text{SP3}) (\text{SP4}-\text{SP5}) (m_b^2+6 m_W^2))\nn\\[0.1cm]
&-&4 ((\text{SP3}\,\text{SP4}-\text{SP2} (\text{SP4}-\text{SP5})) D_{222}^{(2)}(m_W^2) (m_b^2+2 m_W^2)\nn\\[0.1cm]
&+&\text{SP5} (4 (\text{SP2}-\text{SP3}) D_{13}^{(2)}(m_W^2) m_W^2+((\text{SP2}-\text{SP3}) (D_{112}^{(2)}(m_W^2)+D_{113}^{(2)}(m_W^2))\nn\\[0.1cm]
&+&2 \text{SP2} D_{123}^{(2)}(m_W^2)-\text{SP3} D_{133}^{(2)}(m_W^2)+\text{SP2} D_{223}^{(2)}(m_W^2)) (m_b^2+2 m_W^2))))\nn\\[0.1cm]
&+&\text{SP3} (T_e(2 \text{SP4} (D_{23}^{(2)}(m_W^2)-D_{3}^{(2)}(m_W^2)) (m_b^2-2 m_W^2)-(4 \text{SP4} (D_{123}^{(2)}(m_W^2)\nn\\[0.1cm]
&+&D_{223}^{(2)}(m_W^2))-4 \text{SP5} (2 D_{122}^{(2)}(m_W^2)+3 D_{123}^{(2)}(m_W^2)+D_{222}^{(2)}(m_W^2)\nn\\[0.1cm]
&+&2 D_{223}^{(2)}(m_W^2))) (m_b^2+2 m_W^2))-T_d(4 (\text{SP4}-2 \text{SP5}) D_{223}^{(1)}(m_W^2) (m_b^2+2 m_W^2)\nn\\[0.1cm]
&+&2 D_{23}^{(1)}(m_W^2) ((\text{SP4}+\text{SP5}) m_b^2+2 (3 \text{SP4}-5 \text{SP5}) m_W^2)-2 D_{3}^{(1)}(m_W^2) ((\text{SP4}\nn\\[0.1cm]
&-&3 \text{SP5}) m_b^2-2 (\text{SP4}-\text{SP5}) m_W^2)-(\text{SP4}-\text{SP5}) (2 D_{13}^{(1)}(m_W^2) (m_b^2-2 m_W^2)\nn\\[0.1cm]
&-&4 D_{123}^{(1)}(m_W^2) (m_b^2+2 m_W^2))))-(\text{SP1} m_b^2 (T_dC_{0}^{(1)}(m_b^2)+T_eC_{1}^{(1)}(m_b^2)\nn\\[0.1cm]
&+&(T_d+T_e) C_{2}^{(1)}(m_b^2)-T_d(2 D_{00}^{(1)}(m_{H^\pm}^2)+4 D_{002}^{(1)}(m_{H^\pm}^2)+D_{0}^{(1)}(m_{H^\pm}^2) (m_b^2-m_{H^\pm}^2))\nn\\[0.1cm]
&-&T_e(4 (D_{001}^{(2)}(m_{H^\pm}^2)+D_{002}^{(2)}(m_{H^\pm}^2))+(D_{1}^{(2)}(m_{H^\pm}^2)+D_{2}^{(2)}(m_{H^\pm}^2)) (m_b^2-m_{H^\pm}^2)))\nn\\[0.1cm]
&-&T_dD_{2}^{(1)}(m_{H^\pm}^2) (2 (\text{SP2}-\text{SP3}) (\text{SP4}-\text{SP5}) m_b^2+\text{SP1} (m_b^4-m_b^2 (m_{H^\pm}^2+m_t^2)))) \text{X}_{dd}\nn\\[0.1cm]
&-&T_d(s_{23} \text{SP1} (D_{1}^{(1)}(m_{H^\pm}^2)+D_{12}^{(1)}(m_{H^\pm}^2)-D_{23}^{(1)}(m_{H^\pm}^2)) \text{X}_{dd} m_b^2\nn\\[0.1cm]
&+&(\text{SP2}-\text{SP3}) (\text{SP4}-\text{SP5}) (2 D_{12}^{(1)}(m_W^2) (m_b^2-2 m_W^2)\nn\\[0.1cm]
&-&4 (D_{122}^{(1)}(m_W^2)+D_{222}^{(1)}(m_W^2)) (m_b^2+2 m_W^2))-4 \text{SP2}\,\text{SP5} (D_{23}^{(1)}(m_W^2) (m_b^2-2 m_W^2)\nn\\[0.1cm]
&-&D_{223}^{(1)}(m_W^2) (m_b^2+2 m_W^2))-D_{2}^{(1)}(m_W^2) (4 (\text{SP2}-\text{SP3}) (\text{SP5} m_b^2+(\text{SP4}-\text{SP5}) m_W^2)\nn\\[0.1cm]
&+&\text{SP1} (m_b^4-2 m_W^4+(m_b^2+4 m_t^2) m_W^2)))-\text{SP1} (-((T_d+T_e) C_{0}^{(1)}(m_b^2)\nn\\[0.1cm]
&-&T_dD_{0}^{(1)}(m_{H^\pm}^2) (m_b^2-m_{H^\pm}^2)+T_eD_{0}^{(2)}(m_{H^\pm}^2) (-m_b^2+m_{H^\pm}^2+m_t^2)) \text{X}_{du} m_b^2\nn\\[0.1cm]
&+&(T_eC_{1}^{(1)}(m_b^2)+(T_d+T_e) C_{2}^{(1)}(m_b^2)) (m_b^2+2 m_W^2)-C_{0}^{(1)}(m_b^2) (T_em_b^2-2 T_dm_W^2)\nn\\[0.1cm]
&-&T_d((4 D_{00}^{(1)}(m_W^2)-6 t_{12} D_{2}^{(1)}(m_W^2)+2 D_{0}^{(1)}(m_W^2) (m_b^2-m_W^2)) m_W^2\nn\\[0.1cm]
&+&4 D_{002}^{(1)}(m_W^2) (m_b^2+2 m_W^2))-T_e(-(4 (D_{00}^{(2)}(m_W^2)-D_{001}^{(2)}(m_W^2))\nn\\[0.1cm]
&-&D_{0}^{(2)}(m_W^2) (-m_b^2+m_W^2+m_t^2)) m_b^2+8 D_{001}^{(2)}(m_W^2) m_W^2+4 D_{002}^{(2)}(m_W^2) (m_b^2+2 m_W^2)\nn\\[0.1cm]
&+&D_{1}^{(2)}(m_W^2) (m_b^4+m_W^2 m_b^2-2 m_W^4)+D_{2}^{(2)}(m_W^2) (m_b^4+(m_t^2+m_W^2) m_b^2-2 m_W^4))\nn\\[0.1cm]
&+&s_{23} (2 T_eD_{3}^{(2)}(m_W^2) m_W^2+T_d(D_{0}^{(1)}(m_W^2) m_b^2+2 (D_{1}^{(1)}(m_W^2)+3 D_{2}^{(1)}(m_W^2)) m_W^2\nn\\[0.1cm]
&+&(D_{12}^{(1)}(m_W^2)-D_{23}^{(1)}(m_W^2)) (m_b^2+2 m_W^2)+D_{3}^{(1)}(m_W^2) (m_b^2-2 m_W^2)))\nn\\[0.1cm]
&+&t (6 T_dD_{2}^{(1)}(m_W^2) m_W^2+T_e((D_{0}^{(2)}(m_W^2)-D_{1}^{(2)}(m_W^2)+D_{11}^{(2)}(m_W^2)+D_{3}^{(2)}(m_W^2)) m_b^2\nn\\[0.1cm]
&+&2 D_{11}^{(2)}(m_W^2) m_W^2+(D_{12}^{(2)}(m_W^2)-D_{13}^{(2)}(m_W^2)-D_{23}^{(2)}(m_W^2)) (m_b^2+2 m_W^2)))), \\[0.3cm]
R_2^{\text{box}}&=&m_b^2((\text{SP4}-\text{SP5}) (T_dD_{0}^{(1)}(m_{H^\pm}^2)-T_eD_{0}^{(2)}(m_{H^\pm}^2)) \text{X}_{du} m_t^2\nn\\[0.1cm]
&+&(((\text{SP4}-\text{SP5}) T_dD_{12}^{(1)}(m_{H^\pm}^2)+T_e((3 \text{SP4}-\text{SP5}) D_{12}^{(2)}(m_{H^\pm}^2)-2 \text{SP5} D_{13}^{(2)}(m_{H^\pm}^2))\nn\\[0.1cm]
&+&\text{SP4} T_dD_{2}^{(1)}(m_{H^\pm}^2)+T_e((4 \text{SP4}-2 \text{SP5}) D_{22}^{(2)}(m_{H^\pm}^2)-\text{SP5} D_{2}^{(2)}(m_{H^\pm}^2))\nn\\[0.1cm]
&-&2 \text{SP5} (T_d(D_{22}^{(1)}(m_{H^\pm}^2)+D_{23}^{(1)}(m_{H^\pm}^2))+T_eD_{23}^{(2)}(m_{H^\pm}^2))) m_t^2+(\text{SP4} (T_d-T_e)\nn\\[0.1cm]
&-&\text{SP5} (3 T_d-T_e)) C_{0}^{(1)}(m_b^2)-(\text{SP5} (T_d-T_e)-\text{SP4} (T_d+T_e)) C_{1}^{(1)}(m_b^2)\nn\\[0.1cm]
&-&2 (\text{SP5} T_d-\text{SP4} T_e) C_{2}^{(1)}(m_b^2)+T_e(2 (3 \text{SP4}+\text{SP5}) D_{00}^{(2)}(m_{H^\pm}^2)\nn\\[0.1cm]
&-&4 (\text{SP4}-2 \text{SP5}) D_{002}^{(2)}(m_{H^\pm}^2))-T_d(2 (\text{SP4}-5 \text{SP5}) D_{00}^{(1)}(m_{H^\pm}^2)\nn\\[0.1cm]
&+&4 (\text{SP4}-\text{SP5}) (D_{001}^{(1)}(m_{H^\pm}^2)+D_{002}^{(1)}(m_{H^\pm}^2))-4 \text{SP5} D_{003}^{(1)}(m_{H^\pm}^2)\nn\\[0.1cm]
&+&((\text{SP4}-3 \text{SP5}) D_{0}^{(1)}(m_{H^\pm}^2)+(\text{SP4}-\text{SP5}) D_{1}^{(1)}(m_{H^\pm}^2))(m_b^2-m_{H^\pm}^2))\nn\\[0.1cm]
&+&T_e(((\text{SP4}-\text{SP5}) D_{0}^{(2)}(m_{H^\pm}^2)-(\text{SP4}+\text{SP5}) D_{1}^{(2)}(m_{H^\pm}^2)) (m_b^2-m_{H^\pm}^2)\nn\\[0.1cm]
&+&\text{SP4} D_{2}^{(2)}(m_{H^\pm}^2) (-2 m_b^2+2 m_{H^\pm}^2+m_t^2))+\text{SP5} T_d(2 D_{3}^{(1)}(m_{H^\pm}^2) (m_b^2-m_{H^\pm}^2)\nn\\[0.1cm]
&+&D_{2}^{(1)}(m_{H^\pm}^2) (2 m_b^2-2 m_{H^\pm}^2-3 m_t^2))) \text{X}_{dd}) +2m_W^2 ((\text{SP4}-\text{SP5}) t_{12} T_dD_{3}^{(1)}(m_W^2)\nn\\[0.1cm]
&+&T_eD_{3}^{(2)}(m_W^2) (2 \text{SP5} m_b^2+(\text{SP4}-\text{SP5}) m_t^2)) +C_{0}^{(1)}(m_b^2) (m_b^2((\text{SP4}-3 \text{SP5}) T_d\nn\\[0.1cm]
&+&(\text{SP5}-\text{SP4}) T_e) -2 (\text{SP5} (T_d-T_e)-\text{SP4} (3 T_d-T_e)) m_W^2)\nn\\[0.1cm]
&-&C_{2}^{(1)}(m_b^2) ( m_b^2(2 \text{SP5} T_d-2 \text{SP4} T_e)-4 (\text{SP4} T_d-\text{SP5} T_e) m_W^2)\nn\\[0.1cm]
&+&C_{1}^{(1)}(m_b^2) ((\text{SP4}-\text{SP5}) T_d(m_b^2+2 m_W^2)+(\text{SP4}+\text{SP5}) T_e(m_b^2-2 m_W^2))\nn\\[0.1cm]
&+&\text{SP5} (4 (2 T_eD_{002}^{(2)}(m_W^2)+T_dD_{003}^{(1)}(m_W^2)) (m_b^2+2 m_W^2)+2 T_dD_{3}^{(1)}(m_W^2) (m_b^4\nn\\[0.1cm]
&-&(m_b^2-m_t^2) m_W^2))-T_d(2 D_{00}^{(1)}(m_W^2) ((\text{SP4}-5 \text{SP5}) m_b^2+2 (3 \text{SP4}-\text{SP5}) m_W^2)\nn\\[0.1cm]
&+&(\text{SP4}-\text{SP5}) (m_b^2+2 m_W^2) (4 (D_{001}^{(1)}(m_W^2)+D_{002}^{(1)}(m_W^2))\nn\\[0.1cm]
&+&D_{1}^{(1)}(m_W^2) (m_b^2-m_W^2))-D_{2}^{(1)}(m_W^2) (4 \text{SP4} m_W^4+2 (\text{SP4}+\text{SP5}) (3 (s_{23}+t)\nn\\[0.1cm]
&+&t_{12}) m_W^2+m_b^2 (\text{SP5} (2 m_b^2-3 m_t^2)-2 (2 \text{SP4}+\text{SP5}) m_W^2))+D_{0}^{(1)}(m_W^2) ((\text{SP4}\nn\\[0.1cm]
&-&3 \text{SP5}) m_b^4-((\text{SP4}-\text{SP5}) m_t^2-(5 \text{SP4}+\text{SP5}) m_W^2) m_b^2-2 (3 \text{SP4}-\text{SP5}) m_W^4))\nn\\[0.1cm]
&-&T_e(2 D_{3}^{(2)}(m_W^2) (2 \text{SP5} m_W^2+(\text{SP4}+\text{SP5}) t_{12}) m_W^2+4 \text{SP4} D_{002}^{(2)}(m_W^2) (m_b^2+2 m_W^2)\nn\\[0.1cm]
&-&2 D_{00}^{(2)}(m_W^2) ((3 \text{SP4}+\text{SP5}) m_b^2-2 (\text{SP4}+\text{SP5}) m_W^2)-(\text{SP4}-\text{SP5}) D_{0}^{(2)}(m_W^2) (m_b^4\nn\\[0.1cm]
&-&(m_t^2-m_W^2) m_b^2-2 m_W^4)+D_{2}^{(2)}(m_W^2) (2 \text{SP4} m_b^4-((\text{SP4}-\text{SP5}) m_t^2\nn\\[0.1cm]
&+&2 (\text{SP4}+2 \text{SP5}) m_W^2) m_b^2-2 m_W^2 (-3 (\text{SP4}-\text{SP5}) m_t^2-2 \text{SP5} m_W^2+(2 \text{SP4}\nn\\[0.1cm]
&-&2 \text{SP5}) t_{12}))+D_{1}^{(2)}(m_W^2) (4 (\text{SP4}-\text{SP5}) (s_{23}+t) m_W^2+(\text{SP4}+\text{SP5}) (2 m_W^4\nn\\[0.1cm]
&+&m_b^2 (m_b^2-3 m_W^2))))+m_t^2 (-2 \text{SP5} (T_eD_{23}^{(2)}(m_W^2) (m_b^2+2 m_W^2)\nn\\[0.1cm]
&+&T_dD_{23}^{(1)}(m_W^2) (m_b^2-2 m_W^2))+T_d(((\text{SP4}-\text{SP5}) D_{12}^{(1)}(m_W^2)\nn\\[0.1cm]
&-&2 \text{SP5} D_{22}^{(1)}(m_W^2)) (m_b^2-2 m_W^2)+\text{SP4} (D_{2}^{(1)}(m_W^2) m_b^2-2 D_{3}^{(1)}(m_W^2) m_W^2))\nn\\[0.1cm]
&+&T_e(D_{12}^{(2)}(m_W^2) ((3 \text{SP4}-\text{SP5}) m_b^2-2 (\text{SP4}+\text{SP5}) m_W^2)-2 (\text{SP5} D_{13}^{(2)}(m_W^2) m_b^2\nn\\[0.1cm]
&-&D_{22}^{(2)}(m_W^2) ((2 \text{SP4}-\text{SP5}) m_b^2-2 \text{SP5} m_W^2))))+s_{23} (T_e(4 (\text{SP4}-\text{SP5}) (D_{1}^{(2)}(m_W^2)\nn\\[0.1cm]
&-&D_{22}^{(2)}(m_W^2))-4 (\text{SP4}-2 \text{SP5}) D_{23}^{(2)}(m_W^2)+4 \text{SP5} (D_{12}^{(2)}(m_W^2)+D_{33}^{(2)}(m_W^2))) m_W^2\nn\\[0.1cm]
&-&4 (\text{SP5} T_dD_{12}^{(1)}(m_W^2) m_b^2+\text{SP4} T_eD_{12}^{(2)}(m_W^2) m_W^2)+\text{SP5} (4 T_eD_{13}^{(2)}(m_W^2) m_W^2\nn\\[0.1cm]
&-&T_dD_{13}^{(1)}(m_W^2) (3 m_b^2-2 m_W^2))+T_d(((\text{SP4}-\text{SP5}) D_{11}^{(1)}(m_{H^\pm}^2)\nn\\[0.1cm]
&+&2 (\text{SP4}-2 \text{SP5}) D_{12}^{(1)}(m_{H^\pm}^2)+(\text{SP4}-3 \text{SP5}) (D_{1}^{(1)}(m_{H^\pm}^2)+D_{13}^{(1)}(m_{H^\pm}^2))\nn\\[0.1cm]
&+&2 (\text{SP4}-\text{SP5}) (D_{2}^{(1)}(m_{H^\pm}^2)+D_{22}^{(1)}(m_{H^\pm}^2)+D_{23}^{(1)}(m_{H^\pm}^2))) \text{X}_{dd} m_b^2\nn\\[0.1cm]
&+&(\text{SP4}-\text{SP5}) D_{11}^{(1)}(m_W^2) (m_b^2+2 m_W^2)+D_{1}^{(1)}(m_W^2) ((\text{SP4}-3 \text{SP5}) m_b^2\nn\\[0.1cm]
&+&2 (3 \text{SP4}-\text{SP5}) m_W^2)+2 D_{23}^{(1)}(m_W^2) ((\text{SP4}-\text{SP5}) m_b^2-2 \text{SP4} m_W^2)\nn\\[0.1cm]
&+&\text{SP4} (2 D_{12}^{(1)}(m_W^2) (m_b^2+2 m_W^2)+D_{13}^{(1)}(m_W^2) (m_b^2-2 m_W^2))+2 ((\text{SP4}-\text{SP5}) \nn\\[0.1cm]
&~&D_{22}^{(1)}(m_W^2) m_b^2+D_{2}^{(1)}(m_W^2) ((\text{SP4}-\text{SP5}) m_b^2-2 (\text{SP4}+\text{SP5}) m_W^2))))\nn\\[0.1cm]
&+&t (T_d(4 (\text{SP4}-\text{SP5}) (D_{12}^{(1)}(m_W^2)+D_{22}^{(1)}(m_W^2))-4 \text{SP5} (2 D_{2}^{(1)}(m_W^2)\nn\\[0.1cm]
&+&D_{23}^{(1)}(m_W^2))) m_W^2+T_e(-(2 (\text{SP4}-\text{SP5}) (D_{22}^{(2)}(m_W^2)+D_{23}^{(2)}(m_W^2))\nn\\[0.1cm]
&-&((\text{SP4}+\text{SP5}) D_{11}^{(2)}(m_{H^\pm}^2)+2 \text{SP5} D_{12}^{(2)}(m_{H^\pm}^2)-(\text{SP4}-\text{SP5}) (D_{1}^{(2)}(m_{H^\pm}^2)\nn\\[0.1cm]
&+&D_{13}^{(2)}(m_{H^\pm}^2)+2(D_{2}^{(2)}(m_{H^\pm}^2)+D_{22}^{(2)}(m_{H^\pm}^2)+D_{23}^{(2)}(m_{H^\pm}^2)))) \text{X}_{dd}) m_b^2\nn\\[0.1cm]
&+&(-(\text{SP4}-\text{SP5}) D_{1}^{(2)}(m_W^2)+(\text{SP4}+\text{SP5}) D_{11}^{(2)}(m_W^2)-\text{SP4} D_{13}^{(2)}(m_W^2)\nn\\[0.1cm]
&+&\text{SP5} D_{13}^{(2)}(m_W^2)-2 (\text{SP4}-\text{SP5}) D_{2}^{(2)}(m_W^2)) (m_b^2-2 m_W^2)+\text{SP5} (4 (D_{23}^{(2)}(m_W^2)\nn\\[0.1cm]
&+&D_{33}^{(2)}(m_W^2)) m_W^2+2 D_{12}^{(2)}(m_W^2) (m_b^2-2 m_W^2)))), \\[0.3cm]
R_3^{\text{box}}&=&- m_b^2((4 T_d((\text{SP2}-\text{SP3}) D_{002}^{(1)}(m_{H^\pm}^2)-\text{SP3} D_{003}^{(1)}(m_{H^\pm}^2))+\text{SP3} (4 T_dD_{00}^{(1)}(m_{H^\pm}^2)\nn\\[0.1cm]
&-&2 T_e(D_{0}^{(2)}(m_{H^\pm}^2)+D_{1}^{(2)}(m_{H^\pm}^2)+D_{3}^{(2)}(m_{H^\pm}^2)) (m_b^2-m_{H^\pm}^2))\nn\\[0.1cm]
&+&T_e((8 \text{SP2}-4 \text{SP3}) D_{00}^{(2)}(m_{H^\pm}^2)+4 ((\text{SP2}-\text{SP3}) D_{001}^{(2)}(m_{H^\pm}^2)\nn\\[0.1cm]
&+&(\text{SP2}-2 \text{SP3}) D_{002}^{(2)}(m_{H^\pm}^2))+2 \text{SP3} (D_{2}^{(2)}(m_{H^\pm}^2) m_{H^\pm}^2+C_{0}^{(1)}(m_b^2)\nn\\[0.1cm]
&+&C_{1}^{(1)}(m_b^2)))-2 (\text{SP3} T_eD_{2}^{(2)}(m_{H^\pm}^2) m_b^2+\text{SP2} (T_dD_{22}^{(1)}(m_{H^\pm}^2)-T_eD_{22}^{(2)}(m_{H^\pm}^2)) m_t^2\nn\\[0.1cm]
&+&(\text{SP3} T_d+\text{SP2} T_e) C_{0}^{(1)}(m_b^2)+(\text{SP2} T_d-\text{SP3} T_e) C_{2}^{(1)}(m_b^2)+T_dD_{2}^{(1)}(m_{H^\pm}^2) \nn\\[0.1cm]
&~&(\text{SP2} m_{H^\pm}^2+\text{SP3} m_t^2))+2 ((\text{SP3} T_dD_{0}^{(1)}(m_{H^\pm}^2)+\text{SP2} T_eD_{0}^{(2)}(m_{H^\pm}^2)) (m_b^2-m_{H^\pm}^2)\nn\\[0.1cm]
&+&\text{SP2} (T_dD_{2}^{(1)}(m_{H^\pm}^2) m_b^2+T_eD_{12}^{(2)}(m_{H^\pm}^2) m_t^2))) \text{X}_{dd}-2  m_t^2((\text{SP2}\nn\\[0.1cm]
&-&\text{SP3})(T_dD_{2}^{(1)}(m_{H^\pm}^2)+T_e(D_{0}^{(2)}(m_{H^\pm}^2)+D_{1}^{(2)}(m_{H^\pm}^2)+D_{2}^{(2)}(m_{H^\pm}^2)))\nn\\[0.1cm]
&-&\text{SP3} (T_dD_{3}^{(1)}(m_{H^\pm}^2)+T_eD_{3}^{(2)}(m_{H^\pm}^2))) \text{X}_{du})-4 m_W^2 (\text{SP3} T_dD_{3}^{(1)}(m_W^2) m_W^2\nn\\[0.1cm]
&+&(s_{23}+t) (\text{SP2} T_eD_{3}^{(2)}(m_W^2)-\text{SP3} T_dD_{33}^{(1)}(m_W^2)))+2 C_{2}^{(1)}(m_b^2) ((\text{SP2} T_d-\text{SP3} T_e) m_b^2\nn\\[0.1cm]
&-&2 (\text{SP3} T_d-\text{SP2} T_e) m_W^2)+\text{SP3} (4 (2 T_eD_{002}^{(2)}(m_W^2)+T_dD_{003}^{(1)}(m_W^2)) (m_b^2+2 m_W^2)\nn\\[0.1cm]
&+&2 T_eD_{3}^{(2)}(m_W^2) (m_b^4-(m_t^2+m_W^2) m_b^2+2 m_t^2 m_W^2))+m_t^2 (2 T_dD_{22}^{(1)}(m_W^2) (\text{SP2} m_b^2\nn\\[0.1cm]
&+&s2 (\text{SP2}-2 \text{SP3}) m_W^2)-2 (2 (\text{SP2}+\text{SP3}) T_dD_{23}^{(1)}(m_W^2) m_W^2+\text{SP2} T_eD_{22}^{(2)}(m_W^2) (m_b^2\nn\\[0.1cm]
&-&2 m_W^2)))-T_e(2 C_{1}^{(1)}(m_b^2) (\text{SP3} m_b^2-2 \text{SP2} m_W^2)+\text{SP2} (2 D_{12}^{(2)}(m_W^2) (m_b^2-2 m_W^2) m_t^2\nn\\[0.1cm]
&+&4 D_{002}^{(2)}(m_W^2) (m_b^2+2 m_W^2))+4 ((\text{SP2}-\text{SP3}) D_{001}^{(2)}(m_W^2) (m_b^2+2 m_W^2)\nn\\[0.1cm]
&+&D_{00}^{(2)}(m_W^2) ((2 \text{SP2}-\text{SP3}) m_b^2-4 \text{SP3} m_W^2))-D_{1}^{(2)}(m_W^2) (2 \text{SP3} m_b^4+2 (m_t^2(\text{SP2}\nn\\[0.1cm]
&-&\text{SP3}) -(2 \text{SP2}+\text{SP3}) m_W^2) m_b^2+4 \text{SP2} m_W^4)-D_{2}^{(2)}(m_W^2) (2 \text{SP3} m_b^4+(2 (\text{SP2}\nn\\[0.1cm]
&-&\text{SP3}) m_t^2-2 (2 \text{SP2}+\text{SP3}) m_W^2) m_b^2-4 m_W^2 ((\text{SP2}+\text{SP3}) t_{12}+\text{SP2} (-2 m_t^2-m_W^2)))\nn\\[0.1cm]
&-&(\text{SP2}-\text{SP3}) (2 C_{0}^{(1)}(m_b^2) (m_b^2+2 m_W^2)-2 D_{0}^{(2)}(m_W^2) (m_b^4-(m_t^2-m_W^2) m_b^2\nn\\[0.1cm]
&-&2 m_W^4)))+T_d(4 (\text{SP2}-\text{SP3}) t_{12} (D_{12}^{(1)}(m_W^2)+D_{23}^{(1)}(m_W^2)) m_W^2+2 C_{0}^{(1)}(m_b^2) (\text{SP3} m_b^2\nn\\[0.1cm]
&-&2 \text{SP2} m_W^2)-D_{00}^{(1)}(m_W^2) (4 \text{SP3} m_b^2-8 (\text{SP2}+\text{SP3}) m_W^2)-4 (\text{SP3} t_{12} D_{13}^{(1)}(m_W^2) m_W^2\nn\\[0.1cm]
&+&(\text{SP2}-\text{SP3}) D_{002}^{(1)}(m_W^2) (m_b^2+2 m_W^2))-2 (\text{SP3} D_{3}^{(1)}(m_W^2) (m_t^2-2 m_W^2) m_b^2\nn\\[0.1cm]
&+&D_{0}^{(1)}(m_W^2) (m_b^2-m_W^2) (\text{SP3} m_b^2-2 \text{SP2} m_W^2))-D_{2}^{(1)}(m_W^2) (4 \text{SP3} m_W^2(-m_b^2 \nn\\[0.1cm]
&+&m_W^2+m_t^2)+\text{SP2} (2 m_b^4-2 (m_t^2+m_W^2) m_b^2+4 (s_{23}+t) m_W^2)))\nn\\[0.1cm]
&-&t (4 T_d((\text{SP2}-\text{SP3}) D_{22}^{(1)}(m_W^2)-\text{SP3} (D_{2}^{(1)}(m_W^2)+D_{23}^{(1)}(m_W^2))) m_W^2\nn\\[0.1cm]
&+&2 T_e((2 \text{SP3} D_{12}^{(2)}(m_W^2)-\text{SP2} D_{13}^{(2)}(m_W^2)-(\text{SP2}-\text{SP3}) D_{22}^{(2)}(m_W^2)) m_b^2\nn\\[0.1cm]
&+&(\text{SP3} D_{11}^{(2)}(m_{H^\pm}^2)-(\text{SP2}-2 \text{SP3}) D_{12}^{(2)}(m_{H^\pm}^2)-(\text{SP2}-\text{SP3}) (D_{1}^{(2)}(m_{H^\pm}^2)\nn\\[0.1cm]
&+&D_{13}^{(2)}(m_{H^\pm}^2)+D_{2}^{(2)}(m_{H^\pm}^2)+D_{22}^{(2)}(m_{H^\pm}^2)+D_{23}^{(2)}(m_{H^\pm}^2))) \text{X}_{dd} m_b^2-((\text{SP2}\nn\\[0.1cm]
&-&\text{SP3}) D_{1}^{(2)}(m_W^2)+\text{SP2} D_{12}^{(2)}(m_W^2)) (m_b^2+2 m_W^2)+D_{13}^{(2)}(m_W^2) (\text{SP3} m_b^2+2 \text{SP2} m_W^2)\nn\\[0.1cm]
&-&D_{23}^{(2)}(m_W^2) ((\text{SP2}-\text{SP3}) m_b^2-2 \text{SP2} m_W^2)+D_{11}^{(2)}(m_W^2) (\text{SP3} m_b^2-2 \text{SP2} m_W^2)\nn\\[0.1cm]
&-&D_{2}^{(2)}(m_W^2) ((\text{SP2}-\text{SP3}) m_b^2-2 (\text{SP2}+\text{SP3}) m_W^2)))+s_{23} (4 (\text{SP2}-\text{SP3}) T_em_W^2\nn\\[0.1cm]
&~&(D_{12}^{(2)}(m_W^2)+D_{22}^{(2)}(m_W^2)) +\text{SP3} (2 T_d(D_{2}^{(1)}(m_W^2)+D_{23}^{(1)}(m_W^2)) m_b^2\nn\\[0.1cm]
&-&4 T_e m_W^2(2 D_{2}^{(2)}(m_W^2)+D_{23}^{(2)}(m_W^2)))-2 T_d((\text{SP2}-\text{SP3}) (D_{22}^{(1)}(m_W^2)\nn\\[0.1cm]
&+&(D_{2}^{(1)}(m_{H^\pm}^2)+D_{22}^{(1)}(m_{H^\pm}^2)+D_{23}^{(1)}(m_{H^\pm}^2)) \text{X}_{dd}) m_b^2-D_{12}^{(1)}(m_W^2) (\text{SP3} m_b^2\nn\\[0.1cm]
&+&2 (\text{SP2}-2 \text{SP3}) m_W^2)+\text{SP2} (D_{2}^{(1)}(m_W^2)+D_{23}^{(1)}(m_W^2)) (m_b^2-2 m_W^2)\nn\\[0.1cm]
&-&D_{1}^{(1)}(m_W^2) (\text{SP3} m_b^2-2 \text{SP2} m_W^2)-\text{SP3} ((D_{1}^{(1)}(m_{H^\pm}^2)+D_{12}^{(1)}(m_{H^\pm}^2)\nn\\[0.1cm]
&+&D_{13}^{(1)}(m_{H^\pm}^2)) \text{X}_{dd} m_b^2+D_{13}^{(1)}(m_W^2) (m_b^2-4 m_W^2)))), \\[0.3cm]
R_4^{\text{box}}&=&-(4 (2 \text{SP3}\,\text{SP5} T_dD_{123}^{(1)}(m_{H^\pm}^2)+(\text{SP3}\,\text{SP4}+(\text{SP2}-2 \text{SP3}) \text{SP5}) T_eD_{123}^{(2)}(m_{H^\pm}^2)\nn\\[0.1cm]
&+&\text{SP3}\,\text{SP5} (T_dD_{13}^{(1)}(m_{H^\pm}^2)-T_eD_{13}^{(2)}(m_{H^\pm}^2)+T_dD_{133}^{(1)}(m_{H^\pm}^2)-T_eD_{133}^{(2)}(m_{H^\pm}^2)))\nn\\[0.1cm]
&+&(4 \text{SP3}\,\text{SP4}-4 \text{SP2}\,\text{SP5}) (T_d(D_{22}^{(1)}(m_{H^\pm}^2)+D_{23}^{(1)}(m_{H^\pm}^2))+T_e(D_{22}^{(2)}(m_{H^\pm}^2)\nn\\[0.1cm]
&+&D_{23}^{(2)}(m_{H^\pm}^2)))+\text{SP1} ((T_d-T_e) (C_{0}^{(1)}(m_b^2)-C_{1}^{(1)}(m_b^2))-2 T_d(D_{00}^{(1)}(m_{H^\pm}^2)\nn\\[0.1cm]
&-&2 D_{001}^{(1)}(m_{H^\pm}^2))+2 T_e(D_{00}^{(2)}(m_{H^\pm}^2)-2 D_{001}^{(2)}(m_{H^\pm}^2))-(T_dD_{0}^{(1)}(m_{H^\pm}^2)\nn\\[0.1cm]
&-&T_eD_{0}^{(2)}(m_{H^\pm}^2)-T_dD_{1}^{(1)}(m_{H^\pm}^2)+T_eD_{1}^{(2)}(m_{H^\pm}^2)) (m_b^2-m_{H^\pm}^2))\nn\\[0.1cm]
&-&T_d((\text{SP4}-\text{SP5}) (4 \text{SP3} D_{113}^{(1)}(m_{H^\pm}^2)-4 (\text{SP2}-\text{SP3}) (D_{112}^{(1)}(m_{H^\pm}^2)\nn\\[0.1cm]
&+&D_{122}^{(1)}(m_{H^\pm}^2)))+4 (\text{SP3}\,\text{SP4}+\text{SP2}\,\text{SP5}) D_{123}^{(1)}(m_{H^\pm}^2)-D_{2}^{(1)}(m_{H^\pm}^2) (\text{SP1} m_t^2\nn\\[0.1cm]
&+&4 \text{SP3}\,\text{SP4}-4 \text{SP2}\,\text{SP5})-D_{12}^{(1)}(m_{H^\pm}^2) (-\text{SP1} m_t^2+4 \text{SP3}\,\text{SP5}+\text{SP2} (4 \text{SP4}\nn\\[0.1cm]
&-&8 \text{SP5})))+T_e((\text{SP2}-\text{SP3}) (4 \text{SP5} D_{113}^{(2)}(m_{H^\pm}^2)-4 (\text{SP4}-\text{SP5}) (D_{112}^{(2)}(m_{H^\pm}^2)\nn\\[0.1cm]
&+&D_{122}^{(2)}(m_{H^\pm}^2)))-D_{12}^{(2)}(m_{H^\pm}^2) (-\text{SP1} m_t^2+(4 \text{SP2}-8 \text{SP3}) \text{SP4}+4 \text{SP3}\,\text{SP5})\nn\\[0.1cm]
&+&D_{2}^{(2)}(m_{H^\pm}^2) (-\text{SP1} m_t^2+4 \text{SP3}\,\text{SP4}-4 \text{SP2}\,\text{SP5}))) \text{X}_{dd} m_b^2\nn\\[0.1cm]
&-&T_d(4 (\text{SP2}-\text{SP3}) (\text{SP4}-\text{SP5}) (D_{112}^{(1)}(m_W^2)+D_{122}^{(1)}(m_W^2)) (m_b^2+2 m_W^2)\nn\\[0.1cm]
&-&4 D_{22}^{(1)}(m_W^2) (\text{SP2}\,\text{SP5} m_b^2+2 (\text{SP3}\,\text{SP4}-\text{SP2}\,\text{SP5}) m_W^2)+D_{12}^{(1)}(m_W^2) ((-\text{SP1} m_t^2\nn\\[0.1cm]
&+&4 \text{SP3}\,\text{SP5}+\text{SP2} (4 \text{SP4}-8 \text{SP5})) m_b^2+2m_W^2(-\text{SP1} m_t^2+(4 \text{SP2}-8 \text{SP3}) \text{SP4}\nn\\[0.1cm]
&+&4 \text{SP3}\,\text{SP5}))+sD_{2}^{(1)}(m_W^2) (\text{SP1} (m_b^2+2 m_W^2) m_t^2+4 (\text{SP3}\,\text{SP4}-\text{SP2}\,\text{SP5}) (m_b^2\nn\\[0.1cm]
&-&2 m_W^2)))-4 (\text{SP3}\,\text{SP4} T_dD_{22}^{(1)}(m_W^2) m_b^2+\text{SP1} t_{12} T_eD_{2}^{(2)}(m_W^2) m_W^2\nn\\[0.1cm]
&+&(\text{SP3}\,\text{SP4}-\text{SP2}\,\text{SP5}) (T_eD_{22}^{(2)}(m_W^2)+T_dD_{23}^{(1)}(m_W^2)+T_eD_{23}^{(2)}(m_W^2)) (m_b^2-2 m_W^2)\nn\\[0.1cm]
&+&\text{SP5} (\text{SP3} (T_dD_{133}^{(1)}(m_W^2)-T_eD_{133}^{(2)}(m_W^2)) (m_b^2+2 m_W^2)-T_eD_{13}^{(2)}(m_W^2) (\text{SP3} m_b^2\nn\\[0.1cm]
&-&2 (2 \text{SP2}-\text{SP3}) m_W^2)))+4 ((m_b^2+2 m_W^2)(\text{SP3}\,\text{SP5} T_eD_{122}^{(2)}(m_W^2)+(\text{SP3}\,\text{SP4}\nn\\[0.1cm]
&+&\text{SP2}\,\text{SP5}) T_dD_{123}^{(1)}(m_W^2))+\text{SP3} ((\text{SP5} T_eD_{112}^{(2)}(m_W^2)+T_d((\text{SP4}-\text{SP5}) D_{113}^{(1)}(m_W^2)\nn\\[0.1cm]
&-&2 \text{SP5} D_{123}^{(1)}(m_W^2))-(\text{SP4}-2 \text{SP5}) T_eD_{123}^{(2)}(m_W^2)) (m_b^2+2 m_W^2)\nn\\[0.1cm]
&-&T_dD_{13}^{(1)}(m_W^2) (\text{SP5} m_b^2-2 (2 \text{SP4}-\text{SP5}) m_W^2)))-T_e(4 ((\text{SP3}\,\text{SP4}-\text{SP2} (\text{SP4}\nn\\[0.1cm]
&-&\text{SP5})) (D_{112}^{(2)}(m_W^2)+D_{122}^{(2)}(m_W^2))+\text{SP5} ((\text{SP2}-\text{SP3}) D_{113}^{(2)}(m_W^2)\nn\\[0.1cm]
&+&\text{SP2} D_{123}^{(2)}(m_W^2))) (m_b^2+2 m_W^2)-D_{12}^{(2)}(m_W^2) ((-\text{SP1} m_t^2+(4 \text{SP2}-8 \text{SP3}) \text{SP4}\nn\\[0.1cm]
&+&4 \text{SP3}\,\text{SP5}) m_b^2+2 (-\text{SP1} m_t^2+4 \text{SP3}\,\text{SP5}+\text{SP2} (4 \text{SP4}-8 \text{SP5})) m_W^2)\nn\\[0.1cm]
&+&D_{2}^{(2)}(m_W^2) (4 (\text{SP3}\,\text{SP4}-\text{SP2}\,\text{SP5}) (m_b^2-2 m_W^2)-\text{SP1} m_t^2 (m_b^2+6 m_W^2))\nn\\[0.1cm]
&+&\text{SP1} t (-(D_{1}^{(2)}(m_{H^\pm}^2)-D_{11}^{(2)}(m_{H^\pm}^2)+D_{13}^{(2)}(m_{H^\pm}^2)) \text{X}_{dd} m_b^2+4 D_{2}^{(2)}(m_W^2) m_W^2\nn\\[0.1cm]
&+&(D_{11}^{(2)}(m_W^2)-D_{13}^{(2)}(m_W^2)) (m_b^2+2 m_W^2)-D_{1}^{(2)}(m_W^2) (m_b^2-2 m_W^2)))\nn\\[0.1cm]
&+&\text{SP1} (2 (s_{23}+t) (T_dD_{3}^{(1)}(m_W^2)-T_eD_{3}^{(2)}(m_W^2)) m_W^2-s_{23} (4 T_e(D_{1}^{(2)}(m_W^2)\nn\\[0.1cm]
&+&D_{2}^{(2)}(m_W^2)) m_W^2+T_d((D_{1}^{(1)}(m_{H^\pm}^2)-D_{11}^{(1)}(m_{H^\pm}^2)+D_{13}^{(1)}(m_{H^\pm}^2)) \text{X}_{dd} m_b^2\nn\\[0.1cm]
&+&(D_{1}^{(1)}(m_W^2)-D_{11}^{(1)}(m_W^2)+D_{13}^{(1)}(m_W^2)) (m_b^2+2 m_W^2))))-\text{SP1} ((T_dD_{0}^{(1)}(m_{H^\pm}^2)\nn\\[0.1cm]
&-&T_eD_{0}^{(2)}(m_{H^\pm}^2)) m_b^2 \text{X}_{du} m_t^2+((T_d-T_e) C_{0}^{(1)}(m_b^2)-T_dC_{1}^{(1)}(m_b^2)) (m_b^2+2 m_W^2)\nn\\[0.1cm]
&-&T_d((m_b^2+2 m_W^2) (2 D_{00}^{(1)}(m_W^2)-4 D_{001}^{(1)}(m_W^2)-D_{1}^{(1)}(m_W^2) (m_b^2-m_W^2))\nn\\[0.1cm]
&+&D_{0}^{(1)}(m_W^2) (m_b^4-(m_t^2-m_W^2) m_b^2-2 m_W^4))+T_e((C_{1}^{(1)}(m_b^2)+2 D_{00}^{(2)}(m_W^2)\nn\\[0.1cm]
&-&4 D_{001}^{(2)}(m_W^2)) (m_b^2+2 m_W^2)+D_{0}^{(2)}(m_W^2) (m_b^4-(m_t^2-m_W^2) m_b^2-2 m_W^4)\nn\\[0.1cm]
&-&D_{1}^{(2)}(m_W^2) (m_b^4+m_W^2 (m_b^2-2 m_W^2+4 (s_{23}+t))))), \\[0.3cm]
R_5^{\text{box}}&=&m_t ((-(T_dD_{2}^{(1)}(m_{H^\pm}^2)-T_eD_{2}^{(2)}(m_{H^\pm}^2)) m_t^2+T_dD_{0}^{(1)}(m_{H^\pm}^2) (m_b^2-m_{H^\pm}^2)\nn\\[0.1cm]
&+&T_eD_{0}^{(2)}(m_{H^\pm}^2) (-m_b^2+m_{H^\pm}^2+m_t^2)) \text{X}_{du} m_b^2+((T_dD_{22}^{(1)}(m_W^2)-T_eD_{22}^{(2)}(m_W^2)) m_t^2\nn\\[0.1cm]
&+&(T_d-T_e) C_{2}^{(1)}(m_b^2)) (m_b^2-2 m_W^2)+T_e((C_{0}^{(1)}(m_b^2)-C_{1}^{(1)}(m_b^2)) m_b^2\nn\\[0.1cm]
&+&D_{0}^{(2)}(m_W^2) (-m_b^2+m_W^2+m_t^2) m_b^2+2 C_{1}^{(1)}(m_b^2) m_W^2-(D_{12}^{(2)}(m_W^2) m_t^2\nn\\[0.1cm]
&+&4 D_{00}^{(2)}(m_W^2)) (m_b^2-2 m_W^2)+D_{1}^{(2)}(m_W^2) (m_b^4-3 m_W^2 m_b^2+2 m_W^4)\nn\\[0.1cm]
&+&D_{2}^{(2)}(m_W^2) (m_b^4+(m_t^2-3 m_W^2) m_b^2+2 m_W^4))-T_d(-2 D_{0}^{(1)}(m_W^2) (m_b^2-m_W^2) m_W^2\nn\\[0.1cm]
&+&2 D_{00}^{(1)}(m_W^2) (m_b^2-2 m_W^2)+D_{2}^{(1)}(m_W^2) (m_b^4-m_W^2 (3 m_b^2-2 (m_W^2\nn\\[0.1cm]
&+&4 (s_{23}+t)+t_{12}))))+(((T_dD_{22}^{(1)}(m_{H^\pm}^2)-T_eD_{22}^{(2)}(m_{H^\pm}^2)) m_t^2+T_dC_{0}^{(1)}(m_b^2)\nn\\[0.1cm]
&-&T_eC_{1}^{(1)}(m_b^2)+(T_d-T_e) C_{2}^{(1)}(m_b^2)+s_{23} T_d(D_{1}^{(1)}(m_{H^\pm}^2)+D_{12}^{(1)}(m_{H^\pm}^2)-D_{23}^{(1)}(m_{H^\pm}^2))\nn\\[0.1cm]
&+&T_e((D_{1}^{(2)}(m_{H^\pm}^2)+D_{2}^{(2)}(m_{H^\pm}^2)) (m_b^2-m_{H^\pm}^2)-t (D_{11}^{(2)}(m_{H^\pm}^2)+D_{12}^{(2)}(m_{H^\pm}^2)\nn\\[0.1cm]
&-&D_{13}^{(2)}(m_{H^\pm}^2)-D_{23}^{(2)}(m_{H^\pm}^2)))) m_b^2-(T_d(2 D_{00}^{(1)}(m_{H^\pm}^2)+D_{0}^{(1)}(m_{H^\pm}^2) (m_b^2-m_{H^\pm}^2))\nn\\[0.1cm]
&+&T_e(D_{12}^{(2)}(m_{H^\pm}^2) m_t^2+4 D_{00}^{(2)}(m_{H^\pm}^2))) m_b^2-T_dD_{2}^{(1)}(m_{H^\pm}^2) (m_b^4-m_b^2 (m_{H^\pm}^2\nn\\[0.1cm]
&+&m_t^2))) \text{X}_{dd}-C_{0}^{(1)}(m_b^2) ((T_d-T_e) \text{X}_{du} m_b^2+2 T_dm_W^2)-s_{23} (2 T_eD_{3}^{(2)}(m_W^2) m_W^2\nn\\[0.1cm]
&+&T_d(-(D_{0}^{(1)}(m_W^2)+(D_{0}^{(1)}(m_{H^\pm}^2)-D_{1}^{(1)}(m_{H^\pm}^2)+D_{3}^{(1)}(m_{H^\pm}^2)) \text{X}_{du}) m_b^2\nn\\[0.1cm]
&+&2 (D_{1}^{(1)}(m_W^2)-3 D_{2}^{(1)}(m_W^2)) m_W^2-(D_{12}^{(1)}(m_W^2)-D_{23}^{(1)}(m_W^2)\nn\\[0.1cm]
&+&D_{3}^{(1)}(m_W^2)) (m_b^2-2 m_W^2)))+t (6 T_dD_{2}^{(1)}(m_W^2) m_W^2+T_e(-(D_{0}^{(2)}(m_W^2)-D_{1}^{(2)}(m_W^2)\nn\\[0.1cm]
&+&D_{11}^{(2)}(m_W^2)+D_{3}^{(2)}(m_W^2)+(D_{0}^{(2)}(m_{H^\pm}^2)-D_{1}^{(2)}(m_{H^\pm}^2)+D_{3}^{(2)}(m_{H^\pm}^2)) \text{X}_{du}) m_b^2\nn\\[0.1cm]
&+&2 D_{11}^{(2)}(m_W^2) m_W^2-(D_{12}^{(2)}(m_W^2)-D_{13}^{(2)}(m_W^2)-D_{23}^{(2)}(m_W^2)) (m_b^2-2 m_W^2)))), \\[0.3cm]
R_6^{\text{box}}&=&m_b^2 (-\text{SP5} (\text{X}_{du} (T_d(D_{0}^{(1)}(m_{H^\pm}^2)+D_{3}^{(1)}(m_{H^\pm}^2))+T_e(D_{0}^{(2)}(m_{H^\pm}^2)+D_{3}^{(2)}(m_{H^\pm}^2)))\nn\\[0.1cm]
&+&T_dD_{0}^{(1)}(m_W^2)+T_eD_{0}^{(2)}(m_W^2))+\text{X}_{du} (\text{SP4}-\text{SP5}) (T_d(D_{1}^{(1)}(m_{H^\pm}^2)+D_{2}^{(1)}(m_{H^\pm}^2))\nn\\[0.1cm]
&+&T_eD_{2}^{(2)}(m_{H^\pm}^2))-\text{X}_{dd} ((\text{SP4}-\text{SP5}) (T_d(D_{12}^{(1)}(m_{H^\pm}^2)+D_{2}^{(1)}(m_{H^\pm}^2)+D_{22}^{(1)}(m_{H^\pm}^2))\nn\\[0.1cm]
&+&T_e(D_{2}^{(2)}(m_{H^\pm}^2)+D_{22}^{(2)}(m_{H^\pm}^2)))-\text{SP5} (T_dD_{23}^{(1)}(m_{H^\pm}^2)+T_eD_{23}^{(2)}(m_{H^\pm}^2))\nn\\[0.1cm]
&+&T_e((\text{SP4}-\text{SP5}) D_{12}^{(2)}(m_{H^\pm}^2)-\text{SP5} D_{13}^{(2)}(m_{H^\pm}^2)))+T_d((\text{SP4}-\text{SP5}) D_{1}^{(1)}(m_W^2)\nn\\[0.1cm]
&-&\text{SP4} D_{12}^{(1)}(m_W^2)))+(m_b^2-2 m_W^2) (\text{SP5} (T_d(D_{23}^{(1)}(m_W^2)-D_{3}^{(1)}(m_W^2))\nn\\[0.1cm]
&+&T_e(D_{22}^{(2)}(m_W^2)+D_{23}^{(2)}(m_W^2)-D_{3}^{(2)}(m_W^2)))+T_e(\text{SP5} (D_{12}^{(2)}(m_W^2)+D_{13}^{(2)}(m_W^2))\nn\\[0.1cm]
&-&\text{SP4} (D_{12}^{(2)}(m_W^2)+D_{22}^{(2)}(m_W^2))))+T_dD_{12}^{(1)}(m_W^2) (\text{SP5} m_b^2+2 m_W^2 (\text{SP4}-\text{SP5}))\nn\\[0.1cm]
&+&(\text{SP4}-\text{SP5}) (2 m_W^2 (T_dD_{2}^{(1)}(m_W^2)+T_eD_{2}^{(2)}(m_W^2))-T_d(m_b^2-2 m_W^2) D_{22}^{(1)}(m_W^2)), \\[0.3cm]
R_7^{\text{box}}&=&m_b^2 (\text{SP3} (\text{X}_{du} (T_d(D_{0}^{(1)}(m_{H^\pm}^2)+D_{3}^{(1)}(m_{H^\pm}^2))+T_e(D_{0}^{(2)}(m_{H^\pm}^2)+D_{3}^{(2)}(m_{H^\pm}^2)))\nn\\[0.1cm]
&-&\text{X}_{dd} (T_dD_{23}^{(1)}(m_{H^\pm}^2)+T_eD_{23}^{(2)}(m_{H^\pm}^2)))+\text{SP3} (T_dD_{0}^{(1)}(m_W^2)+T_eD_{0}^{(2)}(m_W^2))\nn\\[0.1cm]
&+&(\text{SP2}-\text{SP3}) (\text{X}_{dd} (T_dD_{22}^{(1)}(m_{H^\pm}^2)+T_e(D_{12}^{(2)}(m_{H^\pm}^2)+D_{2}^{(2)}(m_{H^\pm}^2)+D_{22}^{(2)}(m_{H^\pm}^2)))\nn\\[0.1cm]
&-&\text{X}_{du} (T_dD_{2}^{(1)}(m_{H^\pm}^2)+T_eD_{1}^{(2)}(m_{H^\pm}^2)+T_eD_{2}^{(2)}(m_{H^\pm}^2)))\nn\\[0.1cm]
&+&T_d(\text{X}_{dd} ((\text{SP2}-\text{SP3}) D_{12}^{(1)}(m_{H^\pm}^2)-\text{SP3} D_{13}^{(1)}(m_{H^\pm}^2)+(\text{SP2}-\text{SP3}) D_{2}^{(1)}(m_{H^\pm}^2))\nn\\[0.1cm]
&+&\text{SP2} D_{12}^{(1)}(m_W^2))-T_e(\text{SP2}-\text{SP3}) D_{1}^{(2)}(m_W^2))+(m_b^2-2 m_W^2) (\text{SP3} (-T_d(D_{13}^{(1)}(m_W^2)\nn\\[0.1cm]
&+&D_{23}^{(1)}(m_W^2))+T_dD_{3}^{(1)}(m_W^2)-T_e(D_{12}^{(2)}(m_W^2)+D_{22}^{(2)}(m_W^2)+D_{23}^{(2)}(m_W^2))\nn\\[0.1cm]
&+&T_eD_{3}^{(2)}(m_W^2))+\text{SP2} T_e(D_{12}^{(2)}(m_W^2)+D_{22}^{(2)}(m_W^2)))-T_dD_{12}^{(1)}(m_W^2) (\text{SP3} m_b^2\nn\\[0.1cm]
&+&2 m_W^2 (\text{SP2}-\text{SP3}))-(\text{SP2}-\text{SP3}) (2 m_W^2 (T_dD_{2}^{(1)}(m_W^2)+T_eD_{2}^{(2)}(m_W^2))\nn\\[0.1cm]
&-&T_d(m_b^2-2 m_W^2) D_{22}^{(1)}(m_W^2)), \\[0.3cm]
R_8^{\text{box}}&=&m_b^2 \text{X}_{dd} (-((T_d+T_e) (C_{0}^{(1)}(m_b^2)-C_{1}^{(1)}(m_b^2))-(m_b^2-m_{H^\pm}^2) (T_dD_{0}^{(1)}(m_{H^\pm}^2)\nn\\[0.1cm]
&-&T_dD_{1}^{(1)}(m_{H^\pm}^2)+T_eD_{0}^{(2)}(m_{H^\pm}^2)-T_eD_{1}^{(2)}(m_{H^\pm}^2))-6 (T_dD_{00}^{(1)}(m_{H^\pm}^2)\nn\\[0.1cm]
&+&T_eD_{00}^{(2)}(m_{H^\pm}^2))+m_t^2 (-(T_dD_{12}^{(1)}(m_{H^\pm}^2)-T_dD_{2}^{(1)}(m_{H^\pm}^2)+T_eD_{12}^{(2)}(m_{H^\pm}^2)\nn\\[0.1cm]
&-&T_eD_{2}^{(2)}(m_{H^\pm}^2)))))-(m_b^2-2 m_W^2) ((T_d+T_e) (C_{0}^{(1)}(m_b^2)-C_{1}^{(1)}(m_b^2))\nn\\[0.1cm]
&-&6 (T_dD_{00}^{(1)}(m_W^2)+T_eD_{00}^{(2)}(m_W^2)))-m_b^2 \text{X}_{du} m_t^2 (T_dD_{0}^{(1)}(m_{H^\pm}^2)+T_eD_{0}^{(2)}(m_{H^\pm}^2))\nn\\[0.1cm]
&+&(-m_b^2 (m_t^2+3 m_W^2)+m_b^4+2 m_W^4) (T_dD_{0}^{(1)}(m_W^2)+T_eD_{0}^{(2)}(m_W^2))\nn\\[0.1cm]
&-&T_d(-3 m_b^2 m_W^2+m_b^4+2 m_W^4) D_{1}^{(1)}(m_W^2)+s_{23} (4 T_em_W^2 D_{1}^{(2)}(m_W^2)\nn\\[0.1cm]
&-&T_d(m_b^2 \text{X}_{dd} (D_{1}^{(1)}(m_{H^\pm}^2)-D_{11}^{(1)}(m_{H^\pm}^2)+D_{13}^{(1)}(m_{H^\pm}^2))+(m_b^2-2 m_W^2) (D_{1}^{(1)}(m_W^2)\nn\\[0.1cm]
&-&D_{11}^{(1)}(m_W^2)+D_{13}^{(1)}(m_W^2))))+m_t^2 ((m_b^2-2 m_W^2) (T_dD_{12}^{(1)}(m_W^2)-T_dD_{2}^{(1)}(m_W^2)\nn\\[0.1cm]
&+&T_eD_{12}^{(2)}(m_W^2)-T_eD_{2}^{(2)}(m_W^2))+2 T_dm_W^2 D_{3}^{(1)}(m_W^2))-m_W^2 (2 t_{12} T_dD_{3}^{(1)}(m_W^2)\nn\\[0.1cm]
&-&2 T_e(s_{23}+t) D_{3}^{(2)}(m_W^2))-t T_e(m_b^2 \text{X}_{dd} (D_{1}^{(2)}(m_{H^\pm}^2)-D_{11}^{(2)}(m_{H^\pm}^2)+D_{13}^{(2)}(m_{H^\pm}^2))\nn\\[0.1cm]
&+&(m_b^2-6 m_W^2) D_{1}^{(2)}(m_W^2)-(m_b^2-2 m_W^2) (D_{11}^{(2)}(m_W^2)-D_{13}^{(2)}(m_W^2)))\nn\\[0.1cm]
&-&T_eD_{1}^{(2)}(m_W^2) (m_W^2 (-3 m_b^2+2 m_W^2+4 (s_{23}+t))+m_b^4).
\end{eqnarray}
Finally, for the self-energy part (cf. Eqs.~\eqref{eq:Amptcggself}), we get
\begin{eqnarray}
R_1^{\text{self}}&=&-((\text{SP4}-\text{SP5}) F_3+2 \text{SP3}\,F_4-\text{SP1}\,F_5+F_{16}) ((B_{0}^{(2)}(m_W^2)+B_{1}^{(2)}(m_W^2)\nn\\[0.1cm]
&+&B_{1}^{(2)}(m_{H^\pm}^2) \text{X}_{dd}+B_{0}^{(2)}(m_{H^\pm}^2) \text{X}_{du}) m_b^2+(2 B_{1}^{(2)}(m_W^2)+1) m_W^2),\\[0.3cm]
R_2^{\text{self}}&=&((\text{SP4}-\text{SP5}) F_3+2 \text{SP3}\,F_4-\text{SP1}\,F_5+F_{16}+(\text{SP1}\,F_1-F_{12}) m_t) (B_{0}^{(1)}(m_W^2)\nn\\[0.1cm]
&+&B_{0}^{(1)}(m_{H^\pm}^2) \text{X}_{du}),\\[0.3cm]
R_3^{\text{self}}&=& ((\text{SP4}-\text{SP5}) F_3-2 \text{SP2}\,F_4-\text{SP1}\,F_5-F_{16}+(\text{SP1}\,F_1+F_{12}) m_t) ((B_{0}^{(2)}(m_W^2)\nn\\[0.1cm]
&+&B_{1}^{(2)}(m_W^2)+B_{1}^{(2)}(m_{H^\pm}^2) \text{X}_{dd}+B_{0}^{(2)}(m_{H^\pm}^2) \text{X}_{du}) m_b^2+(2 B_{1}^{(2)}(m_W^2)+1) m_W^2),\\[0.3cm]
R_4^{\text{self}}&=&-((\text{SP4}-\text{SP5}) F_3-2 \text{SP2}\,F_4-\text{SP1}\,F_5-F_{16})(B_{0}^{(1)}(m_W^2)+B_{0}^{(1)}(m_{H^\pm}^2) \text{X}_{du}),\\[0.3cm]
R_5^{\text{self}}&=&-(2 (\text{SP4}-\text{SP5}) F_3-2 ((\text{SP2}-\text{SP3}) F_4+\text{SP1}\,F_5)+\text{SP1}\,F_1 m_t) ((2 B_{1}^{(2)}\nn\\[0.1cm]
&~&(m_W^2)+1) m_W^2-m_b^2 (B_{0}^{(1)}(m_W^2)-B_{0}^{(2)}(m_W^2)-B_{1}^{(2)}(m_W^2)-B_{1}^{(2)}(m_{H^\pm}^2) \text{X}_{dd}\nn\\[0.1cm]
&+&(B_{0}^{(1)}(m_{H^\pm}^2)-B_{0}^{(2)}(m_{H^\pm}^2)) \text{X}_{du})),\\[0.3cm]
R_6^{\text{self}}&=& -(t+t_{12}) ((\text{SP4}-\text{SP5}) F_3+2 \text{SP3}\,F_4-\text{SP1}\,F_5+F_{16}) (B_{0}^{(3)}(m_W^2)\nn\\[0.1cm]
&+&B_{0}^{(3)}(m_{H^\pm}^2) \text{X}_{du}) m_b^2-s_{23} ((\text{SP4}-\text{SP5}) F_3+2 \text{SP3}\,F_4-\text{SP1}\,F_5+F_{16}\nn\\[0.1cm]
&+&(\text{SP1}\,F_1-F_{12}) m_t) ((B_{0}^{(3)}(m_W^2)+B_{1}^{(3)}(m_W^2)+B_{1}^{(3)}(m_{H^\pm}^2) \text{X}_{dd}\nn\\[0.1cm]
&+&B_{0}^{(3)}(m_{H^\pm}^2) \text{X}_{du}) m_b^2+(2 B_{1}^{(3)}(m_W^2)+1) m_W^2),\\[0.3cm]
R_7^{\text{self}}&=& m_b^2 m_t^2 ((\text{SP4}-\text{SP5}) F_3-2 \text{SP2}\,F_4-\text{SP1}\,F_5-F_{16}+(\text{SP1}\,F_1+F_{12}) m_t)\nn\\[0.1cm]
&~& (B_{0}^{(4)}(m_W^2)+B_{0}^{(4)}(m_{H^\pm}^2) \text{X}_{du})+t (((\text{SP4}-\text{SP5}) F_3-2 \text{SP2}\,F_4-\text{SP1}\,F_5-F_{16})\nn\\[0.1cm]
&~& (B_{1}^{(4)}(m_{H^\pm}^2) \text{X}_{dd} m_b^2+m_W^2+B_{1}^{(4)}(m_W^2) (m_b^2+2 m_W^2))\nn\\[0.1cm]
&-&(\text{SP1}\,F_1+F_{12}) m_b^2 m_t (B_{0}^{(4)}(m_W^2)+B_{0}^{(4)}(m_{H^\pm}^2) \text{X}_{du})).
\end{eqnarray}
Here the scalar functions $B_{\alpha}^{(k)}$, $C_{\beta}^{(k)}$, and $D_{\lambda}^{(k)}$ are defined, respectively, as
\begin{eqnarray}
B_{\alpha}^{(1)}(m^2) &=& B_{\alpha}(0,m_b^2,m^2),\;\;\;
B_{\alpha}^{(2)}(m^2) = B_{\alpha}(m_t^2,m_b^2,m^2),\;\;\;
B_{\alpha}^{(3)}(m^2) = B_{\alpha}(s_{23},m_b^2,m^2),\nn\\[0.1cm]
B_{\alpha}^{(4)}(m^2) &=& B_{\alpha}(t,m_b^2,m^2),\;\;\;\;
B_{\alpha}^{(5)}(m^2) = B_{\alpha}(t_{12},m_b^2,m^2),\\[0.2cm]
C_\beta^{(1)}(m^2) &=& C_\beta(0,0,t_{12},m_b^2,m_b^2,m^2),\;\;\;\;\;
C_\beta^{(2)}(m^2) = C_\beta(0,s_{23},0,m_b^2,m_b^2,m^2),\nn\\[0.1cm]
C_\beta^{(3)}(m^2) &=& C_\beta(0,s_{23},m_t^2,m_b^2,m_b^2,m^2),\;\;
C_\beta^{(4)}(m^2) = C_\beta(0,t,0,m_b^2,m_b^2,m^2),\nn\\[0.1cm]
C_\beta^{(5)}(m^2) &=&C_\beta(0,t,m_t^2,m_b^2,m_b^2,m^2),\;\;\;\;\;
C_\beta^{(6)}(m^2) = C_\beta(t_{12},m_t^2,0,m_b^2,m_b^2,m^2),\\[0.2cm]
D_\lambda^{(1)}(m^2) &=& D_\lambda(0,0,m_t^2,0,t_{12},s_{23},m_b^2,m_b^2,m_b^2,m^2),\nn\\[0.1cm]
D_\lambda^{(2)}(m^2) &=& D_\lambda(0,0,m_t^2,0,t_{12},t,m_b^2,m_b^2,m_b^2,m^2),
\end{eqnarray}
where
\begin{eqnarray}
\alpha&=&0,1,\nn\\[0.1cm]
\beta&=&0,00,1,12,2,22,\nn\\[0.1cm]
\lambda &=& 0,00,001,002,003,1,11,112,113,12,122,123,13,133,2,22,222,223,23,233,3,33. \nn
\end{eqnarray}
These functions can be further decomposed into the basic scalar one-loop integrals $A_0$, $B_0$, $C_0$, and $D_0$; see Refs.~\cite{Denner:2019vbn,Denner:1991kt,Passarino:1978jh,tHooft:1978jhc} for details. 

\bibliographystyle{JHEP}
\bibliography{reference}

\end{document}